\documentclass[a4paper,fleqn,usenatbib,useAMS]{mnras}

\usepackage{graphicx}	% Including figure files
\usepackage{amsmath}	% Advanced maths commands
\usepackage{amssymb}	% Extra maths symbols
\usepackage{multicol}        % Multi-column entries in tables
\usepackage{bm}		% Bold maths symbols, including upright Greek
\usepackage{pdflscape}	% Landscape pages
\usepackage{subcaption}
\usepackage{nicefrac}
\usepackage{caption}
\usepackage{subcaption}
\usepackage{dcolumn}
\usepackage{array,booktabs}
\usepackage{bbm}
\usepackage{color, colortbl}
\definecolor{Gray}{gray}{0.9}

\newcommand{\vect}[1]{\boldsymbol{#1}}
\newcolumntype{P}[1]{>{\raggedright\arraybackslash}p{#1}}

\usepackage[T1]{fontenc}
\usepackage{ae,aecompl}

\usepackage{newtxtext,newtxmath}
\usepackage{tabularx}
\usepackage{makecell} % multi-row cells
\usepackage{newtxmath} % fancy 1 via \vmathbb{}

\title[Position angle standards]{Polarization position angle standard stars: a reassessment of {$\theta$} and its variability for seventeen stars based on a decade of observations\vspace{-0.35cm}}
\author[D. V. Cotton \textit{et al.}]{Daniel V. Cotton$^{1,2}$\thanks{Contact e-mail: \href{mailto:dc@mira.org}{dc@mira.org}}, Jeremy Bailey$^{3,2}$, Lucyna Kedziora-Chudczer$^{4}$, 
\newauthor Kimberly Bott$^{5,6,7}$, Ain De Horta$^{2}$\thanks{Contact e-mail: \href{mailto:dc@mira.org}{a.dehorta@westernsydney.edu.au}}, Normandy Filcek$^{1,8,9}$, Jonathan P. Marshall$^{10}$,
\newauthor Graeme Melville$^{11}$, Derek L. Buzasi$^{12}$, Ievgeniia Boiko$^{1,13,14}$, Nicholas W. Borsato$^{15,16}$,
\newauthor Jean Perkins$^{1}$, Daniela Opitz$^{17}$, Shannon Melrose$^{18, 3}$, Gesa Gr\"uning$^{19}$,
\newauthor Dag Evensberget$^{20,4}$ and Jinglin Zhao$^{21}$ \\
{}\\
{$^1$Monterey Institute for Research in Astronomy, 200 Eighth Street, Marina, CA, 93933, USA.}\\
{$^2$Western Sydney University, Locked Bag 1797, Penrith-South DC, NSW 1797, Australia.}\\
{$^3$School of Physics, UNSW Sydney, New South Wales, 2052, Australia.}\\
{$^4$Centre for Astrophysics, University of Southern Queensland, Toowoomba, Queensland 4350, Australia.}\\
{$^5$Department of Earth and Planetary Science, University of California, Riverside, CA, 92521, USA.}\\
{$^6$NASA Nexus for Exoplanet System Science, Virtual Planetary Laboratory, Bldg. 3910, 15th Ave NE, University of Washington, Seattle, WA 98195, USA.}\\
{$^7$NASA Nexus for Exoplanet System Science, Terrestrial Polarisation Team, 4111 Libra Drive, University of Central Florida, Orlando, FL 32826, USA.}\\
{$^{8}$York School, 9501 York Rd, Monterey, CA 93940, USA.}\\
{$^{9}$Willamette University, 900 State St, Salem, OR 97301, USA.}\\
{$^{10}$Academia Sinica Institute of Astronomy and Astrophysics, 11F of AS/NTU Astronomy-Mathematics Bldg, No. 1, Sect. 4, Roosevelt Rd, Taipei 106216, Taiwan.}\\
{$^{11}$School of Physics, University of Wollongong, NSW, 2522, Australia.}\\
{$^{12}$Department of Chemistry \& Physics, Florida Gulf Coast University, 10501 FGCU Boulevard S., Fort Myers, FL 33965, USA.}\\
{$^{13}$Monterey Peninsula College, 980 Fremont Street, Monterey, CA 93940, USA.}\\
{$^{14}$California State University, Long Beach, 1250 Bellflower Blvd, Long Beach, CA 90840, USA.}\\
{$^{15}$Lund Observatory, Division of Astrophysics, Department of Physics, Lund University, Box 43, SE-221 00 Lund, Sweden.}\\
{$^{16}$School of Mathematical and Physical Sciences, Macquarie University, Sydney, NSW 2151, Australia.}\\
{$^{17}$Data Science Institute, Faculty of Engineering, Universidad del Desarrollo, Av. Plaza 680, Las Condes, Santiago, Chile}\\
{$^{18}$UNSW College, Building L5, UNSW Sydney Campus, 223 Anzac Parade, Kensington, NSW 2033, Australia.}\\
{$^{19}$Department of Physics, Carl von Ossietzky University Oldenburg, Carl-von-Ossietzky-Str. 9-11, 26129 Oldenburg, Germany.}\\
{$^{20}$Leiden Observatory, Leiden University, PO Box 9513, 2300 RA Leiden, The Netherlands.}\\
{$^{21}$Department of Astronomy \& Astrophysics, The Pennsylvania State University, 525 Davey Lab, University Park, PA 16802, USA.}\vspace{-0.5cm}}

\date{Last updated \today; in original form \today \vspace{-0.35cm}}

\pubyear{2024}
\begin{document}
\label{firstpage}
\pagerange{\pageref{firstpage}--\pageref{lastpage}}
\maketitle

% Abstract of the paper
\begin{abstract}
Observations of polarization position angle ($\theta$) standards made from 2014 to 2023 with the High Precision Polarimetric Instrument (HIPPI) and other HIPPI-class polarimeters in both hemispheres are used to investigate their variability. Multi-band data were first used to thoroughly recalibrate the instrument performance by bench-marking against carefully selected literature data. A novel Co-ordinate Difference Matrix (CDM) approach -- which combines pairs of points -- was then used to amalgamate monochromatic ($g^\prime$ band) observations from many observing runs and re-determine $\theta$ for 17 standard stars. The CDM algorithm was then integrated into a fitting routine and used to establish the impact of stellar variability on the measured position angle scatter. The approach yields variability detections for stars on long time scales that appear stable over short runs. The best position angle standards are $\ell$~Car, $o$~Sco, HD\,154445, HD\,161056 and $\iota^1$~Sco which are stable to $\leq$ 0.123$^\circ$. Position angle variability of 0.27--0.82$^\circ$, significant at the 3-$\sigma$ level, is found for 5 standards, including the Luminous Blue Variable HD~160529 and all but one of the other B/A-type supergiants (HD~80558, HD~111613, HD~183143 and 55~Cyg), most of which also appear likely to be variable in polarization magnitude ($p$) -- there is no preferred orientation for the polarization in these objects, which are all classified as $\alpha$ Cygni variables.  Despite this we make six key recommendations for observers -- relating to data acquisition, processing and reporting -- that will allow them to use these standards to achieve \mbox{$<$ 0.1$^\circ$} precision in the telescope position angle with similar instrumentation, and allow data sets to be combined more accurately. 

\end{abstract}

% Select between one and six entries from the list of approved keywords.
% Don't make up new ones.
\begin{keywords}
techniques: polarimetric; stars: supergiants; instrumentation: polarimeters \vspace{-0.85cm}
\end{keywords}

%%%%%%%%%%%%%%%%%%%%%%%%%%%%%%%%%%%%%%%%%%%%%%%%%%

%%%%%%%%%%%%%%%%% BODY OF PAPER %%%%%%%%%%%%%%%%%%
%\clearpage
%\pagebreak

\section{Introduction}
\label{sec:intro}

The 21st Century has seen the advent of broadband optical polarimeters capable of a precision of 10 parts-per-million or better. Their development was sparked by the hunt for exoplanet signatures (e.g. \citealp{Hough06, Wiktorowicz08, Piirola14, Bailey15}) but instead lead to the discovery of new and predicted stellar polarigenic mechanisms, such as rapid rotation \citep{Cotton17, Bailey20, Lewis22, Howarth23}, binary photospheric reflection \citep{Bailey19, Cotton20}, linear polarization from global magnetic fields \citep{Cotton17, Cotton19b}, and non-radial pulsations \citep{Cotton22a}. Precise maps of interstellar polarization close to the Sun are now possible \citep{Cotton16a, Piirola20}, and inferences have been made about the nature of hot dust \citep{marshall16}, debris disks \citep{Marshall20, Marshall23}, and even the heliosphere \citep{Frisch22}. Higher precision studies of known phenomena are also revealing new details about such diverse topics as asteroids \citep{Wiktorowicz15}, gas entrained between binary stars \citep{Berdyugin18}, the nature of the interstellar medium \citep{Cotton19}, and extreme variable stars \citep{Bailey24}. Alongside this progress, the dream of detecting and characterising exoplanet atmospheres with polarimetry remains a live ambition \citep{Bailey21, Bott22, Wiktorowicz24}. The development of new instruments continues at pace, both for medium to very large sized telescopes \citep{Wiktorowicz15, Bailey20, Piirola21} and even amateur-sized telescopes \citep{Bailey17, Bailey23}. 

Despite the ground-breaking improvements in instrumental precision, polarimetric observations of objects at increasing distance are naturally affected by the interstellar polarization background. The detection of small polarization signals from distant objects is therefore critically dependant upon the accurate calibration of the polarization position angle -- a craft that has not progressed at the same rate. We aim to address this issue here.

Linear polarization is defined either in terms of normalised Stokes parameters $q=Q/I$ and $u=U/I$ (typically measured in per cent: $10^{-2}$, or parts-per-million, ppm: $10^{-6}$), or as polarization magnitude, \begin{equation}p=\sqrt{q^2+u^2},\label{eq:p}\end{equation} and position angle, \begin{equation}\theta=\tfrac{1}{2}\tan^{-1}(u/q),\label{eq:theta}\end{equation} measured North over East, relative to the North Celestial Pole ($\theta_0$), i.e. in the Equatorial system. Polarimetric data is almost universally reported in either or both of these co-ordinate frames, but collected in an instrument frame, ($q_i$, $u_i$), and then rotated according to, \begin{equation}q=q_i \cos(2\theta_t) + u_i \sin(2\theta_t),\end{equation} and \begin{equation}u=u_i \cos(2\theta_t)-q_i \sin(2\theta_t),\end{equation} where $\theta_t$, usually called the telescope position angle, is the difference between the instrument reference axis and $\theta_0$ -- which is readily accessible in astrometry but not polarimetry \citep{vdKamp67, Hsu82}\footnote{\citet{Serkowski74b} summarises some alternative methods of finding $\theta_0$, mostly involving polarizers carefully aligned to the horizon mounted external to the telescope, however \citet{Hsu82} infer the accuracy of these methods is not better than 1$^\circ$.}. Instead polarimetrists often have to determine $\theta_t$ by reference to high polarization standard stars \citep{Serkowski74, Serkowski74b}. For this purpose, $\theta_t$ must be re-determined for every observing run (and whenever the equipment is disturbed) to reflect the current condition of the instrument and telescope. It is also a difficult task to perform with precision and accuracy, since the available calibration stars vary with observing location and season. Indeed, there can sometimes be no established standards in the sky bright enough for polarimetry on the smallest telescopes (e.g. the $<$10-inch telescopes used by \citealp{Bailey23} and \citealp{Bailey24}).

Despite some standards apparently having $\theta$ determined to 0.2$^\circ$ accuracy \citep{Hsu82}, the accuracy is usually considered to be only 1$^\circ$ (e.g. \citealp{Wiktorowicz15, Bailey20}). With recent advances, 1$^\circ$ accuracy is not always good enough for the intended science (e.g. \citealp{Cotton20}).

A good high polarization standard has two qualities: (i) it is non-variable (especially in $\theta$), and (ii) it has a high polarization relative to its brightness, since position angle uncertainty, $e_\theta$, is related to polarization magnitude uncertainty, $e_p$, \citep{Serkowski68, Hsu82}: \begin{equation}e_{\theta}\approx 28.65\,e_{p}/p, \label{eq:del_th}\end{equation} where $\theta$ is in degrees, and $e_p$ is a function of photon count when not limited by instrumentation or seeing.

Most ordinary stars have little intrinsic polarization. Instead the dominant polarizing mechanism is the interstellar medium (ISM) \citep{Hiltner49, Hall49, Serkowski68}. As light travels from a star to the observer it interacts with oblate dust grains within the ISM aligned by large scale magnetic fields; these act like a wire grid polarizer. The interstellar polarization imparted is dependent on the uniformity of the ISM as well as the quantity of dust on the sight line -- and hence, indirectly, on distance. Within about 100~pc of the Sun -- i.e. within the Local Hot Bubble -- interstellar polarization is imparted at a rate of about 0.2 to 2.0~ppm/pc \citep{Bailey10, Cotton16a, Cotton17}, beyond that it is 20 ppm/pc \citep{Behr59}.

The ISM is assumed to be unchanging on relevant astrophysical timescales, which leads to choosing standards that are relatively distant and bright. Typically, the best small telescope standards have polarizations of several percent, have $m_V \lesssim 6$, and have parallaxes $<$ $\sim$2-4 mas -- these are necessarily some of the most extreme stars. The standards used today were mostly chosen in the 1960s and 1970s \citep{Serkowski68, Serkowski74, Serkowski74b, Serkowski75, Clarke10}, with much of the work establishing wavelength dependence and refining $\theta$ taking place from the 1970s to 1990s \citep{Serkowski75, Whittet80, Wilking82, Whittet92, Wolff96, Martin99}. The most comprehensive modern re-examination of the wavelength dependence of interstellar polarization was provided by \citet{Bagnulo17}, but there are scant recent works\footnote{\citet{Wiktorowicz23} and \citet{Bailey23} make a cursory examination of a few standards as part of much broader works. And although \citet{Blinov23} are conducting a monitoring campaign with the RoboPol instrument, this seems to include few, if any, bright standards.} looking at the long term stability of the most important stars. 

In the earlier literature there was an important debate about which standards might be variable. \citet{Hsu82}, \citet{Dolan86}, \citet{Lupie87}, \citet{Bastien88} and \citet{Clemens90} all, often contrastingly, identified standards they considered to be variable. Of these, the most thorough analyses were performed by \citet{Hsu82} and \citet{Bastien88}. However, these works have all been criticised as not statistically rigorous by \citet{NaghiZadeh91}, who pointed out that in most cases only partial data was presented and the data sets were small. The work of \citet{Bastien88} was the most comprehensive, yet came in for particular criticism by \citet{Clarke94}, who in reanalysing their data were convinced of the variability of only one star out of the eleven claimed. There, the main objection was that the data were drawn from different sets without this being properly accounted for, and the reanalysis used only a subset of the observations. Some time later \citet{Bastien07} revisited their work. They applied the Cumulative Distribution Function (CDF) test ``in a very conservative manner'' that was used and recommended by \citet{Clarke94}, concluding that 7 of the 11 stars they originally declared variable were, and that the other 4 ``may be.'' This does not seem a particularly satisfactory resolution. Consequently, a pall hangs over the question of which polarization standards are variable on long timescales, and the caution implied by \citet{Bastien88}'s findings has gone substantially unheeded by observers. % with it never having been satisfactorily concluded.

Putting aside the controversy, more broadly there are three motivations that provoke further study of these stars:
\vspace{-6pt}
\begin{enumerate}
\item Interstellar polarization may not be constant on 50-yr timescales. On any given sight line there will be many different dust clouds, which are in motion with respect to our standard stars. Significant movement of the clouds would cause the observed value of $\theta$ to vary over time \citep{Bastien88, Clarke10}.
\item Extreme stars are the most likely to have large intrinsic polarizations -- intrinsic polarization is more common in stars of B-type and earlier\footnote{Furthermore, the more massive a star the more likely it is to have a close companion, which results in variable polarization, scattered either from material entrained betwixt the binary, or the photospheres of the components (see \citealp{Cotton20} for a historical overview of both mechanisms).} and K-type and later \citep{Cotton16a, Clarke10}, and in more luminous stars \citep{Dyck71, Clarke10, Lewis22}. Polarization variability could have a very long period, un-captured by prior shorter duration studies, or be episodic as in the case of Be stars (e.g \citealp{Carciofi07}) or LBV stars \citep{Gootkin20}. So, stars seemingly non-variable decades ago may not be so now. 
\item Modern high precision polarimeters (\citealp{Wiktorowicz15, Bailey20, Piirola21, Bailey23}) are up to 100$\times$ more precise than those used to establish the standards. Consequently, new stellar polarigenic mechanisms are now being detected \citep{Cotton17, Bailey19, Cotton22b}. Yet, polarimetric variation associated with these phenomena is usually small, so its study is limited to the nearest stars -- without precise $\theta$ calibration large interstellar polarization overwhelms small intrinsic signals investigated over many observing runs.
\end{enumerate}

\begin{figure}
    \centering
    \includegraphics[width=\columnwidth]{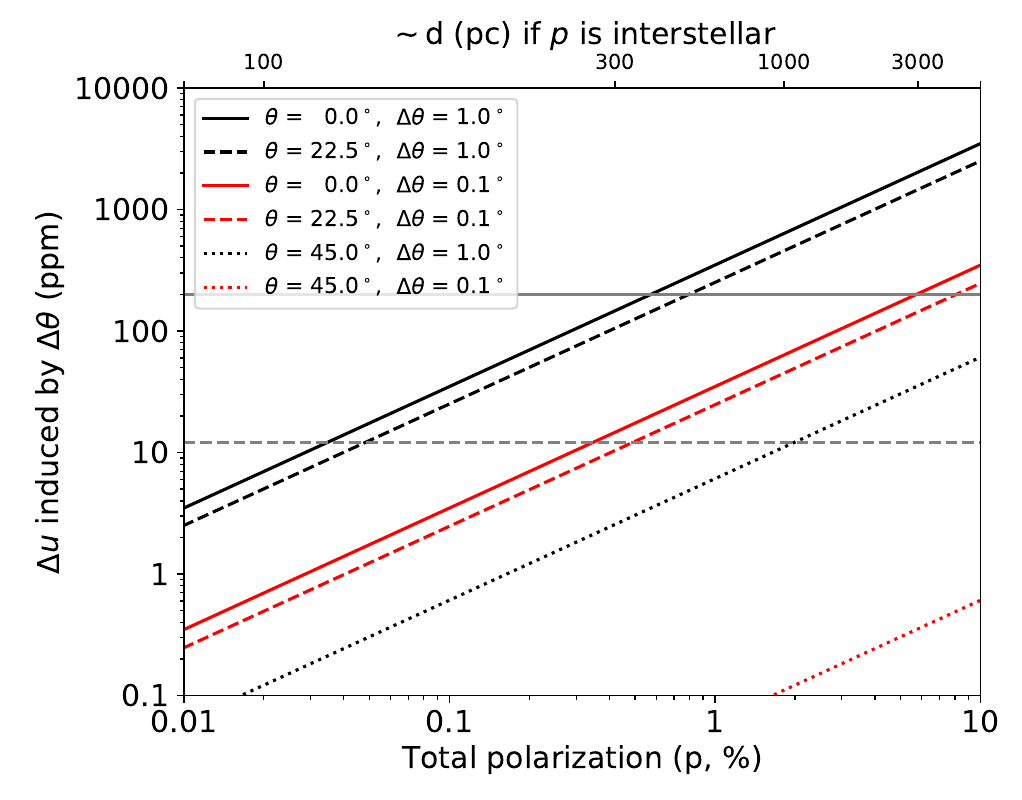}
    \caption{Effect of position angle error ($\Delta\theta$) on polarization in Stokes parameters (e.g. $u$). The size of the induced error depends on $p$ (likely interstellar polarization) as well as $\theta$. The solid and dashed grey lines correspond to 200~ppm and 12~ppm respectively, representative of science cases described in the text. Note: key order as per vertical order of lines.}
    \label{fig:TheoreticalPAErr}
\end{figure}

Because interstellar polarization increases with distance, the number of objects that can be studied long term at high precision is severely limited and many rarer stellar types are completely unavailable. To enable the discovery of new polarigenic mechanisms this must be remedied. To understand the scale of the problem, consider polarization due to binary reflection: in the Spica system this has an amplitude of 200~ppm \citep{Bailey19} -- represented by the solid grey horizontal line in Fig.~\ref{fig:TheoreticalPAErr}. A 1$^\circ$ error in $\theta$ can produce errors in the Stokes parameters at that level at a distance of 300~pc ($p_{\rm ISM} \sim$ 0.55\%). The predicted Rayleigh scattering signal from hot-Jupiter exoplanet atmospheres in the combined light of star and planet is, at best, of order 10-20~ppm \citep{Bott16, Bott18, Bailey21}. Similarly, the pulsation-driven polarization produced in the $\beta$ Cep variable $\beta$ Cru is just 12~ppm \citep{Cotton22b} (dashed grey line). For these signal levels a 1$^\circ$ error can be significant even within 100~pc of the Sun. Improving precision in $\theta$ to 0.1$^\circ$ displaces the threshold for hot-Jupiter or $\beta$ Cru like polarization to 300~pc, and Spica like polarization to 3000~pc.

Our first objective in this paper is to establish mean $\theta$ offsets between the standards. As it stands, varying the mix of standards changes the calibration. Presumably, zero point differences between different observers are a source of imprecision. The second objective is to provide an updated assessment of the position angle variability of established polarization standards -- especially long-term variability -- and in so doing determine which, if any, are suitable for achieving 0.1$^\circ$ precision. 

This paper is structured as follows: Sec.~\ref{sec:objects} provides background on each of the high polarization standard stars studied. Sec.~\ref{sec:obs} describes our observations; the analysis of which is carried out in Sec.~\ref{sec:analysis}. In Sec.~\ref{sec:discuss} we discuss the implications of the results. Of particular note, Sec. \ref{sec:impact} shows the impact of each correction we made. While, Sec. \ref{sec:recommendations} lists six specific recommendations for observers relating to the acquisition, processing and reporting of position angle data. The conclusions are presented in Sec.~\ref{sec:conclude}. Appendices \ref{apx:lit_PA}, \ref{apx:lit_homog}, and \ref{apx:mod_recal} detail literature data and calibration details. For easy reference, Appendix \ref{apx:symbols} lists a selection of symbols used through the paper.

%\vspace{-0.5cm}
\section{High polarization standard stars}
\label{sec:objects}

%
% Table 1: 
% 1. Change RA/Dec to HMS.S HMS
% 2. Standardize parallax values
% 3. Add theta_g from Table A1
% 4. Add p_max, l_max, K from Table B1
% 5. Add E(B-V), R_V from Table B2
\begin{table*}
\caption{Properties of High Polarization Standard Stars}
\tabcolsep 3.6 pt
%\centering
\begin{tabular}{rlccrccccccccrcc}
\toprule
\multicolumn{2}{c}{Standard}    &   \multicolumn{1}{c}{R.A.}    &   \multicolumn{1}{c}{Dec.}     &   \multicolumn{1}{c}{Plx.} &  SpT &   $B$   &   $V$ & $E_{\rm (B-V)}$ & $R_{\rm V}$  & $p_{\rm max}$ & $\lambda_{\rm max}$ & $K$ & \multicolumn{1}{c}{$\theta_{\rm g^\prime}$} &    \multicolumn{1}{c}{$\Delta\theta/\Delta\lambda$}  & GCVS\\     
(HD)        &   (Alt.)      & \multicolumn{2}{c}{(ICRS J2000)}  & \multicolumn{1}{c}{(mas)} &  & (mag) & (mag) &  (mag) &   (mag) & (\%) & ($\mu$m) & & \multicolumn{1}{c}{($\degr$)} & \multicolumn{1}{c}{($\degr/\mu$m)} &\\
\midrule
7927        &   $\phi$ Cas      & 01 20 04.9 & $+$58 13 54   & 0.21            &   F0\,Ia      & 5.66  & 4.98   & 0.51 & 3.11 & 3.31 & 0.507 & 0.85 &  93.0  & $-$5.7 & \\
23512       &   BD$+$23\,524    & 03 46 34.2 & $+$23 37 26   & 7.33            &   A0\,V       & 8.44  & 8.09   & 0.37 & 3.27 & 2.29 & 0.600 & 1.01 &  30.4  & $-$3.6 & \\
43384       &   9 Gem           & 06 16 58.7 & $+$23 44 27   & 0.55            &   B3\,Iab     & 6.70  & 6.25   & 0.57 & 3.06 & 3.06 & 0.566 & 0.97 &  170.0 & $+$2.6 & $\alpha$~Cyg\phantom{:} \\
80558       &   LR Vel         & 09 18 42.4 & $-$51 33 38   & 0.54            &   B6\,Ia      & 6.47  & 5.93    & 0.59 & 3.25 & 3.34 & 0.597 & 1.00 &  163.3  & $+$1.4 & $\alpha$~Cyg\phantom{:} \\
84810       &   $\ell$ Car      & 09 45 14.8 & $-$62 30 28   & 1.98            &   G5\,Iab     & 5.09  & 3.75   & 0.18 & 3.72 & 1.62 & 0.570 & 0.96 &  100.0  & \phantom{$+$}0.0 & $\delta$~Cep\phantom{:} \\
111613      &   DS Cru         & 12 51 18.0 & $-$60 19 47   & 0.45            &   A1\,Ia      & 6.10  & 5.72    & 0.40 & 3.72 & 3.14 & 0.560 & 0.94 &  80.8  & \phantom{$+$}0.0 & $\alpha$~Cyg: \\
147084      &   $o$ Sco         & 16 20 38.2 & $-$24 10 10   & 3.71  &   A4\,II      & 5.40  & 4.57   & 0.75 & 3.67 & 4.41 & 0.684 & 1.15 & 31.8    & \phantom{$+$}0.0 & \\
149757      &   $\zeta$ Oph     & 16 37 09.5 & $-$10 34 02   & 8.91  &   O9.5\,Vn    & 2.58  & 2.56   & 0.32 & 2.93 & 1.45 & 0.602 & 1.17 & 127.2   & $-$5.0 & $\gamma$~Cas\phantom{:} \\
154445      &   HR 6353         & 17 05 32.3 & $-$00 53 31   & 4.02            &   B1\,V       & 5.73  & 5.61   & 0.40 & 3.03 & 3.66 & 0.569 & 0.95 &  90.0   & \phantom{$+$}0.0 & \\
160529      &   V905 Sco     & 17 41 59.0 & $-$33 30 14 & 0.54            &   A2\,Ia      & 7.87  & 6.66      & 1.29 & 2.94 & 7.31 & 0.543 & 0.91 & 20.0  & $+$3.5 & $\alpha$~Cyg:\\
161056      &   HR 6601         & 17 43 47.0 & $-$07 04 47   & 2.44            &   B1.5\,V     & 6.68  & 6.32   & 0.60 & 3.11 & 4.01 & 0.584 & 0.96 &   67.3  & $-$1.5 & \\
161471      &   $\iota^1$ Sco   & 17 47 35.1 & $-$40 07 37  & 1.69  &   F2\,Ia      & 3.49  & 2.99   & 0.26 & 2.42 & 2.28 & 0.560 & 0.94 & 2.4   & $-$1.1 & \\
183143      &   HT Sge     & 19 27 26.6 & $+$18 17 45   & 0.43            &   B7\,Iae     & 8.08  & 6.86       & 1.24 & 3.16 & 6.16 & 0.550 & 1.15 & 179.2 & \phantom{$+$}0.0 & $\alpha$~Cyg: \\
187929      &   $\eta$ Aql      & 19 52 28.4 & $+$01 00 20   & 3.67            &   \phantom{$^+$}F6\,Ib$^+$      & 4.61  & 3.80   & 0.16  & 3.10 & 1.73 & 0.552 & 0.93 &   93.7  & $-$7.3 & $\delta$~Cep\phantom{:} \\
198478      &   55 Cyg          & 20 48 56.3 & $+$46 06 51   & 0.54            &   B3\,Ia      & 5.28  & 4.86   & 0.54 & 2.89 & 2.75 & 0.515 & 0.88 &  3.0  & \phantom{$+$}0.0 & $\alpha$~Cyg\phantom{:} \\
203532      &   HR 8176         & 21 33 54.6 & $-$82 40 59   & 3.44            &   B3\,IV      & 6.51  & 6.38   & 0.32 & 3.05 & 1.39 & 0.574 & 0.86 &   126.9  & $+$2.4 & \\
210121      &   HIP 109265      & 22 08 11.9 & $-$03 31 53   & 3.00            &   B7\,II      & 7.84  & 7.68   & 0.35 & 2.22 & 1.38 & 0.434 & 0.73 &   155.1  & $+$8.6 & \\
\bottomrule
\end{tabular}
\begin{flushleft}
\underline{Notes} -- $+$ $\eta$ Aql has an SB companion classified computationally as B9.8\,V. Photometric data and astrometric data, presented in sexagesimal IRCS J2000, are taken directly from SIMBAD. For the origin/derivation of position angle data see Appendix \ref{apx:lit_PA}. Note that $\theta$ is given for the SDSS $g^\prime$ band and a 2020 equinox. For the origin of Serkowski fit parameters, reddening data and spectral type references see Appendix \ref{apx:lit_homog}. The final column has the variability type as given in the General Catalog of Variable Stars (GCVS, \citealp{Samus17}), where a colon indicates some uncertainty; HD~160529 is elsewhere classified as a Luminous Blue Variable (LBV) star (e.g. \citealp{Stahl03}), and HD~149757 as an Oe star (e.g. \citealp{Negueruela04}) and a $\beta$~Cep star (e.g. \citealp{Hubrig11}).
\end{flushleft}
\label{tab:basic}
\vspace{-12pt}
\end{table*}

Very bright high polarization standards are rare. The large distances required for significant interstellar polarization mean that only stars with small absolute magnitudes are bright enough. As a result, most standards trace their lineage to the first decades of stellar polarimetric study when the first bright star surveys were being conducted. In particular, the most used standards are drawn from a recommended list first published by \citet{Serkowski74}. The parameters for those stars were all refined in \citet{Serkowski75}. Other observers have occasionally added to (or subtracted from) this list, according to their needs, but have largely applied the same selection criteria. There are perhaps as many as two dozen standards in irregular use, depending on what brightness criteria are applied. These stars are far from evenly distributed across the sky. Overwhelmingly the standards are located in dusty regions fairly close to the Sun, such as the Sco-Cen association; the few that aren't can be very important. For instance, \citet{Matsumura97} described reports of variability in HD~43384 as a ``serious problem,'' stressing that there was no bright alternative within $\sim$6\,h right ascension in the northern hemisphere.

We have largely worked from southern mid-latitudes, and so most stars we report on here are accessible primarily from there, but the transportation of an instrument to the Monterey Institute for Research in Astronomy (MIRA), has allowed us to add a number of northern stars. The standard stars in this study all appear in the catalogs of \citet{Serkowski75}, \citet{Hsu82}, and/or \citet{Bagnulo17}; their properties are summarised in Table~\ref{tab:basic}. They are all either well established standards or have been used as such in making observations with the High Precision Polarimetric Instrument (HIPPI) and other HIPPI-class polarimeters. Appendices \mbox{\ref{apx:lit_PA} and \ref{apx:lit_homog}} provide references and describe, in meticulous detail, how we came to favour the tabulated polarization and reddening properties. The co-ordinates and magnitudes for each standard given here -- that define which telescopes they are accessible to -- are taken directly from SIMBAD. Below is an account of other pertinent details, including variability found by other methods that might portent polarimetric variability, as well as a detailed account of claims and counter-claims of polarimetric variability for each star.

\subsection{HD~7927}
\label{sec:phiCas}

HD\,7927 ($\phi$~Cas) is a bright yellow supergiant star of spectral type F0\,Ia \citep{Gray01} that is likely, though not conclusively, a member of the NGC 457 moving group \citep{Eggen82, Rosenzweig93}. It has two notable visual companions, the brightest companion ($\phi^2$ Cas) is $m_V=$ 7.04, 132.8$\arcsec$ away, and the closest companion is a $m_V=$ 12.3 at 48.4$\arcsec$ separation \citep{Mason01}. Small amplitude variations with no defined period have been found in RV \citep{Adams24, Ferro88}\footnote{\citet{Ferro88} state HD\,7927 is not in \citet{Adams24}. However, \citet{Adams24} list it according to its catalogue number in \citet{Boss10}'s \textit{Preliminary General Catalogue of 6,188 Stars}. His son's later \textit{General Catalogue of 33,342 Stars} \citep{Boss36} uses different catalogue numbers for the same stars, however both catalogues are generally referred to by the prefix ``Boss.'' We believe this to be the source of confusion.} and in photometry by \citet{Percy89}, who note that the photometric variations are too small compared to RV to indicate Cepheid-like behaviour. 

First measurements of HD\,7927's polarization were made by \cite{Hiltner51}. The star was not found to be variable by \citet{Coyne71} but he did note its $p(\lambda)$ as anomalous. No variability was found by \citet{Hsu82}, whose claimed detection thresholds are 0.01 per cent in $p$ and 0.2$^\circ$ in $\theta$. Wavelength dependence of $\theta$ in HD\,7927 has been observed on multiple occasions \citep{Gehrels65, Coyne66, Hsu82} but only \citet{Dolan86} found that the wavelength dependence varied from night to night; they emphasize this as critically problematic for a position angle standard. \citet{Dolan86} also found $\theta$ variable. Furthermore, \citep{Bastien88} found HD\,7927 to be variable in both $p$ and $\theta$, although the results of this paper are heavily criticized and this result refuted by \citet{Clarke94}. Earlier \citet{NaghiZadeh91} had described this star as displaying ``definite polarization variability'' both in $\theta$, and in $p$ in $R$ band (but not in $p$ in $B$ band) based on his own observations. \vspace{-0.3cm}

\subsection{HD~23512}
\label{sec:HD23512}

HD~23512 (BD$+$23 524) is an A0\,V type star \citep{Fitzpatrick07} and is a member of the Pleiades cluster \citep{abt78}. The star has a companion, discovered by lunar occultation, with a brightness difference of 2~mag and a separation of $0.1\arcsec$ \citep{Mason01} or $0.05\arcsec$ \citep{Torres21}. It has been a candidate for having a variable RV \citep{smith44} but this was not confirmed by \citet{abt65}. The star has also been a double line candidate \citep{liu91} but this was not corroborated by \citet{Torres20}. The polarization of HD\,23512 was found not to be variable by \citet{Hsu82}. It is described as ``clearly'' variable in both $p$ and $\theta$ by \citet{Bastien88}, which was refuted by \citet{Clarke94}. \vspace{-0.3cm}

\subsection{HD~43384}
\label{sec:9Gem}

HD\,43384 (9~Gem) is of spectral type B3\,Ib \citep{Rachford09} classified as an $\alpha$~Cyg variable star \citep{ESA97}. \cite{Hsu82} found that the star's polarization angle is variable at a level of 0.8 $\pm$ 0.2$^\circ$ on the short term, with larger long term variations apparent ($\Delta\theta\sim2^\circ$; $\Delta p= 0.25$\% over a decade). \citet{Coyne71} had previously described variability around thrice as much in both $p$ and $\theta$.  \cite{Matsumura97} found that the polarization variability ($\Delta\theta\sim1^\circ$; $\Delta p=0.2\%$) was phase locked with the 13.70 day period observed in \textit{Hipparcos} photometry \citep{ESA97}. In contrast \citet{Dolan86}, though noting an extreme $\Delta\theta/\Delta\lambda$ found neither that parameter to be complex nor $\theta$ to be variable. \vspace{-0.3cm}

\subsection{HD~80558}

HD~80558 (LR~Vel) is a B6\,Ia supergiant \citep{Houk78} with prominent photometric variability \citep{vanGenderen89}. The polarization of HD~80558 was first studied by \cite{Serkowski69} and it has been used as a high polarization standard since then. \citet{Dolan86}, in comparing their data to \cite{Serkowski74}'s, found no significant difference in $p$ or $\theta$. \citet{Hsu82} also reported no variability. \citet{Bastien88} found HD~80558 to have variable polarization over 35 nights of observation. This result was refuted by \citet{Clarke94}'s reanalysis of \citet{Bastien88}'s data. \vspace{-0.3cm}

\subsection{HD~84810}
 
HD\,84810 ($\ell$~Car) is a classical Cepheid variable with a spectral type that ranges from F8--G9 \citep{Albrecht21} and a period of $\approx$35.5 days \citep{Trahin21}. Owing to its brightness and proximity it has been extensively observed from ultraviolet (UV) to infrared (IR) wavelengths for more than a century (e.g. \citealp{Bohm1994, Kervella06}). In principle the purely radial pulsations of a Cepheid variable should produce no polarization change \citep{Odell79}, and HD\,84810 has been found to be invariable in $p$ and $\theta$ by \citet{Hsu82}, \citet{Bastien88} and \citet{Clarke94}. Sensitive measurements by \citet{Bailey23} show only small variations in $p$ of 0.023 $\pm$ 0.005 per cent from 48 observations over about a year. \vspace{-0.3cm}

\subsection{HD~111613}

HD\,111613 (DS~Cru) is a supergiant of spectral type A2\,Iab \citep{Ebenbichler22} and a member of NGC 4755 \citep{Humphreys78}. \citet{Hsu82} find no variability for HD\,111613 in $p$ or $\theta$. \citet{Dolan86} saw no change in $p$ over a four-year period (1980-84), but found $\theta$ and its wavelength dependence to be inconsistent between observing runs. \citet{Bastien88} observed for 41 nights and saw significant variations in both $\theta$ and $p$ ($\Delta\theta=2.4^\circ$, $\Delta p=0.105$ per cent) on a timescale of $\approx$32 days, a result confirmed by \citet{Clarke94}'s reanalysis. \citet{Bastien88} noted that the polarization was seen to vary slowly, and supposing a binary system, derived an inclination of $84 \pm 1 \degr$ based on an assumed a 64-day period. \vspace{-0.3cm} 

\subsection{HD~147084}

HD\,147084 ($o$~Sco) is an A4\,II bright giant \citep{Martin99} in Upper-Scorpius \citep{deGeus89}. Small amplitude RV variations were measured by \citet{Levato87} who state that this range may be due to intrinsic motions in the atmosphere. HD\,147084 is noteworthy for being a standard for circular as well as linear polarization. It has a maximum fractional circular polarization of approximately 0.04 per cent at $2.32 ~\mu m$ \citep{Kemp1972}, indicating that the light passes through at least two regions of the interstellar medium with differently aligned dust particles.

HD\,147084 has substantial coverage in polarization data spanning ultraviolet to infrared wavelengths, owing to its large $\lambda_{\rm max}$, making it particularly useful as a standard \citep{Kemp1972, Martin99}. No variability in $p$ or $\theta$ was found by \citet{Hsu82} nor \citet{Dolan86}. In contrast, \citet{Bastien88} find it to be variable in both $p$ and $\theta$, a result refuted by \citet{Clarke94}. A small potential variability in $p$ of 0.028 $\pm$ 0.008 per cent has been found by \citet{Bailey23} from 108 observations over more than a year. \vspace{-0.3cm}

\subsection{HD~149757}
\label{sec:zetOph}

HD~149757 ($\zeta$~Oph) is a well-studied single star with an O9.5\,V spectral type \citep{Hubrig11}. Its rapid rotation velocity of 400 km/s causes it to lose mass through a strong wind \citep{Hubrig11}, and gives rise to a variable surface temperature through its oblateness \citep{Balona99a}. Periodic variability for this star has been noted in both photometry and spectroscopy (helium line profiles) consistent with a $\beta$ Cephei type classification \citep{Hubrig11}. The spectral variability is likely the result of non-radial pulsations \citep{Balona99b}, where these modes may be excited periodically by lower order modes \citep{Walker05}. HD~149757 was one of ten O-type stars included in a study of polarimetric variability by \citet{Hayes75}, from 12 observations over many weeks, he did not find it to be variable. \citet{Lupie87} describe the star as nonvariable but caution they have few observations. \citet{McDavid00} carried out a study of nine O-type stars with variable winds, including HD\,149757, using agglomerated data from 1949 to 1997; none exhibited statistically significant variability, but small amplitude, short term variability amongst the targets was hinted at by a multi-technique campaign. \vspace{-0.3cm}

\subsection{HD~154445}

HD~154445 (HR~6353) is a B1\,V spectral type star \citep{Reed03}; it has no identified companions \citep{Eggleton08}. The first reported polarimetric observations of HD~154445 were made by \citet{Hiltner51}. Repeated observations at optical \citep[e.g.][]{vanPSmith56,Serkowski75,Cikota18} and near-infrared wavelengths \citep{Dyck71b,Dyck78} have demonstrated consistency in $p$ and $\theta$. \citet{Hsu82} find the HD\,154445 to be invariable in $p$ and $\theta$. The star was claimed as variable in $p$ (but not $\theta$) at the 2-$\sigma$ level by \citet{Bastien88}, but this was not borne out by reanalysis \citep{Clarke94}. Recently reported observations by \citet{Wiktorowicz23} present little evidence for variability. \vspace{-0.3cm}

\subsection{HD~160529}

HD~160529 (V905~Sco) is a Luminous Blue Variable (LBV) of spectral type A2Ia \citep{Stahl03}. It has a prolific history as a photometric and spectroscopic variable star \citep[e.g.][]{Wolf74,Sterken91}. Decades of photometry from \citet{Sterken91} show that the star's magnitude dimmed by $\sim$0.5mag over 18 years. More recent AAVSO data spanning the last 20 years shows similar timescales of variability, with as much as a magnitude in brightness changes. A spectroscopic study by \citet{Wolf74} highlighted many signatures that could likely be attributed to strong mass loss including line profile variations, line splitting, P-Cygni and inverse P-Cygni profiles. This large photometric and spectroscopic variability has likely led to the difficulties in classifying the spectral type; the presence of strong, sharp emission lines and H$\alpha$ excess likely complicated it as well. Early classifications of HD~160259 included, A4\,se$\alpha$ \citep{Merrill33}, A2-3\,Ia \citep{Wallerstein70}, and A9\,Ia \citep{Houk82}. 

Polarization measurements of HD~160529 reach back as far as the early 1950's \citep{Hiltner51,Markowitz51}. No polarimetric variability has been ascribed to the star (e.g. \citealp{Treanor63}, \citealp{Hsu82}, \citealp{Dolan86}), but \citet{Dolan86} do note a complex $\Delta\theta/\Delta\lambda$. \vspace{-0.3cm}

\subsection{HD~161056}

HD\,161056 (HR~6601) is a B1/2V star \citep{ODonnell94}. \citet{Telting06} included it in a study looking for line profile variations associated with pulsation; none were indicated, albeit from a single observation. HD\,161056 was first observed polarimetrically by \citet{vanPSmith56} as part of her survey of interstellar polarization in the Southern Milky Way and it is often included in polarimetric studies of the interstellar medium (e.g. \citealp{Piccone22}). \citet{Bastien88} only suspected variability in $p$ but reported $\theta$ variability of 0.5$^\circ$, however later reanalysis calls into question this conclusion \citep{Clarke94}. \citet{Berdyugin95} constrain any variability to $<$~1$^\circ$. \vspace{-0.3cm}

\subsection{HD~161471}

HD~161471 ($\iota^1$~Sco) is a luminous red supergiant star of spectral type F2Ia \citep{Houk78, Gray89}. It is a spectroscopic binary \citep{Pourbaix04} and has a 13th mag companion at 37$\arcsec$ separation. It's H$\alpha$ line width probably indicates a weak stellar wind \citep{Danks94}. It is not a widely used position angle standard, but has been so utilized by \citet{Bailey23} to calibrate the position angle of polarization in 20-cm PICSARR observations. 
They find potential variability in $p$ of 0.020~$\pm$~0.004~per cent from 18 observations spanning more than a year. \vspace{-0.3cm}

\subsection{HD~183143}

HD\,183143 (HT~Sge) is an extremely luminous hypergiant star of spectral type B7Iae \citep{Chentsov2004}. It was first found to have a high broadband polarization by \citet{Hall50}, who described it as a ``star of special interest.'' \citet{Serkowski74} later named it as a standard. \citet{Clemens90} found their observations of it to be consistent with those of \citet{Serkowski75} and \cite{Schulz1983} claim its polarization as a function of wavelength is consistent with an interstellar origin. Spectropolarimetric data from \citet{Lupie87} marks it as their least variable standard, but shows $\sigma_\theta=$ 0.40$^\circ$. However, \cite{Hsu82} convincingly showed HD\,183143 exhibited polarimetric variability ($\Delta\theta\sim1^\circ$, $\Delta p = 0.2$ per cent) on a timescale of days to weeks, and \citet{Dolan86} found that its $\Delta\theta/\Delta\lambda$ character varied from night-to-night, along with $\theta$ itself. \vspace{-0.3cm}

\subsection{HD~187929}

HD\,187929 ($\eta$~Aql) is a classical Cepheid with spectral type F6\,Ib-G4\,Ib and a pulsational period of 7.18~d \citep{Benedict22}. % V = 3.90, and . 
As the first Cepheid discovered \citep{Pigott85}, it has been well-studied over the years, and perhaps most particularly during the era of space observations \citep{Evans91, Benedict07, Evans13}. The current understanding is that it is a triple system containing, in addition to the Cepheid, a late-B close-in companion as well as an F-type companion lying some 0.66$\arcsec$ from the primary. Polarimetric measurements of the star have focused on attempts to detect a magnetic field using spectropolarimetry to varying degrees of success \citep{Borra81,Plachinda00,Wade02,Grunhut10}. In linear polarization \citet{Hsu82} report a particularly large $\Delta\theta/\Delta\lambda$ of $-7.3 \pm .3 ^\circ/\mu m$, but no variability. \citet{Dolan86} report measurements that differ by $1.5^\circ$ from \citeauthor{Hsu82}'s, along with a complex $\Delta\theta/\Delta\lambda$ behaviour not explainable by a two-cloud model, and they name intrinsic polarization as a possibility. \citet{Bastien88} categorized HD\,187929 as a suspected variable, but later retract this assessment \citep{Bastien07}. \vspace{-0.3cm}

\subsection{HD~198478}

HD\,198478 (55~Cyg) is a blue supergiant star of type B3\,Ia and a prominent $\alpha$~Cyg variable with asymmetric contraction varying over hours to days \citep{Wilson53}. Periods of variability (in pressure, gravity, and modes) appear to correlate with---and are well-modelled by---mass loss episodes \citep{Yadav16, Kraus15}. HD\,198478 may also experience macroturbulence from convection significant enough to contribute to measurable line broadening beyond that from rotation \citep{Jurkic11}, which may further influence the consistency of some parameters like surface gravity.
 
Although it is widely used as a standard polarization star \citep[e.g.][]{Cox07}, large changes in the polarization of HD\,198478 have been observed previously by \citet{Hsu82} and \citet{Wiktorowicz23}. In particular, \citet{Hsu82} saw changes in $\theta$ and $p$ of 1$^\circ$ and $\lesssim$0.06 per cent, respectively, within a short run -- several days. They associated this variability with emission variability seen to occur on the same short time scale as reported by \citet{Underhill60} and \citet{Granes72}. \citet{Dolan86} also suspected variability, partly on the basis that their $\theta$ determination differed substantially from earlier literature but also because \citet{Treanor63} noted $p$ was unusually high for its location. \citet{NaghiZadeh91} describes its variability in $p$ and $\theta$ as ``very obvious,'' in particular reporting $\Delta\theta\sim$~4.8$^\circ$, but observing that $\Delta\theta/\Delta\lambda$ is consistent from night-to-night. \citet{NaghiZadeh91} was critical of the statistical approach of some of the early polarimetric studies, and advocated use of the CDF to aid in matching the polarimetric mechanism to the observed variability. In this specific case he noted the similarity of the variability of this star to that of other supergiants, ascribing it to mass loss and the presence of a stellar wind. \vspace{-0.3cm}

\subsection{HD~203532}

HD\,203532 (HR~8176) is a B3\,IV subgiant in the constellation Octans. It is the southernmost standard in the current study with a declination of $-82.683^{\circ}$. With this latitude, it is placed close to the molecular clouds south of the Chamaeleon complex which are associated with the Galactic plane \citep{Larson00}. Due to coordinate precession being larger for coordinates close to the celestial poles the position angle changes more over time than for the lower latitude stars (see Table \ref {tab:lit_pa}). HD\,203532 has no known companions nor is it a known variable star \citep{Samus17}. 

The first polarization measurement was made by \cite{Mathewson70}. Later \cite{Serkowski75} made four measurements yielding position angles between $126.2 ^{\circ}$ and $127.6^{\circ}$ in the $V$ band. It has not been reported to be variable, but measurements made by \citet{Bagnulo17} and \citet{Bailey20} with modern equipment disagree in $\theta$ by 2.6 $\pm$ 0.9$^\circ$. \vspace{-0.3cm}

\subsection{HD~210121}

HD\,210121 (HIP~109265) is a B-type star, sharing a line of sight with a single, high latitude cloud sitting $\sim 150$ pc from the Galactic plane \citep{Welty92}. The star is of uncertain spectral type, with several incongruent classifications having been assigned at different points in the literature. For example, its spectral type has been listed as B3~V \citep{Welty92, Larson96}, B3.5~V \citep{Siebenmorgen20, Krelowski21}, B5-6~V \citep{deVries88}, B7~II \citep{Valencic04, Fitzpatrick07, Bagnulo17, Piccone22}, and B9~V \citep{Voshchinnikov12}. On the whole, a critical reading of the literature suggests to us that an earlier spectral type is more likely, however we have opted to use the B7 classification in this work, since this is what was used for determining the Serkowski parameters. With the foreground cloud characterised by a high abundance of small grains, HD\,210121 is often cited with reference to its extremely low $R_V$ and high UV extinction. 

Initial polarimetric measurements were made by \citet{Larson96}, which show no significant rotation with wavelength. \citet{Bagnulo17}, with greater sensitivity, identified a gradient of $\Delta\theta / \Delta\lambda = 0.86 \pm 0.07 ^{\circ}/100$ nm, but no variability has been implied.

\section{Observations}
\label{sec:obs}

The data for this work comes from 88\footnote{There have been 91 observing runs with the HIgh Precision Polarimetric Instrument (HIPPI) and its derivatives (referred to as HIPPI-class instruments) to date. One test run with Mini-HIPPI on the Penrith telescope used Achernar in emission (HD\,10144; $\theta=31.5^\circ$) for position angle calibration and is not included. Another two constituted the first re-commissioning run for the second HIPPI-2 at MIRA's OOS and are not considered reliable due to issues with the modulator and rotation stage (see \citealp{Cotton22b}).} observing runs (or sub-runs\footnote{A new mounting of the instrument within an otherwise contiguous run.}) on six different telescopes using four different HIPPI-class polarimeters of three different designs, spanning 10 years of operation. It includes every observation we have made of the 17 different standards listed in Table \ref{tab:basic} during these runs. For 5 standards our data spans the full (almost) 10 years, two stars (HD\,7927 and HD\,198478) were observed from the northern site for only a year, the HD\,43384 data span is only slightly longer than a year, every other star has a multi-year data set, most of which are at least 5 years. Many of the observations were made solely for the purpose of $\theta$ calibration, and some multi-band observations were made to check the modulator efficiency (see Appendix \ref{apx:mod_recal}). However, from June 2020 a number of observing runs were made specifically for this work.

The six telescopes observed with were the 3.9-m Anglo-Australian Telescope (AAT) at Siding Spring Observatory, the Penrith 60-cm (24-in) telescope at Western Sydney University (WSU), a 14-inch Celestron C14 at UNSW Observatory (UNSW), a Celestron 9$\nicefrac{1}{4}$-inch telescope at Pindari Observatory in suburban Sydney (PIN), the 36-in telescope at MIRA's Oliver Observing Station (OOS), and the 8.1-m Gemini North Telescope (GN).

The four HIPPI-class polarimeters include the original HIPPI \citep{Bailey15}, Mini-HIPPI \citep{Bailey17}, and two different HIPPI-2 units (differentiated by the hemispheres they operated in, \citealp{Bailey20, Cotton22a}). They are dual-beam\footnote{Several high polarization standard observations from one run (N2018JUN) were made with only one channel due to a cabling failure.} aperture photo-polarimeters that share common design elements, namely a ferro-electric crystal modulator operating at 500~Hz and optimised for blue wavelengths; a Wollaston (or Glan-Taylor) prism analyzer; and modern, compact photo-multiplier tube detectors (PMTs). The PMTs are manufactured by Hamamatsu; we mostly used blue (B) sensitive units \citep{Nakamura10}, but a few observations were also made with PMTs with a redder response curve (designated R). Most observations were made in the SDSS $g^\prime$ filter or unfiltered (Clear), but a range of other filters were used as described in Appendix \ref{apx:mod_recal}. 

Each of these instruments measure only a single Stokes parameter at once. To measure the other Stokes parameter, the instrument is rotated through 45$^\circ$. With HIPPI this was accomplished with the AAT's Cassegrain rotator. The other instruments instead used their own instrument rotator. In practice an observation was actually made up of four measurements: at angles of 0, 45, 90 and 135$^\circ$, where the perpendicular measurements are combined in a way that minimises any instrumental contribution to the polarisation.

A database containing a summary of the instrument and telescope set-up for each run, and the details of every high polarization standard observation in machine readable format is available at CDS via anonymous ftp to \href{cdsarc.u-strasbg.fr}{cdsarc.u-strasbg.fr}
(130.79.128.5) or via \href{https://cdsarc.cds.unistra.fr/viz-bin/cat/J/MNRAS}{https://cdsarc.cds.unistra.fr/viz-bin/cat/J/MNRAS} or on MIRA's website at \href{http://www.mira.org/research/polarimetry/PA}{http://www.mira.org/research/polarimetry/PA}. The catalogue contains both the $g^\prime$ and Clear (unfiltered) observations analysed in the main part of the paper, as well as those made in other bands used for calibration in Appendix \ref{apx:mod_recal} (583 observations in total). Note that only the errors derived from the internal statistics of the observation are reported in the catalogue, for an assessment of accuracy see initially, Appendix \ref{apx:accuracy}, then Sections \ref{sec:varPA} and \ref{sec:varP}.

\subsection{Reduction}
\label{sec:reduction}

Standard reduction procedures for HIPPI-class polarimeters are described by \citet{Bailey20}. These include a bandpass model that takes account of the spectral type of the target. For this work we have re-evaluated and updated the modulator calibrations to improve the accuracy of this procedure (see Appendix \ref{apx:mod_recal}). Here we improve the bandpass model by including the expected polarization and reddening of the standards (given in Appendix \ref{apx:lit_homog}), the computer code is adapted from the PICSARR reduction \citep{Bailey23}.

The telescope polarization (TP) is measured by many observations of low polarization standard stars assumed to be unpolarized. The full list of standards used is: $\alpha$~Aql, $\alpha$~Boo, $\gamma$~Boo, $\eta$~Boo, $\tau$~Cet, $\alpha$~Cen, $\alpha$~CMa, $\alpha$~CMi, $\beta$~Hyi, $\alpha$~Lac, $\beta$~Leo, $\alpha$~Ser, $\beta$~Tau and $\beta$~Vir. Aside from $\alpha$~CMi, all of the additions from those listed in \citet{Bailey20} have been restricted to northern hemisphere use. The standard bandpass model is applied prior to the straight average of the standard measurements in each band being taken, then these values are vector subtracted from all the other observations. Aside from run N2018JUN (see \citealp{Bailey20}), the TP never exceeded 170 ppm in the $g^\prime$ band, and varied by typically 10s of ppm or less between bands. Typical TP levels for the telescopes we observed with can be seen in \citet{Bailey17, Bailey20} and \citet{Cotton22b}. The nominal errors of this process are mostly less than 10~ppm with larger uncertainties occurring only for some of the smaller telescope runs. These errors are wrapped into the observational errors on an RMS basis -- and for the most part are negligible.

The telescope position angle is usually calibrated as the straight average difference between the expected and observed position angles of standards observed in a $g^\prime$ or Clear filter. The nominal position angles, $\theta_{g^\prime}$ of the standards have been redetermined from the literature in Appendix~\ref{apx:lit_PA}. The expected position angles include some corrections to these. The two most important are related to the change in position angle with wavelength, $\Delta\theta/\Delta\lambda$, and the precession of the co-ordinate system -- these are also described in Appendix~\ref{apx:lit_PA}. 

The precession turns out to be particularly important. Stars in the North precess positively in $\theta$, and vice versa for those in the South -- with typical values for $\Delta\theta/\Delta t$ being several tenths of a degree per century. The literature values for $\theta$ were first established for some standards more than 50 years ago. Failing to account for precession can artificially induce a degree or more difference between some pairs of standards.

Here we also apply a correction for the Faraday rotation of polarization \citep{Faraday1846} by the atmosphere in the presence of the Earth's magnetic field. This step is recommended by \citet{Clarke10} when $p/e_p\gtrsim1000$, which many of our measurements nominally surpass. Though never observed, the possible impacts of Faraday rotation on ground-based astronomical polarimetry have been discussed for nearly a century \citep{Lyot29, Serkowski74b, Hsu82, Clarke10}. Here we make the same simplifying assumptions as \citet{Hsu82}, namely that the Earth's magnetic field is aligned to geographic poles, and employ,
\begin{equation}\Delta\theta=V(\lambda)B_{||}hX,\end{equation}
where $B_{||}$ is the component of the magnetic field ($\approx$0.5~Gauss) parallel to the line of sight, $h$ the scale height of the atmosphere equal to 80000~cm, $X$ the airmass, and $V(\lambda)$ the Verdet constant, which is dependant on wavelength. \citet{West63} give $V$ as 6.27$\times10^{-6}$ arcmin/Gauss/cm for the air at standard temperature and pressure at 567~nm, and we derive the values at other wavelengths by scaling and extrapolating from Figure 2 in \citet{Finkel64} – at 470~nm this gives 9.53$\times10^{-6}$ arcmin/Gauss/cm. At the airmasses of our observations, the correction never comes to much more \mbox{than 0.01$^\circ$}.

\section{Analysis}
\label{sec:analysis}

\subsection{Position Angle Precision by Instrument}
\label{sec:instrument_performance}

\citet{Clarke94}'s primary criticism of \citet{Bastien88} was that measurements from different set-ups were combined in an \textit{unweighted} way. To conduct a long term analysis we need to combine data from multiple (albeit similar) instruments across many sub-runs, where each corresponds to a new mounting of the instrument on the telescope. This task requires some care. So, before we attempt it, we first seek an understanding of the $\theta$ variability attributable to different telescope/instrumental set-ups. 

In Table \ref{tab:stdevsIT} the standard deviation of $\theta$, $\sigma_\theta$, of repeat observations in the same filter (limited to $g^\prime$ or Clear) of the same star within a sub-run is determined; such observations are directly comparable with each other. The table reports the median standard deviation, $\eta(\sigma_\theta)$, of each set of observations, where the number of observations, $N_o\ge2$; and the median of $\epsilon_\theta$, a metric designed to determine the scatter independent of known noise sources,
\begin{equation}\epsilon_{\rm i\theta}=\sqrt{\sigma_{\rm i\theta}^2-e_{\rm m\,\theta}^2}.\label{eq:err1}\end{equation}
where $e_{\rm m\,\theta}$ is the measured uncertainty derived from the internal statistics of our measurements, it is largely photon-shot noise but also incorporates other noise sources associated with seeing and the detector\footnote{There is also a very small contribution ascribed to centering imprecision (see \citealp{Bailey20}), which we include throughout this paper in $e_{\rm m}$ but otherwise neglect to mention in order to simplify discussion.}.

It should be noted that this approach is strictly only valid for a Gaussian distribution, and $\theta$ is not Gaussian. However, it approaches Gaussian at high signal-to-noise, i.e $p/e_{p} \gtrapprox 5$, and in our case $p/e_{\rm m\,p} >> 100$, typically.

\begin{table}
\caption{Median standard deviations of observation sets by instrument and telescope}
\tabcolsep 8.5 pt
%\centering
\begin{tabular}{llrrr}
\toprule
Instrument     &   Telescope    &   $N_S$ &	\multicolumn{1}{c}{$\eta(\sigma_{\rm i\,\theta})$}	&  \multicolumn{1}{c}{$\eta(\epsilon_{\rm i\,\theta})$}\\
        &           &     &   \multicolumn{1}{c}{($^\circ$)}    &   \multicolumn{1}{c}{($^\circ$)}  \\
\midrule
HIPPI       & AAT       & 5	    & 0.070	  & 0.068 \\ % g/Clear only.
HIPPI-2     & AAT       & 33	& 0.082	  & 0.077	\\
HIPPI-2     & WSU-24    & 3	    & 0.144	  & 0.139	\\
HIPPI-2*    & MIRA-36   & 8     & 0.196   & 0.186   \\
M-HIPPI     & UNSW-14   & 5     & 0.292   & 0.282  \\
M-HIPPI     & PIN-9$\nicefrac{1}{4}$ & 25 & 0.483 & 0.474  \\
\midrule
HIPPI-2*    &   WSU/MIRA & 11   & 0.168   & 0.153  \\
HIPPI-2$^\dagger$ &Gemini Nth& 4   & 0.240   & 0.173  \\
\bottomrule
\end{tabular}
\begin{flushleft}
\underline{Notes} -- $N_S$ is the number of sets. * HD\,198478 removed; includes observations only until to 2023-09-01. $\dagger$ Observation sets are combined $g^\prime$, 500SP and Clear observations.\\
\end{flushleft}
\label{tab:stdevsIT}
\vspace{-0pt}
\end{table}

The different instruments had different rotation mechanisms, which is likely to contribute to scatter in $\theta$. HIPPI used the Cassegrain rotator of the AAT to move position angle, while \mbox{HIPPI-2} and Mini-HIPPI have their own instrument rotator. \mbox{HIPPI-2's} rotator is a heavy-duty precision component, and on the AAT the value for $\epsilon_\theta$ is small. Mini-HIPPI's rotator is not as well made, so the lower precision for Mini-HIPPI was expected. Additionally, the Pindari telescope, unlike the others, is not on a fixed mount so is susceptible to external forces, e.g. wind, incidental operator intervention. 

The smaller telescopes do not have as good a value for $\epsilon_\theta$ as the AAT. Potential mechanical explanations would be a slight polar misalignment or play in the mounting. However, the number of observation sets is small, and so the difference could just be a result of which stars were used in the measurements. For the MIRA-36 this is likely to be a factor since, proportionally more standards were observed which are robustly reported as variable in the literature (and we have excluded HD\,198478 from this part of the analysis because it especially biases the results). More reliable standards make up a larger majority of the sets for the other telescope/instrument combinations. To get a more robust measure for the two smaller HIPPI-2 compatible telescopes, we combine their data below the midline of Table \ref{tab:stdevsIT}.

\begin{figure*}
    \centering
    \includegraphics[width=\textwidth, trim={0.55cm 0.5cm 0.85cm 0cm}, clip]{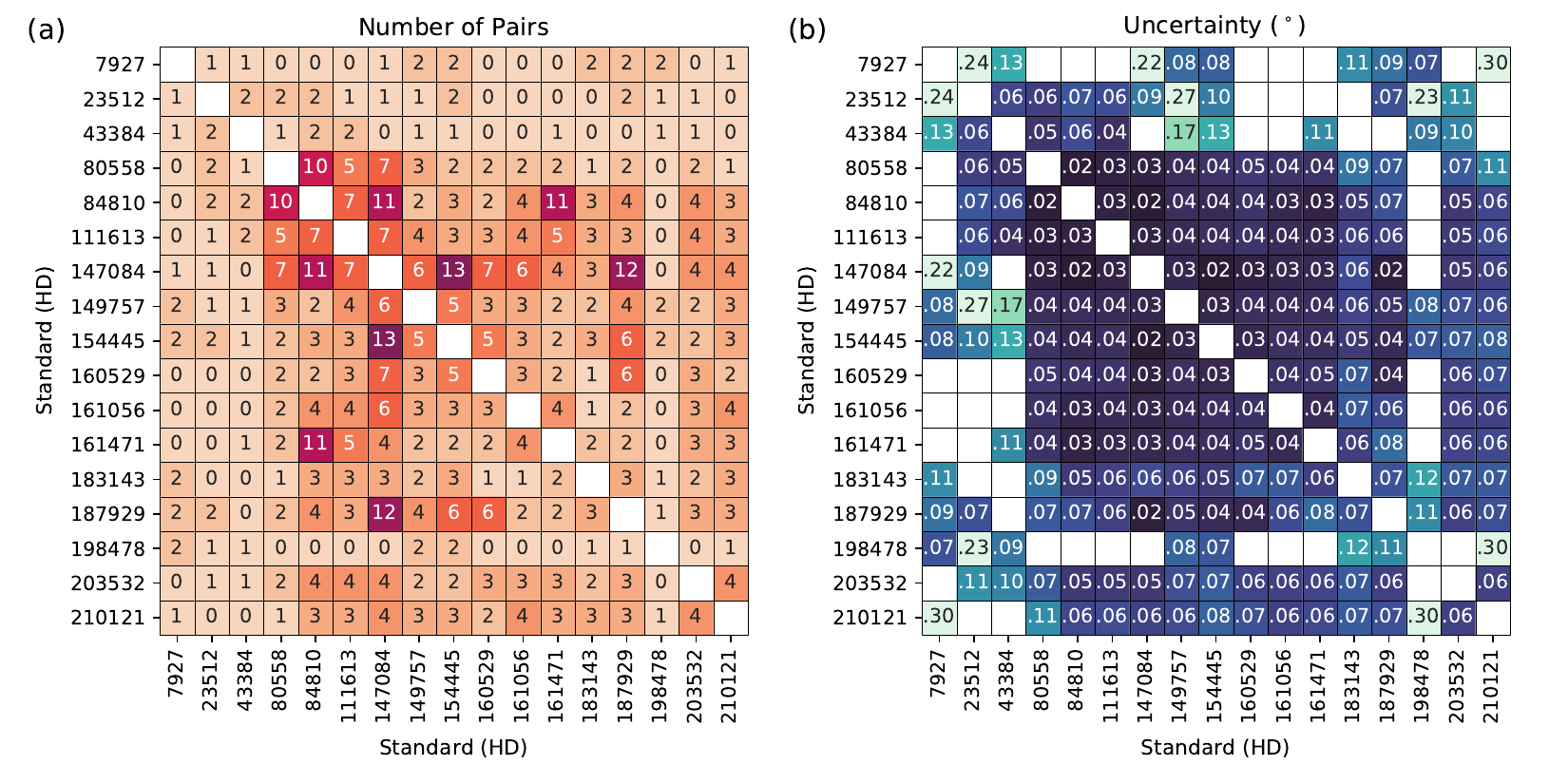}
    \caption{The number of standard pairs (a) and resulting uncertainty in degrees (b) for the CDM analysis. Colour gradients corresponding to the numerical values are used to aid visualisation: (a) low: beige, high: crimson; (b) high: light green, low: deep blue.} When a pair of standards are both observed during the same sub-run, that counts as one pair. Only $g^\prime$ observations made with the B PMT were used. In both panels the information is mirrored for easy reference.
    \label{fig:pairs}
\end{figure*}

We do not have repeat observations of any standards during the Gemini North run (N2018JUN). Hence, to provide a representative figure we make observation sets for four stars by combining $g^\prime$, 500SP and one Clear observation, wavelength corrected using our bandpass model according to the relations in Appendix \ref{apx:lit_PA}. This is less than ideal because the observations were taken sequentially, but they do at least probe a small range of paralactic angles for the Alt-Az mount. This is important because the telescope polarization during the run was large -- probably owing to an inhomogenously aluminized secondary \citep{Wiktorowicz23} -- and the position angle of its wavelength dependence was not well determined \citep{Bailey20}.

\subsection{Relative position angles using the Co-ordinate Difference Matrix}
\label{sec:RPA}

Informed by the relative performance of the instruments, we now seek to re-establish $\theta$ values for our standards using our data to facilitate literature comparisons. We use a co-ordinate difference matrix (CDM, \citealp{Baechler20}) approach to combine $\theta$ data from sub-runs where at least two standards were observed in $g^\prime$ (with the B PMT). Using a single filter reduces the data available, but ensures uncertainties in wavelength effects are minimised. For reasons explained in Sec.~\ref{sec:varPA}, we have also removed the Pindari observations from this calculation.

A relatively new tool, CDMs are similar to a Euclidian Distance Matrices (EDM, \citealp{Dattorro05}) and are finding applications in a number of different fields (e.g. \citealp{Mozaffari19, Krekovic20, Liu23, Chen24}). The CDM algorithm optimally calculates the relative mean differences between objects by combining such information from pairs of points. The algorithm is applicable even when the matrix is incomplete, i.e. when measurements do not exist for every object pair. Our application requires a weighted 1-dimensional CDM. From \citet{Baechler20}, algebraically, the CDM, $\vect{C}$ is made up of elements $C_{jk} = x_j - x_k$, i.e. the differences between points $x_j$ and $x_k$. We have a noisy CDM, \begin{equation} \widetilde{\vect{C}}=(\vect{C}+\vect{Z})\circ\vect{W}, \end{equation} where $\vect{Z}$ is a noise matrix and $\vect{W}$ the weight matrix. To optimally recover the set of points $\{x_{j}\}_{j=1}^{N}$ that generated it, the solution is \begin{equation}\vect{A}\vect{x} = \widetilde{\vect{\nu}},\end{equation} where $\vect{A} = \vect{\Lambda} - \vect{W}$ and $\widetilde{\vect{\nu}} = (\vect{C}\circ\vect{W})\vect{\vmathbb{1}}$, where $\vect{\vmathbb{1}}$ is the all-one vector and \begin{equation}\Lambda_{jk} = \left\{\begin{matrix}
\sum^N_{l=1} W_{jl} & j=k \\
0 & j\neq k \\
\end{matrix}\right..\end{equation} The first point is then fixed to zero in the algorithm by removing the first point in $\vect{x}$ and $\widetilde{\vect{\nu}}$ along with the first row and column of $\vect{A}$ and likewise for all the corresponding matrices.
From this process we take the nominal errors on the recovered $x_j$ points to be, \begin{equation}x_{e\,j}=\sqrt{1/\sum^N_{l=1} W_{jl}}.\label{eqn:CDMe}\end{equation}

Here, our $x$ terms are $\theta$ values, and we calculate the weights using the root-mean-square (RMS) sum of $e_{\rm m\,\theta}$ and $\eta(\epsilon_{\rm i\,\theta})$ for each measurement. Where multiple observations of a standard have been made during a sub-run, we take the error-weighted-average, and thus also the resulting RMS-error for the weight calculation. The runs are combined in the same way. The result therefore makes no account of stellar variability -- only mean position angles \mbox{are calculated}.

Figure \ref{fig:pairs}(a) shows the number of object pairs used in the calculation and \ref{fig:pairs}(b) the associated standard errors (where the weights are the inverse of the error squared). It should be noted that this does not match the number of observations because only a single agglomerated measurement is made per sub-run for a given standard, single-standard sub-runs are discarded, and because more standards were observed in some runs than others. 

The results of the CDM procedure are presented in columns 6--10 of Table \ref{tab:CDM} where the standard error, $\theta_e$, given in column 7 is the error in the mean $\theta$ given in column 6; this does not account for the error in the zero point of the co-ordinate system nor does it describe the distribution of $\theta$ values. Columns 9 and 10 are the weighted standard deviation of $\theta_{\rm obs} - \theta_{\rm pred}$, $\sigma_\theta$, and the average error, $\bar{e}_\theta$ (which is equal to $\theta_e\sqrt{N_o}$) after calculation of $\theta_t$ using the newly determined values of $\theta$ for each standard. Column 9 may be compared to column 5, which is the unweighted standard deviation after calibration of $\theta_t$ instead using $\theta_{\rm lit}$ -- the way calibration of $\theta_t$ has previously been done for HIPPI-class instruments. It can be seen that the CDM-derived $\sigma_\theta$ is reduced for most stars compared to the previous method. Our $\sigma_\theta$ and $\bar{e}_\theta$ terms are the equivalent of \citet{Bastien88}'s $\sigma_2(\theta)$ and $\sigma_1(\theta)$ respectively.

\begin{table*}
\caption{Position angle determinations from the CDM and iterative fitting methods}
\tabcolsep 4.4 pt
%\centering
\begin{tabular}{rrrrr|rrrrr|rrrrrrrr}
\toprule
\multicolumn{1}{c}{\textit{1}} & \multicolumn{1}{c}{\textit{2}} & \multicolumn{1}{c}{\textit{3}} & \multicolumn{1}{c}{\textit{4}} & \multicolumn{1}{c}{\textit{5}} & \multicolumn{1}{c}{\textit{6}} & \multicolumn{1}{c}{\textit{7}} & \multicolumn{1}{c}{\textit{8}} & \multicolumn{1}{c}{\textit{9}} & \multicolumn{1}{c}{\textit{10}} & \multicolumn{1}{c}{\textit{11}} & \multicolumn{1}{c}{\textit{12}} & \multicolumn{1}{c}{\textit{13}} & \multicolumn{1}{c}{\textit{14}} & \multicolumn{1}{c}{\textit{15}} & \multicolumn{1}{c}{\textit{16}} & \multicolumn{1}{c}{\textit{17}} &\multicolumn{1}{c}{\textit{18}} \\
\midrule
\multicolumn{5}{c}{} & \multicolumn{5}{l}{\textit{Co-ordinate Difference Matrix}}  & \multicolumn{8}{l}{\textit{Iterative fitting result}}   \\
\multicolumn{1}{c}{Standard} & \multicolumn{1}{c}{$N_o$} & \multicolumn{1}{c}{$N_r$} & \multicolumn{1}{c}{$\theta_{\rm lit}$} & \multicolumn{1}{c}{$\sigma_{\theta}$} &\multicolumn{1}{c}{$\theta$}  &  \multicolumn{1}{c}{$\theta_e$} & \multicolumn{1}{c}{$\Delta\theta$} & \multicolumn{1}{c}{$\sigma_{\theta}$} & \multicolumn{1}{c}{$\bar{e}_{\theta}$} & \multicolumn{1}{c}{$\theta$}  &  \multicolumn{1}{c}{$\theta_e$} & \multicolumn{1}{c}{$\Delta\theta$} & \multicolumn{1}{c}{$\sigma_{\theta}$}   & \multicolumn{1}{c}{$\bar{e}_{\theta}$}  &   \multicolumn{1}{c}{Sig.}  &   \multicolumn{1}{c}{$e_{\star\theta}$}  &   \multicolumn{1}{c}{$e_{\star\,u^\prime}$}\\

\multicolumn{1}{c}{(HD)}    &   & \multicolumn{1}{c}{}  &   \multicolumn{1}{c}{($^\circ$)}  & \multicolumn{1}{c}{($^\circ$)}    &   \multicolumn{1}{c}{($^\circ$)} &   \multicolumn{1}{c}{($^\circ$)} &   \multicolumn{1}{c}{($^\circ$)}  &   \multicolumn{1}{c}{($^\circ$)} &   \multicolumn{1}{c}{($^\circ$)} &   \multicolumn{1}{c}{($^\circ$)}   &   \multicolumn{1}{c}{($^\circ$)}  &   \multicolumn{1}{c}{($^\circ$)} &   \multicolumn{1}{c}{($^\circ$)}  &   \multicolumn{1}{c}{($^\circ$)}  &   \multicolumn{1}{c}{($\sigma$)}   &   \multicolumn{1}{c}{($^\circ$)}     & \multicolumn{1}{c}{(ppm)}\\
\midrule

  7927  &  10   &   3   &   93.0    &   0.331   &   93.187  &   0.032   &   $+$0.187    &   0.341   &   0.161   &   93.165  &   0.069   &   $+$0.165    &   0.344   &   0.170   &   2.1 &   0.321   &    367 \\ 
 23512  &   7   &   5   &   30.4    &   0.335   &   30.719  &   0.024   &   $+$0.319    &   0.262   &   0.104   &   30.706  &   0.064   &   $+$0.306    &   0.273   &   0.128   &   2.1 &   0.256   &    191   \\
 43384  &   11  &   3   &   170.0   &   0.269   &   170.309 &   0.021   &   $+$0.309    &   0.234   &   0.099   &   170.295 &   0.049   &   $+$0.295    &   0.199   &   0.115   &   1.7 &   0.167   &    171   \\
 80558  &   21  &   17  &   163.3   &   0.325   &   162.484 &   0.011   &   $-$0.816    &   0.360   &   0.086   &   162.512 &   0.045   &   $-$0.788    &   0.442   &   0.102   &   4.4 &   0.384   &    420   \\
 84810  &   33  &   20  &   100.0   &   0.321   &   99.989  &   0.010   &   $-$0.011    &   0.191   &   0.086   &   99.997  &   0.019   &   $-$0.003    &   0.132   &   0.098   &   1.3 &   0.110   &    60   \\
111613  &   17  &   10  &   80.8    &   0.366   &   80.835  &   0.011   &   $+$0.035    &   0.372   &   0.083   &   80.836  &   0.034   &   $+$0.036    &   0.317   &   0.096   &   3.1 &   0.266   &    281   \\
147084  &   53  &   30  &   31.8    &   0.315   &   32.015  &   0.009   &   $+$0.215    &   0.195   &   0.091   &   32.028  &   0.018   &   $+$0.228    &   0.141   &   0.110   &   1.4 &   0.118   &    155   \\
149757  &   15  &   8   &   127.2   &   0.374   &   126.200 &   0.012   &   $-$1.000    &   0.212   &   0.092   &   126.218 &   0.032   &   $-$0.982    &   0.252   &   0.104   &   1.9 &   0.211   &    98   \\
154445  &   24  &   16  &   90.0    &   0.342   &   89.976  &   0.011   &   $-$0.024    &   0.190   &   0.092   &   89.985  &   0.022   &   $-$0.015    &   0.132   &   0.109   &   1.1 &   0.110   &    134   \\
160529  &   11  &   8   &   20.0    &   0.637   &   18.749  &   0.013   &   $-$1.251    &   0.582   &   0.083   &   18.748  &   0.083   &   $-$1.252    &   0.659   &   0.099   &   6.7 &   0.640   &    1604   \\
161056  &   9   &   6   &   67.3    &   0.236   &   67.982  &   0.013   &   $+$0.682    &   0.090   &   0.084   &   68.034  &   0.024   &   $+$0.734    &   0.098   &   0.091   &   1.1 &   0.082   &    108   \\
161471  &   10  &   6   &   2.4     &   0.314   &   2.060   &   0.013   &   $-$0.340    &   0.153   &   0.081   &   2.087   &   0.027   &   $-$0.313    &   0.147   &   0.089   &   1.7 &   0.123   &    94   \\
183143  &   7   &   5   &   179.2   &   0.713   &   179.323 &   0.018   &   $+$0.123    &   0.777   &   0.096   &   179.299 &   0.121   &   $+$0.099    &   0.827   &   0.111   &   8.0 &   0.819   &    1710   \\
187929  &   27  &   17  &   93.7    &   0.457   &   93.711  &   0.013   &   $-$0.011    &   0.234   &   0.096   &   93.703  &   0.037   &   $+$0.003    &   0.288   &   0.127   &   2.2 &   0.241   &    142   \\
198478  &   12  &   2   &   3.0     &   0.660   &   2.474   &   0.031   &   $-$0.526    &   0.643   &   0.161   &   2.417   &   0.095   &   $-$0.583    &   0.647   &   0.170   &   3.8 &   0.628   &    595   \\
203532  &   7   &   6   &   126.9   &   0.263   &   124.328 &   0.017   &   $-$2.572    &   0.253   &   0.095   &   124.360 &   0.039   &   $-$2.540    &   0.201   &   0.107   &   1.8 &   0.175   &    81   \\
210121  &   8   &   6   &   155.1   &   0.282   &   153.836 &   0.019   &   $-$1.264    &   0.255   &   0.123   &   153.903 &   0.049   &   $-$1.197    &   0.287   &   0.132   &   2.0 &   0.264   &    126   \\

\bottomrule
\end{tabular}
\begin{flushleft}
\underline{Notes} -- Top row gives column numbers for easy reference. Analyses in Sec's \ref{sec:RPA} and \ref{sec:varPA} correspond to columns 6--10 and 11--18 respectively, whereas columns 4 and 5 correspond to ordinary calibration procedures. The symbols $N_o$ and $N_r$ denote the number of observations and subruns, respectively, from which data is drawn for each standard. The literature values of position angle are denoted $\theta_{\rm lit}$, whereas $\theta$ are the determined values from our analyses and, as elsewhere, given for a 2020 equinox; correspondingly $\theta_e$ is the nominal error on the determination from Equation~\ref{eqn:CDMe} (effectively the error on the mean), $\Delta\theta$ is the difference between $\theta$ and $\theta_{\rm lit}$; $\sigma_\theta$ is either the unweighted (column 5) or error-weighted (columns 9 and 14) standard deviations of position angle measurements after recalibration, and $\bar{e}_\theta$ is the average of all the nominal position angle errors for each standard. The scatter attributed to variability on each star is fit value $e_{\star\theta}$, Sig. the significance of that value (note that observations with larger errors are down-weighted in its calculation), and $e_{\star u^\prime}$ the minimum polarization needed to shift $\theta$ by $e_{\star\theta}$ if it acted perpendicular to the interstellar polarization. Note that $e_{\star\theta}^2 \approx \sigma_\theta^2 - \bar{e}_\theta^{\,2}$ (see the text of Sec.~\ref{sec:varPA} for details). The absolute co-ordinate system uncertainty, not included in the reported uncertainties is 0.177$^\circ$. \\
\end{flushleft}
\label{tab:CDM}
\vspace{-12pt}
\end{table*}

The CDM only calculates the \textit{relative} position angles of the standards (the value of the first listed object in the matrix is arbitrarily set to zero, so that $\theta_{\rm diff}$ represents the difference from it for each standard). For the absolute value we have calculated and applied an offset, $\zeta$, based on the literature values (as given in Table \ref{tab:lit_pa}). Here $\zeta$ is calculated by finding the median difference between the determined, $\theta_{\rm diff}$, and the literature, $\theta_{\rm lit}$ values,

\begin{equation}\zeta = \eta(\theta_{\rm diff\,\textit{1}} - \theta_{\rm lit\,\textit{1}}, \theta_{\rm diff\,\textit{2}} - \theta_{\rm lit\,\textit{2}}, ..., \theta_{\rm diff\,\textit{n}} - \theta_{\rm lit\,\textit{n}})\label{eq:med}\end{equation}

We estimate the error in $\theta_0$ by this method to be 0.177$^\circ$. In this case we calculated errors as the RMS sum of the values in Table \ref{tab:CDM} and 0.5$^\circ$ which we consider an appropriate typical uncertainty in the original literature measurements\footnote{The claimed accuracy is sometimes better than this. \citet{Hsu82} state 0.2$^\circ$ precision for instance, but we have taken an average of their work and others, and none estimate an error associated with the zero point offset.}. HD\,203532 is removed from the calculation as an obvious outlier. 

An alternative method for determining $\zeta$ would be to use the error-weighted average difference,
\begin{equation}\zeta = \frac{\sum_{j=1}^{n} (\theta_{\rm diff\,\textit{j}} - \theta_{\rm lit\,\textit{j}}) w_j}{\sum_{j=1}^{n}w_j}, \label{eq:wa}\end{equation} where the $w_j$ is the weight of the $j$th element, equal to the inverse of the error squared. \citet{Hsu82} used this method to tie their measurements to those of \citet{Serkowski74}.

Calculating $\zeta$ using Equation~\ref{eq:wa} rather than Equation~\ref{eq:med} results in all the values in Table \ref{tab:CDM} being shifted by $-0.210^\circ$, which is not so different from the zero point uncertainty calculated above. We prefer the median approach to mitigate sparse sampling of potentially variable objects. These values describe the possible offset of our entire network of standards but bare no relevance for their relative position angles.

\subsection{Estimating stellar variability in \(\theta\)}
\label{sec:varPA}

In Sec~\ref{sec:instrument_performance} we estimated the uncertainty associated with each instrument/telescope combination, we now wish to do the same for each standard star, $e_{\star\theta}$. A first step is to determine the (error weighted) standard deviation of position angles from their expected values for each standard -- after re-calibrating $\theta_t$ for each run -- these are the values in column 9 of Table \ref{tab:CDM}. The general formula for weighted standard deviation is \citep{Heckert03},
\begin{equation}{\sigma = \sqrt{\frac{\sum_{j=1}^{N} w_{j}(x_{j}-\bar{x})^2}{(\frac{N^\prime-1}{N^\prime})\sum_{j=1}^{N}w_j}},}\end{equation}
where $\bar{x}$ is the weighted mean, $w_j$ the weights on each element $x_j$, $N^\prime$ is the number of non-zero weights\footnote{$N^\prime=N-N_0$, where $N_0$ is the number of zero weighted terms.}, and the term $(N^\prime-1)/N^\prime$ comes about from Bessel's correction to the variance, applied when the population mean is being estimated from the sample mean.

However, estimating $e_{\star\theta}$ in this way is very crude because $\theta_t$ depends on the weighted means of each observation for each run. Typically, each run contains only a handful of measurements where we have considered each standard to be as good as any other. If just one star is variable and significantly different to its assumed value of $\theta$, that will shift the measurements of every other star in the run. The very parameters we aim to determine are corrupted by assuming them to be zero in the first instance.

To overcome this problem we employ a scheme that iteratively fits for $e_{\star\theta}$: For this we use \textsc{SciPy}'s optimize function, employing the Sequential Least SQuares Programming Optimizer (SLSQP), to minimise a function $f$,
\begin{equation}f=\left|\chi_{r}^{2}-1 \right|,\end{equation}
where $\chi_r^2$ is given by,
\begin{equation}\chi_r^2 = \frac{1}{d}\sum_{j}\frac{(O_j-E_j)^2}{\sigma_j^2},\end{equation}
where $d$ is the degrees of freedom, equal to $N_o$ less the number of fit parameters, $(O_j - E_j)$ is the difference between the observed and expected $\theta$ value and $\sigma_j$ the uncertainty associated with each measurement after re-calibration, which has four contributions,
\begin{equation}\sigma^2 = e_{\rm m\,\theta}^2 + e_{\rm i\,\theta}^2 + e_{\star\theta}^2 + e_{t\,\theta}^2,\end{equation}
$e_{\rm m\,\theta}$ is the measured uncertainty, as in equation \ref{eq:err1}; $e_{i\,\theta}$ and $e_{\star\theta}$ are the standard deviations of the variability (errors) associated with the telescope/instrument set-up and the star being observed, respectively, and $e_{t\,\theta}$ is the RMS error in the determined telescope zero point for the run (which, for each run, is derived from the other three terms). %It is the $e_{\star\theta}$ values that are fit to minimise $f$. 

However, $(O_j - E_j)$ depends on the values we assign $\theta$ for each star. The assignments are made using the CDM algorithm, which in turn depends upon the uncertainty assigned to each observation. Our assignments in Sec~\ref{sec:RPA} are made, essentially, with \mbox{$e_{\star\theta}=0$}. Therefore our fit function incorporates a recalculation of the CDM-derived assignments based on the current values of the fit parameters. 

We found that the fit function was prone to getting stuck in local minima, leading to some values of $e_{\star\theta}$ being inconsistent with the subsequently calculated values of $\sigma_\theta$. This occurs because the CDM calculates only one $\theta$ value per star per run, and the fit function just demands a model where the sample variance is appropriately described, it does not care where the uncertainty is placed. To overcome this problem, we used the $\sigma_\theta$ values determined after the fit as initial parameters for $e_{\star\theta}$ in subsequent fits and iterated until the $\sigma_\theta$ values converged (i.e. did not change by more than 5$\times$10$^{-5\,\circ}$ between subsequent iterations) though the changes are small after about the third iteration.

Initially we performed the analysis of Sec~\ref{sec:RPA} using the same set of observations as in Sec.~\ref{sec:instrument_performance}, but found that the three standards observed repeatedly at Pindari gave elevated values of $e_{\star\theta}$ compared to otherwise. We presume this to be due to the complications associated with not using a fixed mount, and so we excluded those observations from further analysis. We also removed the Clear observations; stellar variability is likely to have a wavelength dependence and we wanted an unbiased comparison between stars. We could not apply the same restriction in Sec.~\ref{sec:instrument_performance} since this would have left too few sets for some set-ups.

It follows that the uncertainty for each set-up could use some fine tuning. We tried fitting $e_{\rm i\,\theta}$ along with $e_{\star\theta}$ but this just transferred uncertainty from the set-ups to the stars, where the proportion was overly sensitive to the initial conditions. Though, if boundary conditions were enforced, the uncertainty transferred was mostly small. However, the ratios of $e_{\rm i\,\theta}$ to each other were typically only slightly changed, suggesting that the results of Sec.~\ref{sec:instrument_performance} are not biasing observations made with one set-up over another. It also means that if most stars are variable, our estimates of $e_{\star\theta}$ will underestimate the short term variability of the most stable stars. Ultimately we decided against fitting $e_{\rm i\,\theta}$, as this represented the most conservative approach, and retain the values assigned in Sec.~\ref{sec:instrument_performance}.

The results are given in columns 11-18 of Table \ref{tab:CDM}. There are only minor changes in $\theta$ (column 11) compared to the pure CDM results (column 6). There are marked decreases in $\sigma_\theta$ for the least variable stars, that are only partly offset by increases in the more variable stars. This result should be obvious because we are down-weighting the contributions of the most variable stars to achieve a more accurate determination of $\theta_t$. The errors are increased because they now include a component associated with $e_{\star\theta}$.

The significance of the determined $\sigma_\theta$ we calculate (in column 16 of Table \ref{tab:CDM}) by scaling each observation to the corresponding error, and then taking the error-weighted standard deviation. This is better than dividing $\sigma_\theta$ by $\bar{e}_\theta$ because it takes proper account of the different uncertainties associated with each observation. In column 17 the fit values of $e_{\star\theta}$ are given. Due to the nature of the fitting function $e_{\star\theta}^2 \approx \sigma_\theta^2 - \bar{e}_\theta^{\,2}$. Five stars are found to be variable with 3$\sigma$ confidence -- all of these are A or B type supergiants. Another four stars are seemingly variable only at the 2$\sigma$ level.

The final column of Table \ref{tab:CDM} is a calculation of polarization, following Equation~\ref{eq:theta}, that would be required in $u^\prime$ to rotate by $e_{\star\theta}$ assuming $q^\prime = p_{\rm lit}$. This allows the variability of the objects to be compared directly, independent of the interstellar polarization imparted by the ISM. 

\begin{figure*}
    \centering
    \includegraphics[angle=0, width=17.5cm, trim={0cm 0.1cm 0cm 0.3cm}, clip]{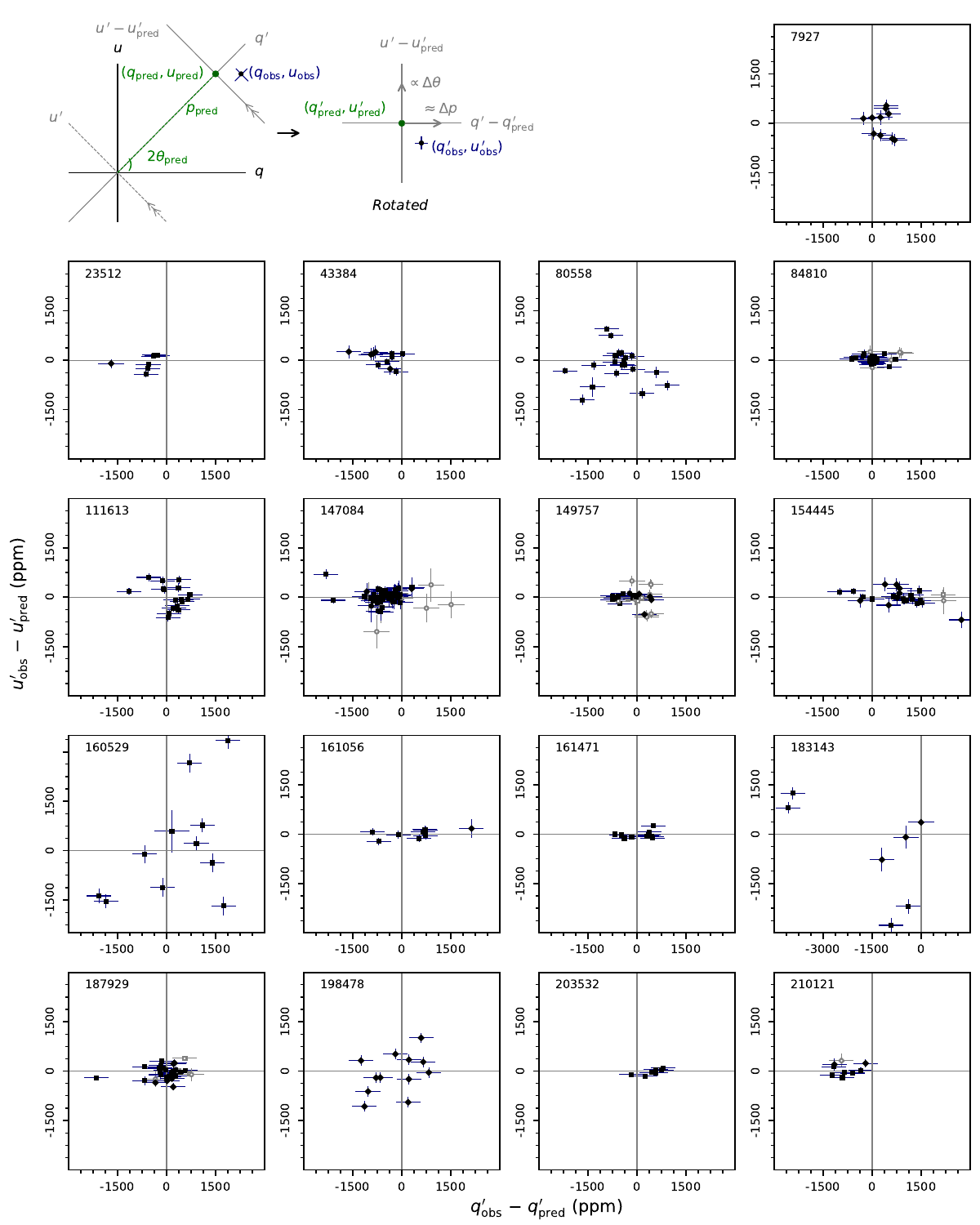}
    \caption{QU diagrams for each standard star showing the difference to $q^{\prime}_{\rm pred}$ and $u^{\prime}_{\rm pred}$ in parts-per-million. The prime indicates indicates the co-ordinate frame has been rotated so that $u^\prime \propto \Delta\theta$ is $0^\circ$ (up) in the diagrams, and $q^\prime \approx \Delta p$, as illustrated in the top left of the figure. In this case the predicted values have been taken from the literature values of $p$ and from $\theta$ as determined by the iterative fitting method, as reported in Table \ref{tab:CDM}. Clear observations are unfilled and in grey, $g^\prime$ are filled black points with navy error bars. The data presented are the same as those in Figures \ref{fig:PA} and \ref{fig:p}. Note that $p_{\rm pred}$ is derived from the literature values of $p$ without correction for the mean of observations. The one Clear point for HD\,160529 is not shown as it is out of range of the diagram.}
    \label{fig:qu}
\end{figure*}

The $u^\prime$ and $q^\prime$ values are depicted graphically as the vertical and horizontal axes, respectively, of QU diagrams in Fig.~\ref{fig:qu} (full explanatory details are in the caption). The error bars plotted for $q^\prime$ are derived below in Sec.~\ref{sec:varP}, but we present this figure here to make an important point: the improvements made in $\theta$ calibration have resulted in this being more accurate than our $p$ calibration. We have used three different modulators, where the performance of at least one of them has evolved over time, this affects modulation efficiency, including as a function of wavelength. Our ability to correct for efficiency changes is limited by the calibration data obtained (see Appendix \ref{apx:mod_recal}). But any change in modulator efficiency applies equally to each Stokes parameter so that the effects on $\theta$ are largely negated -- especially in the bands most closely corresponding to the optimum operating wavelength of the modulator (i.e. the $g^\prime$ band, see Table \ref{tab:eff_chk}) -- the instrumental errors in $\theta$ come from other sources (largely mechanical), as already discussed in Sec.~\ref{sec:instrument_performance}, and these are independently parameterised and accounted for. The consequences are easy to see in Fig.~\ref{fig:qu} -- where the co-ordinate system is rotated, for each object, so that $q^\prime$ corresponds to changes in $p$ and $u^\prime$ to changes in $\theta$ -- where the horizontal scatter is almost always greater than the vertical scatter. This means we cannot employ a CDF test reliably. However, where we see variability in $\theta$ we can be confident it is real, but if there is only scatter in $p$ this probably represents only instrumental variability.

\subsection{Variability in $p$}
\label{sec:varP}

The $p$ statistics for all of the available $g^\prime$ observations are given in Table \ref{tab:varP}, except for those acquired at the Pindari observatory, which are given in Table \ref{tab:PinVarP}. The Pindari observations have larger nominal (photometric) errors, but otherwise should be comparable.

\begin{table}
\caption{Mean and standard deviation of standards in $p$}
\tabcolsep 2.25 pt
%\centering
\begin{tabular}{lrrrrrrrrl}
\toprule
\multicolumn{1}{l}{Standard} &  \multicolumn{1}{c}{$N_o$}   &   \multicolumn{1}{c}{$p_{\rm pred}$}   &  \multicolumn{1}{c}{$\bar{e}_{\rm m\,\textit{p}}$} &   \multicolumn{2}{c}{$\bar{\Delta p}$} &   \multicolumn{2}{c}{$\sigma_{\rm \Delta p}$}  &   Sig.\\
\multicolumn{1}{c}{(HD)}    &       &   \multicolumn{1}{c}{(ppm)}           &   \multicolumn{1}{c}{(ppm)}       &  \multicolumn{1}{c}{(ppm)} & \multicolumn{1}{c}{(/$p_{\rm pred}$)}&  \multicolumn{1}{c}{(ppm)} & \multicolumn{1}{c}{(/$p_{\rm pred}$)}    &   \multicolumn{1}{c}{($\sigma$)}\\
\midrule
\phantom{00}7927    &  10  & 32817 &    39  &   $+$295  & $+$0.0090 &   285 & 0.0087 &  0.8\\
\phantom{0}23512    &   7  & 21387 &    53  &   $-$645  & $-$0.0301 &   443 & 0.0206 &  1.2\\
\phantom{0}43384    &  11  & 29385 &    41  &   $-$592  & $-$0.0201 &   438 & 0.0149 &  1.2\\
\phantom{0}80558    &  25  & 31285 &    26  &   $-$482  & $-$0.0154 &   696 & 0.0222 &  1.9\\
\phantom{0}84810    &  34  & 15576 &    13  &    $+$39  & $+$0.0025 &   297 & 0.0191 &  0.9 & *\\
111613              &  17  & 30228 &    19  &   $+$139  & $+$0.0046 &   440 & 0.0146 &  1.2\\
147084              &  53  & 37512 &    29  &   $-$541  & $-$0.0144 &   454 & 0.0120 &  1.2 & *\\
149757              &  15  & 13310 &     9  &   $-$202  & $-$0.0153 &   383 & 0.0288 &  1.1\\
154445              &  25  & 35047 &    27  &   $+$777  & $+$0.0222 &   765 & 0.0219 &  2.1\\
160529              &  11  & 71769 &    59  &   $-$307  & $-$0.0043 &  1299 & 0.0181 &  3.7\\
161056              &   9  & 37924 &    25  &   $+$418  & $+$0.0110 &   853 & 0.0225 &  2.4\\
161471              &  10  & 21977 &     6  &    $+$75  & $+$0.0034 &   430 & 0.0196 &  1.5 & *\\
183143              &   7  & 59797 &    61  &  $-$1551  & $-$0.0260 &  1582 & 0.0265 &  4.6\\
187929              &  28  & 16828 &    22  &    $-$79  & $-$0.0047 &   495 & 0.0293 &  1.2 & *\\
198478              &  12  & 27162 &    35  &   $-$189  & $-$0.0070 &   715 & 0.0263 &  2.0\\
203532              &   7  & 13287 &    23  &   $+$464  & $+$0.0350 &   308 & 0.0232 &  0.9\\
210121              &   8  & 13688 &    48  &   $-$809  & $-$0.0591 &   365 & 0.0267 &  1.0\\
\midrule
\multicolumn{3}{l}{\textit{GN run excl.}}\\
147084              &  52  & 37512 &    28  &   $-$557  & $-$0.0148 &   443 & 0.0117 &  1.2 & *\\
154445              &  24  & 35049 &    26  &   $+$695  & $+$0.0198 &   666 & 0.0190 &  1.8\\
161056              &   8  & 37914 &    22  &   $+$204  & $+$0.0054 &   637 & 0.0168 &  1.8\\
210121              &   7  & 13688 &    48  &   $-$896  & $-$0.0655 &   302 & 0.0221 &  0.9\\
\bottomrule
\end{tabular}
\begin{flushleft}
\underline{Notes} -- * Indicates the significance has been calculated using both the data from this table and that from Table \ref{tab:PinVarP}.
\end{flushleft}
\label{tab:varP}
\vspace{-12pt}
\end{table}

\begin{table}
\caption{Pindari mean and standard deviation of standards in $p$}
\tabcolsep 2.5 pt
%\centering
\begin{tabular}{lrrrrrrr}
\toprule
\multicolumn{1}{l}{Standard} &  \multicolumn{1}{c}{$N_o$}   &   \multicolumn{1}{c}{$p_{\rm pred}$}   &  \multicolumn{1}{c}{$\bar{e}_{\rm m\,\textit{p}}$} &   \multicolumn{2}{c}{$\bar{\Delta p}$} &   \multicolumn{2}{c}{$\sigma_{\rm \Delta p}$}\\
\multicolumn{1}{c}{(HD)}    &       &   \multicolumn{1}{c}{(ppm)}           &   \multicolumn{1}{c}{(ppm)}       &  \multicolumn{1}{c}{(ppm)} & \multicolumn{1}{c}{(/$p_{\rm pred}$)}&  \multicolumn{1}{c}{(ppm)} & \multicolumn{1}{c}{(/$p_{\rm pred}$)}\\
\midrule
\phantom{0}84810    &  45  & 15659 &    87  &   $+$245  & $+$0.0157 &   381 & 0.0243 \\
147084              &   1  & 37857 &   130  &  $-$1122  & $-$0.0297 &    &  \\
161471              &  57  & 22112 &    56  &   $-$533  & $-$0.0241 &   516 & 0.0234 \\
187929              &   7  & 16856 &    96  &   $+$106  & $+$0.0063 &   206 & 0.0123 \\
\bottomrule
\end{tabular}
\begin{flushleft}
%\underline{Notes} * .
\end{flushleft}
\label{tab:PinVarP}
\vspace{-12pt}
\end{table}

Some standards do not agree as well with the predictions as others. The mean disagreement is quantified in Table \ref{tab:varP}, in terms of the unweighted mean difference, $\bar{\Delta p}$. HD\,210121 is almost 6 per cent under prediction as a ratio. Other stars that differ by more than two per cent are HD\,23512 and HD\,183143 which are also underpolarized; and HD\,203532 and HD\,154445 which are overpolarized. 

Conservatively, a long term change in $p$ from the literature value would be indicated for any star with $\bar{\Delta p} > 3\sigma_{\rm \Delta p}$. None of the standards listed in Table \ref{tab:varP} meet this condition. The nearest is HD\,210121 which is significant only at $2\sigma$. Some correction is, however, probably still justified, but because the measurements are monochromatic it is not possible to say if it is $p_{\rm max}$ or $\lambda_{\rm max}$ that is different. Indeed, we have calibrated the modulator polarization efficiency to all the standard observations collectively, so all that one can say is that several stars deviate in $p$ compared to others based on the source literature, we can't say which are inaccurate. 

\begin{figure}
    \centering
    \includegraphics[width=\columnwidth, trim={0.5cm 0.5cm 0.5cm 0cm}]{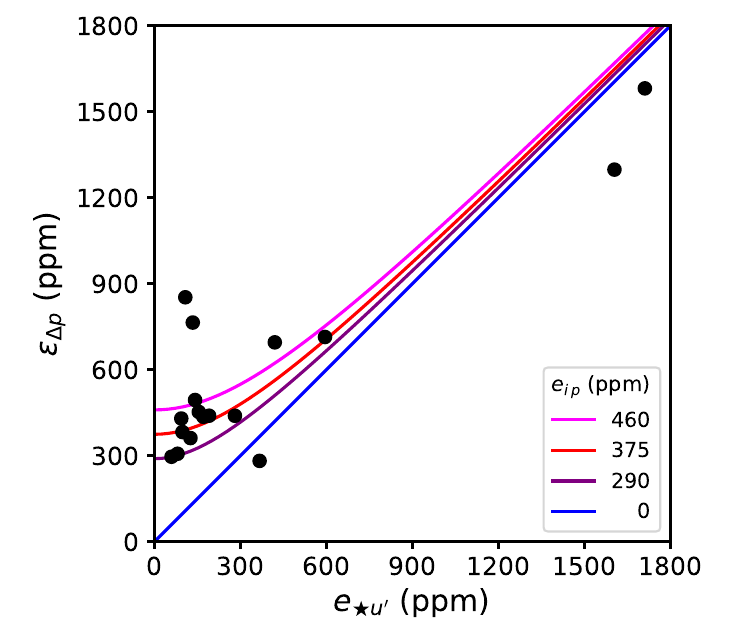}
    \caption{A consistency check for our determination of stellar variability in $\theta$, where $\epsilon_{\Delta p}$ is derived from $p$ and $e_{\star\,u^\prime}$ from $\theta$ measurements (from Table \ref{tab:CDM}) we expect objects to be evenly distributed around equality, as given by the blue line when all errors are properly accounted for. The purple, red and magenta lines represented an additional RMS error, $e_{i\,p}$, according to the legend. The red line, 375~ppm, best divides the data points. Note: see Fig. \ref{fig:qu} for an explanation of the relationship between $u^\prime$ and $\theta$.}
    \label{fig:check}
\end{figure}

Table \ref{tab:varP} also reports the unweighted standard deviation in $\Delta p$ compared to the mean, and the mean measured error, $\bar{e}_{\rm m\,\textit{p}}$. If there are no further contributions to the error in polarization, $e_p$, then the scatter in $p$ attributed to unaccounted for noise sources, $\epsilon_{\Delta p}=\sqrt{\sigma_{\Delta p}^2-\bar{e}_{\rm m\,\textit{p}}^2}$ represents the stellar variability in $p$, i.e. $e_{\star\,p}$ or equivalently $e_{\star\,q^\prime}$ if we rotate the reference frame in the same way as first done in Sec.~\ref{sec:RPA} (i.e. this is the counter-part to column 18 in Table~\ref{tab:CDM}). In Figure \ref{fig:check} we have plotted $\epsilon_{\Delta p}$ against $e_{\star\,u^\prime}$. Owing to the large interstellar polarizations of these stars, whether $e_{\star\,u^\prime}$ or $e_{\star\,q^\prime}$ is larger, should be purely a matter of chance. We would therefore expect an equal number of points to fall either side of the blue line if $p_e$ is dominated by $e_{\rm m\,\textit{p}}$. This is clearly not the case, so there is an additional instrumental error, $e_{i\,p}$, that contributes to $\epsilon_{\Delta p}$ that needs to quantified, i.e. \begin{equation}\epsilon_{\Delta p}=\sqrt{\sigma_{\Delta p}^2-\bar{e}_{\rm m\,\textit{p}}^2-e_{i\,p}^2}.\end{equation} 

Many more points are above the blue line in Fig.~\ref{fig:check} at low values of $e_{\star\,u^\prime}$, so a fixed error is more appropriate than one that scales with $p$. In re-calibrating the modulators (Appendix \ref{apx:accuracy}), we found a typical disagreement between $p_{\rm obs}$ and $p_{\rm pred}$ of 460~ppm. Using this figure overestimates the error (magenta line in Fig.~\ref{fig:check}) because it incorporates inaccuracies in the literature values of $p$ into the metric. If instead we take the median difference to the mean value of $\Delta p$ for each star, the result is $\approx$290~ppm for the observations in Table \ref{tab:varP} and $\approx$155~ppm for the Pindari observations. The fact the precision is better in $p$ for the Pindari observations, which used only a small telescope but a single instrument/telescope set-up, is evidence that variation in instrument and set-up is the dominant source of the extra error. The purple line in Fig.~\ref{fig:check} represents 280~ppm of added error; this appears to be an underestimate -- too many points fall above the line, probably as a consequence of taking the mean from small datasets with uneven temporal sampling. The red line in the same figure is the intermediate value of 375~ppm, which more evenly divides the points, and which we adopt as $e_{i\,p}$\footnote{Adopting this value is still consistent with an assumption of Gaussian behaviour in both $\theta$ and $p$, since $p/e_p \gtrapprox$ 500 for all observations.}.

Having established a value for $e_{i\,p}$, we can now calculate error-weighted figures for $\sigma_p$ and $\bar{e}_p$ to determine the significance of the result, this is given in the final column of Table \ref{tab:varP}. Two stars are found to be variable at 3$\sigma$ sigma significance in $p$; both are also 3$\sigma$ significant variables in $\theta$. Another four stars are 2$\sigma$ significant;  two of these are not similarly significant in $\theta$: HD\,154445 and HD\,161056; they both have early B spectral types. HD\,111613 is a 3$\sigma$ significant variable in $\theta$, but does not approach this level of significance in $p$.

\vspace{-0.15cm}
\subsection{Variability time series}

\begin{figure*}
    \centering
    \includegraphics[angle=0, width=17.5cm]{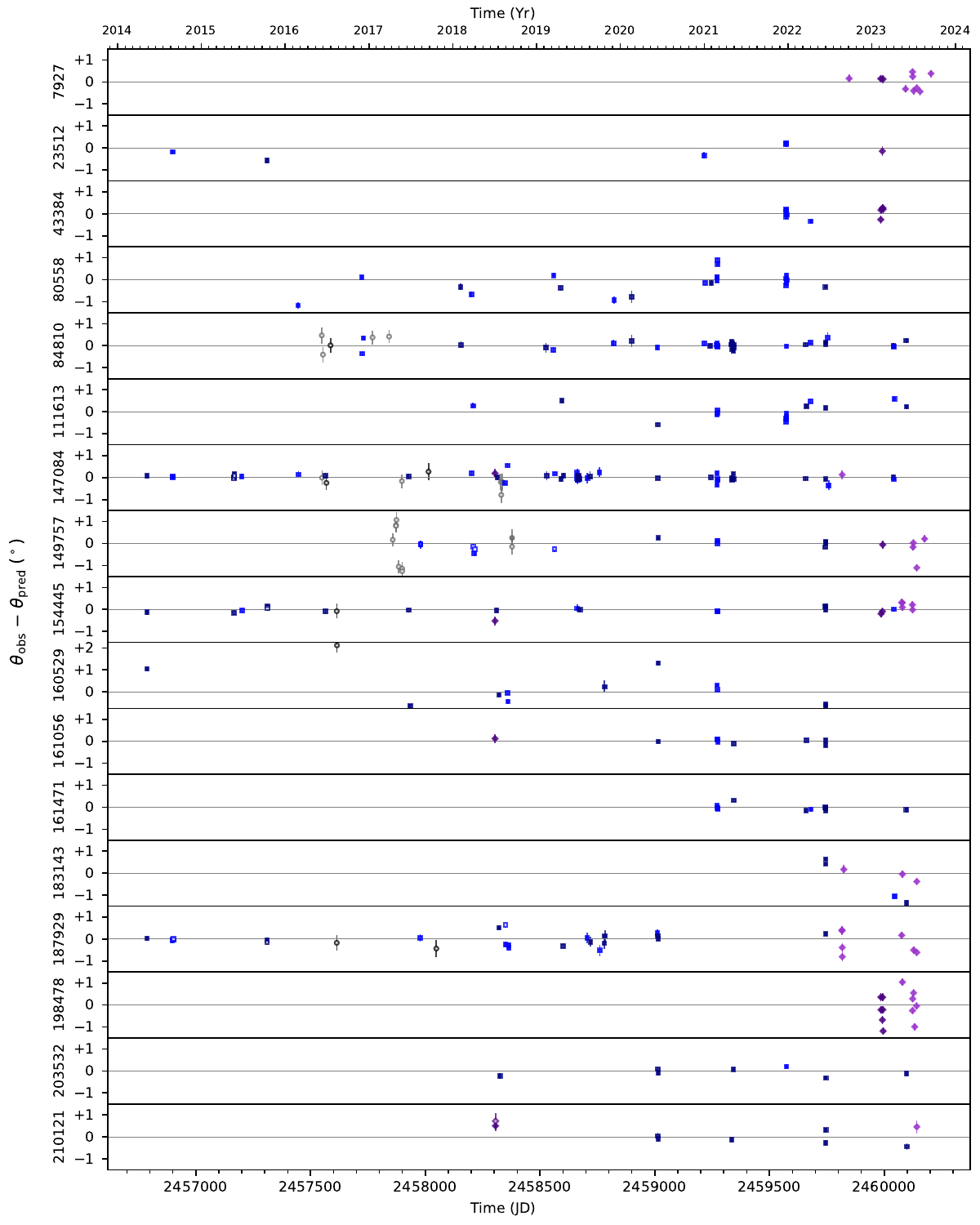}
    \caption{Position angle variation with time for each standard listed by its HD\,catalogue number. Filled circles are $g^\prime$, open symbols are Clear. Runs are displayed alternately in light/dark shading, with purple diamonds denoting northern HIPPI-2 runs, blue squares southern HIPPI/-2 runs and Mini-HIPPI runs with black circles. Errors incorporate both the individual observation errors, $e_{o\,\theta}$, and errors in the zero-point determination for the run, $e_{t\,\theta}$.}
    \label{fig:PA}
\end{figure*}

\begin{figure*}
    \centering
    \includegraphics[angle=0, width=17.5cm]{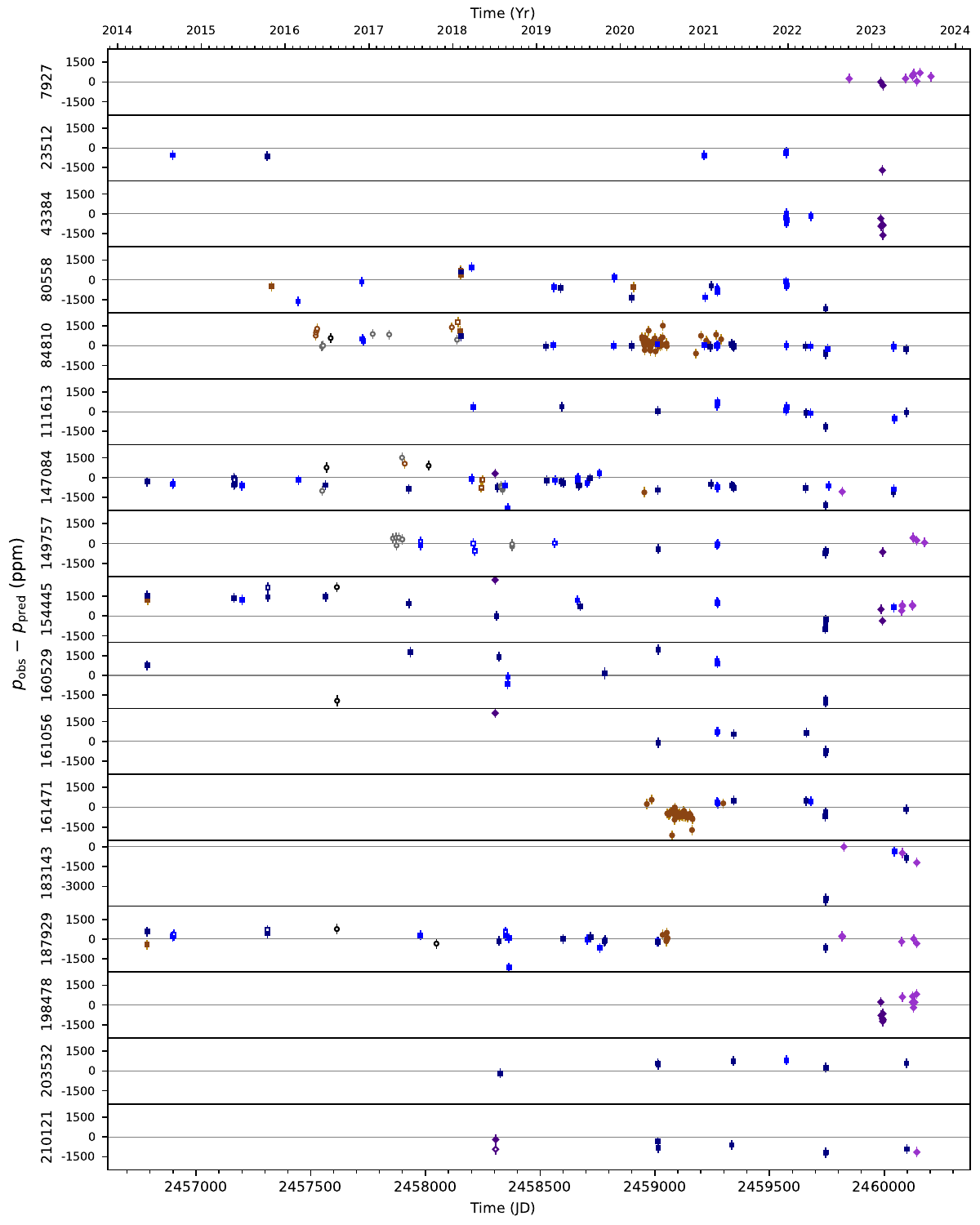}
    \caption{Polarization variation with time for each standard listed by its HD\,catalogue number. Symbols for data common with Fig.~\ref{fig:PA} are the same, i.e. filled circles are $g^\prime$, open symbols are Clear. Runs are displayed alternately in light/dark shading, with purple diamonds denoting northern HIPPI-2 runs, blue squares southern HIPPI/-2 runs and Mini-HIPPI runs with black circles. Additional data, either from Pindari or single standard runs are coloured brown.}
    \label{fig:p}
\end{figure*}

Here we present the $\theta$ and $p$ data for each star as a time series in Figs. \ref{fig:PA} and \ref{fig:p}, respectively. The latter includes data from the Pindari observatory, whereas the former does not, but otherwise excludes any point not on the first plot. Both figures include not just the $g^\prime$ data from runs with multiple stars observed, as used in Sec's \ref{sec:RPA} and \ref{sec:varPA}, but also Clear data (open symbols), and $g^\prime$ and Clear data which were re-calibrated based on Clear observations (where there are insufficient $g^\prime$ standard observation for the run). The additional data may be less accurate but is useful to fill in gaps in the time series (e.g. the 2016 HD\,160529 datum). We have colour coded the observations by run as described in the caption. This is important because despite our precautions it is still a more precise matter to compare observations made within the same run. The error bars incorporate the error in $\theta_t$, but if comparing observations intra-run the uncertainty is less than this.

Long term trends, or regular periodicity -- where sufficiently sampled -- would be revealed in Figs. \ref{fig:PA} and \ref{fig:p}, but neither behaviour is obvious. Considering both Figs. \ref{fig:PA} and \ref{fig:p}, the timescales of variability in $\theta$ and $p$ mostly appear correlated. For instance, the fast variability of HD\,198478 is very easy to see within a run. Whereas the slower, but no less pronounced, variability seen in HD\,160529 or HD\,80558 occurs on longer timescales. 

A noteworthy discrepancy is seen in Fig.~\ref{fig:p} when comparing data from two early 2018 runs for HD\,154445 -- this is the only abrupt change in polarization observed for this star and is most likely not real. The Gemini North run (2018JUN) has the least reliable calibration, and this represents the first of these data points. Closer inspection reveals that all four observations from this run appear over-polarized. If we remove these points from the analysis in Sec.~\ref{sec:varP} (see below the mid-rule in Table \ref{tab:varP}) then the significance of variability in $p$ for both HD\,154445 and HD\,161056 falls below 2-$\sigma$, and so we regard neither as a variability candidate.

\vspace{-0.15cm}
\section{Discussion}
\label{sec:discuss}

\subsection{Changes in {$\theta$} over decades?}
\label{sec:decades}

Four stars have refined $\theta$ values (column 11 in Table \ref{tab:CDM}) significantly different to the literature using the criteria $|\Delta \theta| > 3\sigma_{\theta}$: HD\,149757, HD\,161056, HD\,203532 and HD\,210121 (other differences might be ascribed to stellar variability). This could indicate a slow change in $\theta$ over many decades. However, inspection of Fig.~\ref{fig:PA} does not support this notion. The data for these four stars do not hint at a long term trend despite spanning 4 years or more. If there is one it must only be apparent on longer time scales. 

We sourced $\theta_{\rm lit}$ from only \citet{Bagnulo17} for two stars: HD\,161056 and HD\,210121. HD\,203532, includes data from \citet{Bagnulo17} and \citet{Serkowski75}. Together, these are three of the five stars where $\theta$ was sourced from \citet{Bagnulo17} (another star sourced from \citet{Bagnulo17} is HD\,80558, it has a large $\Delta \theta$ value, but is apparently also more variable). This is curious, because \citet{Bagnulo17} is our most recent reference source, leading us to discount a slow drift in $\theta$ over time. Of the stars mentioned, three were observed by \citeauthor{Bagnulo17} in 2015 and have large negative $\Delta \theta$ values, HD\,161056 was observed later in 2017 and has a large positive $\Delta \theta$ value. We conclude the difference is probably due to inaccurate $\theta_t$ calibration with FORS2 on the VLT. 

This leaves HD\,149757 as the only star of interest; for it we sourced $\theta_{\rm lit}$ from \citet{Serkowski75}, modified for $\Delta\theta/\Delta\lambda$ with data from \citet{Wolff96} to give 127.2$^\circ$. Our determination is almost 1$^\circ$ below this. The spectropolarimetry of \citet{Wolff96} produces a figure of $\approx$125$^\circ$ for the $g^\prime$ band, while that of \citet{Wolstencroft84} gives $\approx$126$^\circ$. $B$ band observations described by \citet{McDavid00} range from $129^\circ\pm0.4$ to $126.2^\circ\pm0.2$; this seems like a lot of variation but it is not too different to that reported for other standards in \citet{McDavid00}'s agglomerated tables. Fig.~\ref{fig:PA} shows that few, if any, of these stars are really variable in $\theta$ over such a range.

More tellingly, perhaps, is the difference in $\Delta\theta/\Delta\lambda$ behaviour between \citet{Wolff96} and \citet{Wolstencroft84} -- the slope is completely opposite in the two cases. This may point to intrinsic polarization that is episodic in nature. This would be in keeping with the irregular variability of the star's photometry and spectroscopy (see Sec.~\ref{sec:zetOph}). If so, it is not captured by the $g^\prime$ observations we analysed, we can assign only 0.211$^\circ$ to stellar variability. However, there is some evidence for a more active era early in 2017 in Clear observations made with Mini-HIPPI at the UNSW observatory -- these can be seen to range over $\sim2^\circ$ in \mbox{Fig.~\ref{fig:PA}}.

\vspace{-0.15cm}
\subsection{Variability in $\theta$ compared to the literature}
\label{sec:litcomp}

\subsubsection{Literature data}

\begin{figure}
    \centering
    \includegraphics[angle=0, width=\columnwidth, trim={1.2cm 0.6cm 1.8cm 0.2cm}, clip]{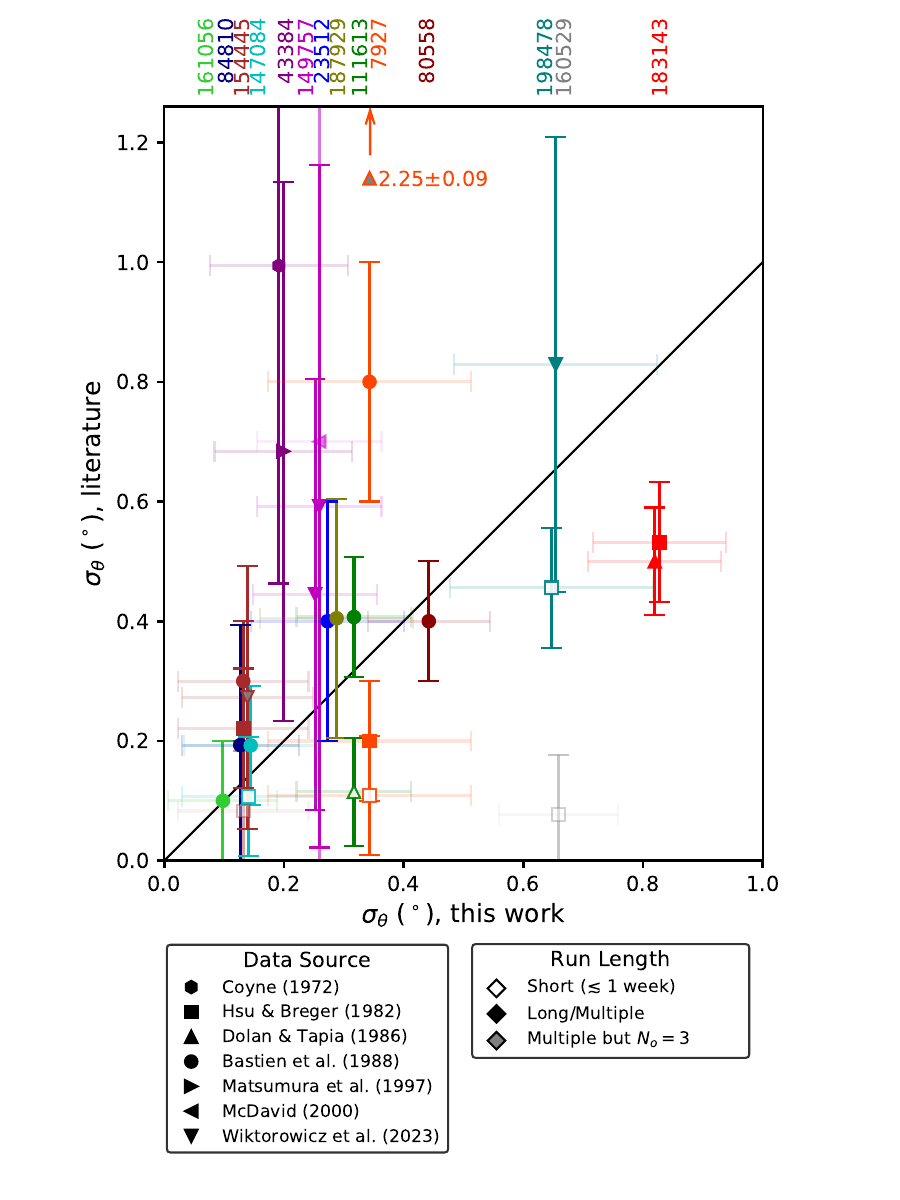}
    \caption{A comparison of standard $\theta$ variability reported in the literature to our determinations, $\sigma_{\theta}$; errors are $\bar{e}_\theta$. The black diagonal line indicates equality. Insignificant literature data is shown as partly transparent. A few data points have been offset by $<0.01^\circ$ and our errors are plotted with high transparency for the sake of clarity. The error bars for \citet{McDavid00}'s HD\,149757 value are not contained within the plot -- the value is 0.7$^\circ\pm$0.9. The positive error-bar for \citet{Coyne71}'s HD\,43384 value is not contained within the plot -- the value is 1.0$^\circ\pm$0.5. Many of the literature values come from very short runs or are derived from only half a dozen or so data points. For a full description of the source data, see the text. In general, there is good agreement between our determinations and others where their observing runs are of a substantial length.}
    \label{fig:litcomp}
\end{figure}

In Fig.~\ref{fig:litcomp} we compare our determinations of intrinsic stellar variability to those available in the literature. As noted by \citet{NaghiZadeh91} only partial data is presented in some of these sources, and others make only a handful of measurements. Since there are so few determinations available, we employ very relaxed criteria to include as many comparisons as possible.

The core of the literature determinations presented come from \citet{Bastien88} who derived variability in $\theta$ from the standard deviation of $q$ and $u$ ($\sigma_2(\theta)$ comparable to our $\sigma_{\theta}$) and compared that to the average error, $\sigma_1(\theta)$ (equivalent to our $\bar{e}_\theta$). Their data includes typically tens of observations per star agglomerated from half a dozen different sources that made use of one of two of the better instruments available between 1983 and 1986.

\citet{Hsu82} present only representative data. They describe the data in detail, but do so in inconsistent ways that are not conducive to analysis. They actually observed each star tens of times between 1979 and 1981 \citep{Hsu82b}, but this is lost to us. All we have been able to do is digitise the data they plotted in their Fig.~1~and~2. We take their typical reported error of 0.1$^\circ$ as representative of $\bar{e}_\theta$ and calculate the standard deviation of the presented data to derive $\sigma_{\theta}$. The data in their Fig.~1 represents a single 6 night observing run from August of 1981, Fig.~2 adds observations from 4 nights in September of the same year. We neglect the $R$ band data available for one target and use only the $V$ band observations.

\citet{Dolan86} observed three of the same stars we did. They were mostly concerned with $\Delta\theta/\Delta\lambda$ but in their \mbox{Table 1} they present an average $\theta$ from all bands for 5 observations of HD\,7927 from 1984, 3 of HD\,111613 from 1980 and 1984, and 8 of HD\,183143 from 1980 and 1984. We calculate standard deviations from this data. They report that their standard error is 0.09$^\circ$ based on a test from injecting a polarized signal into their system.

From \citet{Wiktorowicz23} we take the $B$ band data for 3 stars, including HD\,154445 which has only 3 observations; the others between 6 and 13. All the data sets span many years. Two data sets, from two different telescopes, are presented for HD\,149757, we have plotted both. They give what is effectively $\bar{e}_\theta$ directly in their Table~17, and we calculate $\sigma_{\theta}$ from their $q_{\rm var}$ and $u_{\rm var}$ values in their Table~18 using \citet{Bastien88}'s formula.

The variability of HD\,43384 has been much discussed (see Sec.~\ref{sec:9Gem}), we have two sources of data for HD\,43384. \citet{Coyne71} presented data from 1969 to 1971 in two bandpasses, we take a high $S/N$ ($e_p\leq0.2\%$) subset of the more extensive dataset ($G$ band) which then totals 9 observations; $\theta$ is only reported to the nearest degree, and so we add an additional 0.5$^\circ$ RMS error.  \citet{Matsumura97} made 17 observations relative to HD\,21291\footnote{Also claimed variable by \citet{Dolan86} on the basis of complex $\Delta\theta/\Delta\lambda$ and by comparison with \citet{Wolf72}.} with a single instrument with a typical error of 0.4$^\circ$ between 1991 and 1996. In the paper these data are plotted against phase assuming a 13.70 day period, with a sine curve fit. We digitised the points and calculated the standard deviation.

Finally, the values for one point for HD\,149757 comes from \citet{McDavid00}, who reports an insignificant variability based on agglomerated data from 1949 to 1996; the figures we have used come from the ``Grand Average'' of multi-band data in their Table 11 (though all the bands are pretty similar).

\subsubsection{Comparison}

The picture that emerges from Fig.~\ref{fig:litcomp} is one of surprisingly little disagreement. In particular \citet{Bastien88}'s determinations are quite a good match for ours; with the only notably discrepant point being HD\,7927. There are actually four literature determinations of HD\,7927's variability, and they disagree wildly. Our data on this target spans only a year. So, if the different estimates are to be reconciled it can only be through an appeal to episodic behaviour.

More broadly, it can be seen that variability determinations made from very short observing runs (open symbols in Fig.~\ref{fig:litcomp}) yield low values (the exception is HD\,198478 which exhibits obvious variability within a few nights). Amongst such determinations, we find good agreement with HD\,147084 and HD\,154445 which are two of the least variable stars in $\theta$. In every other case there is disagreement with our data that spans years.

We do find agreement, largely within 1-$\sigma$, with almost everything that has been observed over a longer period. The biggest disagreements, still within 2-$\sigma$, are for HD\,43384 which has some of the largest literature uncertainties, and HD\,183143 for which we have only 7 observations over a year's duration.

Despite the misgivings expressed by \citet{Clarke94} and \citet{NaghiZadeh91}, our analysis is consistent with the findings of earlier authors that many standards are in fact variable. A casual perusal of stellar polarimetry papers from the last 40 years will reveal that observers have largely ignored these reports of variability. This is partly because bright high polarization stars are not that common and alternative calibration methods not readily available. Therefore, the point is not to be able to declare a star variable or not based on statistical criteria, since this will not help observers much. There is enough evidence to suggest variability is common that we should consider it so. Yet, without quantification, this is an equally unhelpful statement. Any departure from spherical symmetry will produce a polarized signal, so it is a truism to write that everything is a polarimetric variable. The important question then is not \textit{which standards are variable?} but rather \textit{what uncertainty should we ascribe to our measurements when using these standards?} Our contention is that a careful analysis will assume the best estimate available, even if it is smaller than the formal detection threshold.

\subsection{Astrophysical inferences}

The data presented in this paper represents a rare opportunity to study the long-term small-scale polarimetric variability of high polarization stars with precision. In Table \ref{tab:qu_var} -- broken down as early-type supergiants, late-type supergiants, and other stars; and ordered by spectral type -- the values of $e_{\star u^\prime}$ and $e_{\star q^\prime}$ are tabulated; The final column combines the two figures in a variability metric, 
\begin{equation}e_\star = \sqrt{e_{\star q^\prime}^2 + e_{\star u^\prime}^2}.\end{equation} The groupings reveal B/A-type supergiants to be more variable as a category than either the F/G-type supergiants or the stars of other classes. In Figures \ref{fig:PA} and \ref{fig:p} time series data is plotted, by star, in $\theta$ and $p$, which we use along with the table and the QU diagrams in Fig.~\ref{fig:qu} to elucidate the nature of the variability below.

A thorough review of the QU patterns associated with different polarigenic mechanisms is given by \citet{Clarke10}. In brief, periodic mechanisms, relating to binarity or persistent surface features -- like magnetic spots -- will be revealed by loops or figure-of-eight-type patterns. Polarization associated with a Be mechanism -- material ejected from the equator followed by slow decretion from a disk -- which has a preferred orientation, will fall predominantly along a straight `intrinsic line'. Polarization generated by randomly distributed clumps of gas within a (perhaps irregularly driven) stellar wind will manifest as a scatter plot. Whereas, an interstellar polarization drift would probably look like a \mbox{random walk}, but predominantly in $p$.

\begin{table}
\caption{Stellar variability by spectral class and type}
\tabcolsep 8 pt
%\centering
\begin{tabular}{lcrrrc}
\toprule
\multicolumn{1}{l}{Standard} &  \multicolumn{1}{c}{SpT}   &   \multicolumn{1}{c}{$e_{\star q^\prime}$}   &  \multicolumn{1}{c}{$e_{\star u^\prime}$} &  \multicolumn{1}{c}{$e_\star$}  & GCVS\\
\multicolumn{1}{c}{(HD)}    &       &   \multicolumn{1}{c}{(ppm)}           &   \multicolumn{1}{c}{(ppm)}      &   \multicolumn{1}{c}{(ppm)}  &\\
\midrule
\multicolumn{2}{l}{\textit{B/A-supergiants}} \\
198478              &  B3\,Ia   & 644    &   595 &   877 & $\alpha$~Cyg \\
\phantom{0}43384    &  B3\,Iab  & 260    &   171 &   311 & $\alpha$~Cyg \\
\phantom{0}80558    &  B6\,Ib   & 604    &   420 &   736 & $\alpha$~Cyg \\
183143              &  B7\,Iae  &1673    &  1710 &  2392 & $\alpha$~Cyg: \\
111613              &  A1\,Ia   & 256    &   281 &   380 & $\alpha$~Cyg: \\
160529              &  A2\,Ia   &1332    &  1604 &  2085 & $\alpha$~Cyg \\
\textit{Mean}       &           &        &       &  1130 \\
\midrule
\multicolumn{2}{l}{\textit{F/G-supergiants}} \\
\phantom{00}7927    &  F0\,Ia   &   0    &   367 &   367 \\
161471              &  F2\,Ia   & 401    &    94 &   412 \\
187929              &  F6\,Ib   & 275    &   142 &   309 & $\delta$~Cep\\
\phantom{0}84810    &  G5\,Iab  &   0    &    60 &    60 & $\delta$~Cep\\
\textit{Mean}       &           &        &       &   287 \\
\midrule
\multicolumn{2}{l}{\textit{Other classes}} \\
149757              &  O9.5\,Vn & 129    &    98 &   162 & $\gamma$~Cas\\
154445              &  B1\,V    & 568    &   134 &   584 \\
161056              &  B1.5\,V  & 569    &   108 &   579 \\
203532              &  B3\,IV   &   0    &    81 &    81 \\
210121              &  B7\,II   &   0    &   126 &   126 \\
\phantom{0}23512    &  A0\,V    & 274    &   191 &   334 \\
147084              &  A4\,II   & 244    &   155 &   289 \\
\textit{Mean}       &           &        &       &   308 \\
\bottomrule
\end{tabular}
\begin{flushleft}
\underline{Notes} -- GN run (2018JUN) excluded from $e_{\star q^\prime}$.\\
\end{flushleft}
\label{tab:qu_var}
\vspace{-12pt}
\end{table}

\subsubsection{$\alpha$~Cyg variables}

\citet{Coyne71} was the first to (statistically) study the polarimetric variability of supergiant stars; which he found twice as variable as his control group. Together Fig.~\ref{fig:qu} and Table \ref{tab:qu_var} are quite revealing in showing that it is the B/A-type supergiants -- which are all $\alpha$~Cyg variables -- that are clearly more variable as a class, with all but one being formally 3-$\sigma$ detected polarimetric variables. By and large their QU diagrams show no favoured direction nor clear pattern, something generally associated with a clumpy stellar wind mechanism (see refs. within \citealp{Bailey23}).

In the era of space-photometry many early-type supergiants have been classified as $\alpha$~Cyg variables on evidence of what is sometimes called stochastic low-frequency variability. In the GCVS catalog $\alpha$ Cyg variables were originally classified as A- or B-type supergiants that displayed semi-regular radial velocity (RV) variability, said to be associated with the beating of many long period non-radial pulsations \citep{Samus17}. Often, it is only assumed that the photometric and RV variability are associated. Nevertheless, many such stars are also known polarimetric variables, including for instance $\lambda$~Cep \citep{Hayes75}, $\chi^2$ Ori \citep{Vitrichenko65}, Rigel \citep{Hayes86} and Deneb itself \citep{Cotton24}. 

\textit{HD\,80558, HD\,111613, HD\,183143, HD\,183143:} All four of these stars have past claims of variability. HD\,183143 and HD\,198478 were found to be variable in position angle by \citet{Hsu82}. HD\,183143 and HD\,198478 are among the most clearly variable stars in our data; they have values of $e_{\star\theta}$ larger than 0.5$^\circ$. \citet{Dolan86} later reported that HD\,111613 and HD\,183143 were variable in both $p$ and $\theta$. HD\,183143 shows perhaps the most dramatic shift in polarization we observed, and we also find it variable in $p$. It is not surprising that it should be a polarimetric variable given the noteworthy emission in its spectral lines. It exhibits P~Cygni profiles in H$_{\alpha - \delta}$ indicative of a strong wind, that if clumped would be an asymmetric scattering medium. \citet{Bastien88} reaffirmed HD\,111613 and added HD\,80558 as $\alpha$~Cyg stars that were polarimetric variables. \citet{Clarke94} criticised \citet{Bastien88} but nevertheless agreed that HD\,111613 was a polarimetric variable. We find HD\,80558 to be variable at 3-$\sigma$ significance in $\theta$ -- easily seen in Fig.~\ref{fig:qu} -- corresponding to an intrinsic variability of almost 0.4$^\circ$ -- and almost 2-$\sigma$ in $p$. Thus, it seems the variability of these four stars is persistent.

\textit{HD\,43384} has broad past evidence for variability (see Sec.~\ref{sec:9Gem}). In particular, it was found to be variable in $\theta$ by \citet{Hsu82}. However, it has the lowest value of $e_\star$ of any of the six early-type supergiants we observed, and is the only one for which there is no formal variability detection. \citet{Matsumura97} claim the star's variability is periodic, but our data represent poor phase coverage for testing the reported period. So this is a potential cause of the disagreement. However, since we have just over a year's data on this star, it is also possible this is simply insufficient for capturing episodic variability.

\textit{HD\,160529} is the only B/A-supergiant without a prior variability claim; it is invariable in the short-run data presented by \citet{Hsu82}. Repeat observations we made of the star in the same run also agree well. However, we see very large changes in $\theta$ and $p$ over years. Deneb, which is of the same spectral class and type, also exhibits polarimetric variability only on longer timescales than other $\alpha$~Cyg variables \citep{Cotton24}. As an LBV star, HD\,160529 is undergoing mass loss that varies on a range of timescales associated with winds, pulsation and rotation \citep{Stahl03}. As such, it is rather surprising that it has not previously been noted as a polarimetric variable. Another LBV star, P~Cygni, has recently been the subject of a decade of spectropolarimetric study by \citet{Gootkin20}, who find that it displays a variety of polarimetric behaviour, in part associated with free-electron scattering off of clumps (mostly) uniformly distributed around the star. We deduce that the large polarimetric variability we see here in HD\,160529 -- amounting to $e_\star =$ 0.21 per cent; $e_{\star\theta} =$ 0.64$^\circ$ -- is associated with similar phenomena. Mass-loss from these stars is greatest during outburst. For HD\,160529 this corresponds to an increase in temperature of $\sim$4000~K and a size increase of 180 $R_\odot$. This process can take a number of years, consistent with the slow variation seen in Fig.~\ref{fig:PA} and Fig.~\ref{fig:p}.

\citet{hayes84} made studies of the two early type $\alpha$~Cyg stars $\alpha$~Cam (O9.5\,Ia) and $\kappa$~Cas (B1\,Ia). These two stars show both slow long term variability and faster short term variability in their polarization, in much the same way as do those early-type supergiants observed here with sufficient regularity. For instance, slow longer term variability is evident in HD\,80558, faster short-term variability can be seen in both HD\,80558 and HD\,198478. We have not sampled all the B-type supergiants sufficiently on both timescales to test that these behaviours are simultaneously displayed, nonetheless it is a fair hypothesis. \citet{Hsu82} found early-type supergiants more variable on short time-scales than other standards and recommended against their use as such on that basis. \citet{hayes84} concluded that the polarization variations he saw were the result of varying mass-loss in the stellar extended atmospheres and that he was witnessing the ``waxing and waning of non-periodic clumpiness in the envelope,'' such as to explain the variability at both timescales.

\begin{figure*}
    \begin{minipage}[l]{0.79\textwidth}
    \includegraphics[clip, trim={0.0cm, 0.4cm, 0.0cm, 0.00cm}, width=\textwidth]{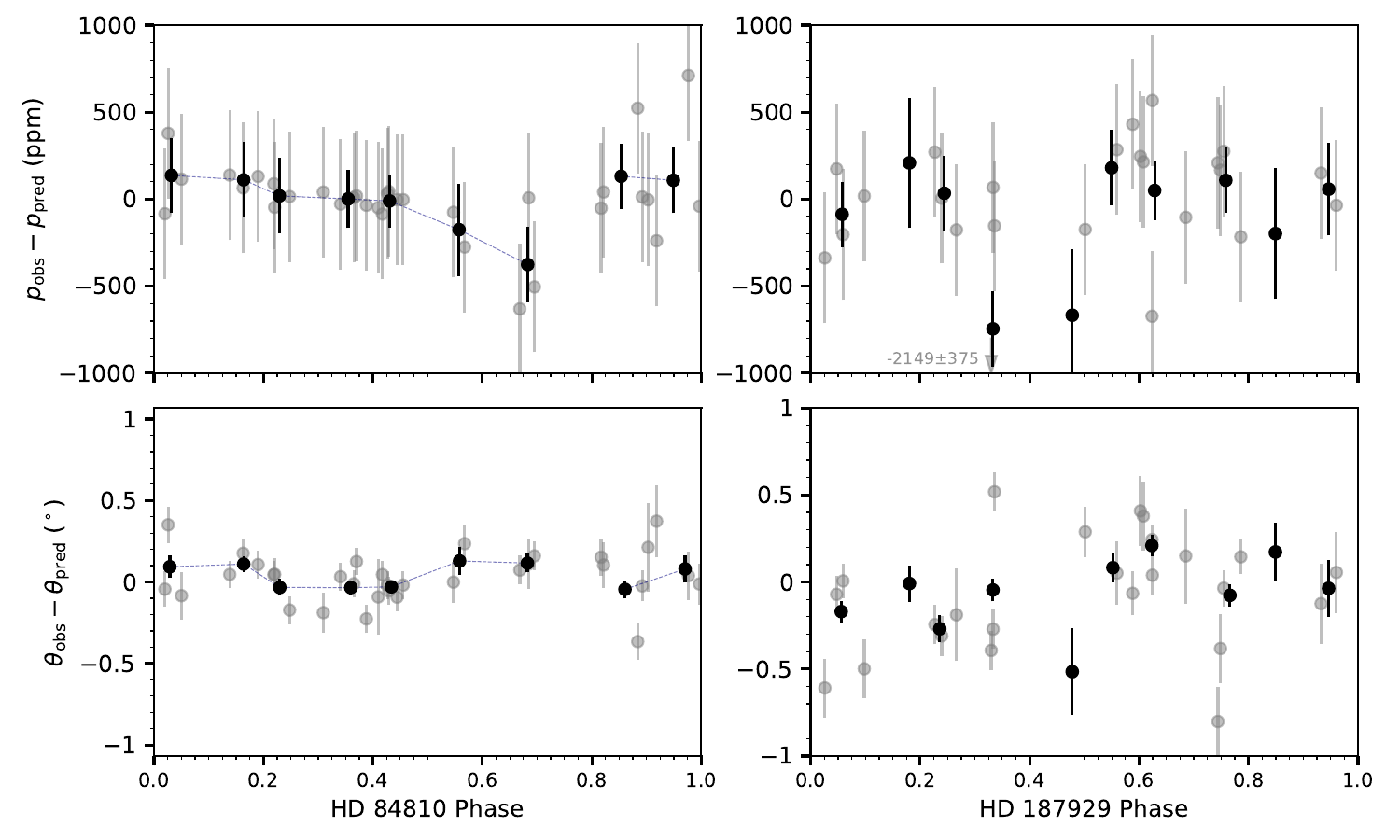}
    \end{minipage}
    \begin{minipage}[r]{0.20\textwidth}
    \caption{Observations (top: $p$, bottom: $\theta$) of the two Cepheid stars (left: HD\,84810, right: HD\,187929) plotted against phase. The grey points show the individual observations, whereas the data binned to 0.1 phase intervals is in black. For HD\,84810 navy dashed lines are shown to guide the eye. The $\theta$ scale has been selected, using Equation~\ref{eq:del_th}, to compensate for the difference in interstellar polarization of the two objects. Only $g^\prime$ data from runs with 2 or more standard observations in $g^\prime$ are shown. Phases have been calculated simply using the elements tabulated by \citet{Trahin21}: for HD\, 84810 $P=35.552$~d, $E_0=2447774.7368$~JD; for HD\,187929 $P=7.177$~d, $E_0=2448069.8905$~JD.}
    \label{fig:cep}
    \end{minipage}
\end{figure*}

That similar polarimetric behaviour might be universal in early-type supergiants is suggested by \citet{Hayes86} similar findings regarding polarimetric variability in the later type (B8\,Iae) Rigel. There, the variability was slower and the conclusion was that it resulted from localised disturbances in the stellar envelope originating at or below the photosphere. \citet{Cotton24} recently observed analogous behaviour for (A2\,Ia) Deneb and came to similar conclusions. The variability of (A2\,Ia) HD\,111613 seen by \citet{Bastien88} is clearly slow in the same way, as is that we see in (A2\,Ia) HD\,160529. Together these data present a picture where the timescale of polarimetric variability in $\alpha$~Cyg stars varies with spectral type. Given the contention that non-radial pulsations might be responsible for RV and photometric variability in similar stars \citep{Bowman19} and the associated debate about the plausibility of that hypothesis \citep{Lenoir-Craig22}, we suggest that the time is right to examine the polarimetric variability of early-type supergiants as a class with a view to informing this question.

\subsubsection{Suspected variables}

Our observations of little intrinsic polarization amongst the \mbox{F/G-supergiants} is in keeping with expectations based on their cooler photospheres, which are less ionized, and produce less prominent winds. The standards of the other classes are of earlier types, but overall their variability is similar. There are no formal 3-$\sigma$ detections of polarization variability amongst any of these other stars. However, the levels of variability we record are generally a match for \citet{Bastien88}'s findings, and four stars show 2-$\sigma$ detections; these might therefore be regarded as \textit{suspected variables}. We briefly comment on each of these below:

\textit{HD\,7927:} We have already briefly discussed HD\,7927 in Sec.~\ref{sec:litcomp}. \citet{Bastien88} found a greater $\theta$ variability for this star than we record, \citet{Hsu82}'s presented data gives a lower value. \citet{Bastien88} also found the star to be significantly variable in $p$, however we do not -- our average error is about double theirs, though this does not wholly account for the discrepancy. It seems likely the star's behaviour is episodic. Being of spectral type F0\,Ia it does not meet the formal definition for an $\alpha$~Cyg classification, though it is adjacent. As noted in Sec.~\ref{sec:phiCas} it displays the characteristics associated with the class, namely irregular RV and stochastic photometric variability, albeit at a low level. If the star does belong to the same class of objects then its polarimetric variability profile looks more like Rigel's in \citet{Hayes86} than the earlier type supergiants studied in \citet{Hayes78, hayes84}.

\textit{HD\,23512:} Variability in $\theta$ for HD\,23512 is 2-$\sigma$ significant, corresponding to $e_{\star\theta}=$ 0.256$^\circ$ -- with which \citet{Bastien88}'s results are consistent. This is surprising since as an ordinary A0\,V star it should be one of the least intrinsically polarized \citep{Cotton16a}, a fact already alluded to by \citet{Hall51}\footnote{At this time HD\,23512 was the closest star with a polarization detection.}. This suggests that its companion may be a polarimetric variable. The companion is responsible for 16 per cent of the combined flux, but being of small separation and discovered by Lunar occultation\footnote{It should be noted that the discovery is relatively recent. Earlier occultation studies did not find a companion (references within \citealp{Guthrie87}).}, little else is known about it. \citet{Breger84} confirmed \citet{Hall51}'s initial finding that HD\,23512 is much more polarized than any of the other bright stars in the cluster, but showed its wavelength dependence was typical. This leads us to speculate that the companion might instead be a bright background star. As discussed by \citet{Guthrie87}, HD\,23512 is unusual in other ways: it displays anomalously weak absorption at 2200~\AA\ and in several other bands for a star of its spectral type, and is depleted in calcium. These factors may provide some clues as to the nature of the other star. If the `B' component is predominantly responsible for the polarization variability, the scale of it would be $\sim$2000~ppm, which is not unreasonable for a supergiant or a Be star.

\textit{HD\,187929:} \citet{Bastien88} describe $\theta$ variability consistent with our measurements. The star has a magnetic field detection, but it is too small to produce significant linear broadband polarization \citep{Barron22}. Classically, $\delta$~Cep variables (Cepheids) should not be polarimetric variables as radial pulsations produce no net polarization change \citep{Odell79}. However, \citet{Polyakova81} and \citet{Polyakova84, Polyakova87, Polyakova90} claim to have measured phase-dependant polarization that is bound by a three-lobed ``rosette'' in the QU diagrams of many such stars. \citet{Polyakova81} ascribe this to an artefact of calibration owing to broadband measurements of colour-variable objects, but \citet{Polyakova90} later suggested an intrinsic polarization mechanism due to a circumstellar envelope 20-30 per cent larger at the equator than the pole -- \citet{Kervella06} thought this worthy of investigation.

We plotted the $p$ and $\theta$ against phase for both $\delta$~Cep variables in Fig.~\ref{fig:cep}, since nothing was obvious from the QU diagrams in Fig.~\ref{fig:qu}. Surprisingly, there is a hint of phase dependant behaviour in the binned data of HD\,84810 consistent with that described in the literature, albeit well below our formal detection thresholds -- investigating this is beyond the scope of the paper. There is no clear phase-dependant behaviour for HD\,187929; therefore, we discard the two aforementioned hypotheses as variable polarization mechanisms for it. Another possibility involves the star's binarity. The unseen B9.8\,V companion \citep{Evans91} has an unknown orbit. Both a long period orbit and one as short as 55~d that is face-on are viable based on current data \citep{Benedict22}. A truly face-on orbit would result in a rotation of $\theta$ with no change in $p$. Detailed calculations are required to determine if either photospheric reflection or entrained gas are good hypotheses in this case, though even a 55~d period seems like it would not produce a lot of polarization. If it does, further observations could reveal the geometry of the orbit \citep{Brown78, Cotton20}.

\textit{HD\,210121:} The final star displaying a $\theta$ variability more significant than 2-$\sigma$ is HD\,210121 -- $e_{\star\theta}=$ 0.264$^\circ$. This star has one of the smallest $p$ values, so this corresponds to an $e_{\star\,u^\prime}$ of only 126 ppm. If the star really is a bright giant, winds might be responsible. However, this polarization is small enough that many mechanisms could explain such variability. Any of non-radial pulsations, a binary mechanism, emission behaviour, or interstellar drift could be responsible, but no clear leads present themselves in the literature.

\begin{table}
\caption{The impact of corrections on telescope position angle determination}
\tabcolsep 15 pt
%\centering
\begin{tabular}{lcc}
\toprule
Correction Removed  & $\eta|\Delta \theta_{t}|$    &   ${\rm max}|{\Delta \theta_{t}}|$ \\
            &  \multicolumn{1}{c}{($^\circ$)}   &   \multicolumn{1}{c}{($^\circ$)}  \\ 
\midrule
Precession (original equinox)                       &   0.229   &   0.396   \\
New $\theta$*                                       &   0.112   &   0.796   \\
$\Delta\lambda/\Delta\theta$ ($B$ to $g^\prime$)    &   0.021   &   0.331   \\
Weighting                                           &   0.011   &   0.510   \\
Precession (2020 equinox)                           &   0.005   &   0.035   \\
$\Delta\lambda/\Delta\theta$ (airmass)              &   0.004   &   0.078   \\
Faraday Rotation                                    &   0.002   &   0.007   \\
\midrule
All                                                 &   0.312   &   1.030   \\
\bottomrule
\end{tabular}
\begin{flushleft}
\underline{Notes} -- * Uses literature values for $\theta$ as per Table \ref{tab:lit_pa}.
\end{flushleft}
\label{tab:corrections}
\vspace{-12pt}
\end{table}

%\vspace{0.5cm}
\subsection{Relative impact of corrections}
\label{sec:impact}

In Sec's \ref{sec:reduction} and \ref{sec:analysis} we have carried out a series of corrections aimed at more accurately calibrating the telescope position angle for each run. In table \ref{tab:corrections} we rank each correction by the median difference between our final determination, and what we'd get neglecting the specified corrections. For this we include all except the Pindari runs. 

Assuming the new values of $\theta$ reported are the real values, the typical improvement in $\theta_t$ determination is 0.3$^\circ$. The most significant corrections are due to precession and the re-determined $\theta$ values. Here we have calculated $\theta$ for each star in a 2020 equinox. Our observations span a decade, from 2014 to 2023. It can be seen that making precession corrections is now essential to achieve 0.1$^\circ$ accuracy if working from older literature. However, using contemporary sources will be suitable for most applications.

The largest $\Delta\theta_{t}$ values for $\Delta\lambda/\Delta\theta$ (airmass) correspond to runs with Clear rather than $g^\prime$ observations. For instruments where $\theta$ rotates with wavelength, narrower bands are better for calibration. If one wants to simplify their reduction routine, this analysis suggests they need not worry about wavelength correction post-fact to account for airmass, but it is a good idea to make band corrections before beginning if the literature data is in a different filter. 

Applying the weightings associated with $e_{i\,\theta}$, $e_{\star\theta}$, and $e_{\rm m\,\theta}$ is only important in a few cases -- those where only a few standard measurements were obtained and one of them was one of the most variable stars. If one observes only the least variable standards, then weighting will generally not be needed. However, if forced to use less reliable standards, one should use the best weightings available, and observe as many \textit{different} standards as possible -- since a standard variable on long timescales may be invariable on short ones and because $e_{t\,\theta}$ scales $\propto \theta_e/\sqrt{N_o}$.

\subsection{Recommendations for observers}
\label{sec:recommendations}

\begin{table}
\caption{Recommended position angle data}
\tabcolsep 10.5 pt
%\centering
\begin{tabular}{rrrrr}
\toprule
Standard    &   $\theta_{g^\prime}$ &   $e_{\star\theta}$   &   $\Delta\theta/\Delta\lambda$&   $\Delta\theta/\Delta t$ \\
(HD)        &   ($^\circ$)          &   ($^\circ$)          &   ($^{\circ}/\mu \rm m$)      &   ($^{\circ}/100\,\rm yr$)\\
\midrule
7927        &   93.165              &   0.321               &   $-$5.7      &   $+$0.4  \\
23512       &   30.706              &   0.256               &   $-$3.6      &   $+$0.5  \\
43384       &   170.295             &   0.167               &   $+$2.6      &   $+$0.6  \\
80558       &   162.512             &   0.384               &   $+$1.4      &   $+$0.6  \\
\rowcolor{Gray}
84810       &   99.997              &   0.110               &\phantom{+}0.0 &   $+$0.7  \\
111613      &   80.836              &   0.266               &\phantom{+}0.0 &   $-$0.2  \\
\rowcolor{Gray}
147084      &   32.028              &   0.118               &\phantom{+}0.0 &   $-$0.6  \\
149757      &   126.218             &   0.211               &   $-$5.0      &   $-$0.5  \\
\rowcolor{Gray}
154445      &   89.985              &   0.110               &\phantom{+}0.0 &   $-$0.5  \\
160529      &   18.748              &   0.640               &   $+$3.5      &   $-$0.7  \\
\rowcolor{Gray}
161056      &   68.034              &   0.082               &   $-$1.5      &   $-$0.6  \\
\rowcolor{Gray}
161471      &   2.087               &   0.123               &   $-$1.1      &   $-$0.7  \\
183143      &   179.299             &   0.819               &\phantom{+}0.0 &   $-$0.5  \\
187929      &   93.703              &   0.241               &   $-$7.3      &   $-$0.5  \\
198478      &   2.417               &   0.628               &\phantom{+}0.0 &   $-$0.6  \\
203532      &   124.360             &   0.175               &   $+$2.4      &   $-$2.7  \\
210121      &   153.903             &   0.264               &   $+$8.6      &   $-$0.3  \\
\bottomrule
\end{tabular}
\begin{flushleft}
\underline{Notes} -- Values of $\theta_{g^\prime}$ are given for a 2020 equinox. Here the $g^\prime$ band nominally corresponds to 470~nm. The best standards are backed in gray.\\
\end{flushleft}
\label{tab:rec}
%\vspace{-12pt}
\end{table}

We reiterate the best determinations for position angle parameters in Table \ref{tab:rec}, as they are important for our recommendations to observers seeking better position angle calibration. Both $\theta$ and its associated variability $e_{\star\theta}$ are derived in Sec.~\ref{sec:varPA}, the other parameters come from the literature as described in Appendix \ref{apx:lit_PA}. Using this data and applying the advice below will allow for better than 0.1$^\circ$ error in $\theta_t$ from a handful of observations in most circumstances. In order of importance our recommendations are:
\begin{enumerate}
    \item Correct $\theta$ for the equinox of observations using the final column in Table \ref{tab:rec} or Equation~\ref{eq:prec}. Report the equinox.
    \item Use the best available determination of $\theta$ for your standards, report the value assigned and provenance of the standards employed. Ideally all determinations should be made with the same $\theta_0$. For the stars studied here, we recommend the values in Table \ref{tab:rec}. If other standards are needed, we prefer the values given in \citet{Hsu82}, and then, if those are unavailable, determinations tied to \citet{Serkowski75}'s $\theta_0$ calibration and determined from long run data.
    \item Use the available $\Delta\theta/\Delta\lambda$ information in determining $\theta_{\rm lit}$ for your bandpass. In many cases a full bandpass model will not be required, but if carrying out one for this purpose, use the spectral types and values of $E_{\rm (B-V)}$, $R_{\rm V}$, $p_{\rm max}$, $\lambda_{\rm max}$, and $K$ given in Table~\ref{tab:basic} for the stars studied here, or follow the methodologies laid out in Appendix \ref{apx:lit_homog} to determine values for others.
    \item Characterise the position angle error associated with your instrument independent of photon shot noise. To achieve this, one should observe at least one, preferably a few, of the least variable standards multiple times within a few nights, calculate the standard deviation of each, subtract the RMS shot-noise contributions, and take the median. Such a determination should be made and reported as part of commissioning an instrument on a new telescope. Without this information combining data between groups or even just runs to a high precision in $\theta$ is fraught. 
    \item Choose the most reliable standards available based on reported $\theta$ variability. For the stars studied here, we recommend the values in Table \ref{tab:rec}. For other standards, taking $\sqrt{\sigma_2(\theta)^2-\sigma_1(\theta)^2}$ from \citet{Bastien88} is recommended. Where $e_{\star\theta}$ of a potential standard is unknown, the following equation might be used as a first approximation: \begin{equation}e_{\star\theta} = \frac{28.65}{\sqrt{2}}\frac{e_\star}{p_{\rm ISM}},\label{eq:etsApprox}\end{equation} where $e_{\star\theta}$ is in degrees, $p_{\rm ISM}$ is the interstellar component of the polarization (equal to $p_{\rm obs}$ for an ideal standard), and $e_\star$ should be taken as 1130~ppm for an $\alpha$~Cyg variable, and 300~ppm otherwise, on the basis of the means in Table \ref{tab:qu_var}.
    \item Especially where the available standards might be less reliable, observe multiple standards (ideally $\ge 3$) for calibration and weight the result according to the RMS sum of measurement uncertainty and the best estimate of variability.
\end{enumerate}

\vspace{-0.2cm}
\section{Conclusions}
\label{sec:conclude}

Long duration studies of high polarization stars are rare leading us to explore the astrophysical implications of our findings: Extreme stars are extremely polarized. The most polarized object in our study, the LBV star HD\,160529, is found to also be one of the largest amplitude polarimetric variables. This star has not previously been noted as a polarimetric variable, seemingly it is variable only on longer timescales.

Large amplitude polarization variability is common and shows no preferred orientation in $\alpha$~Cygni variables (HD\,80558, HD\,111613, HD\,183143, HD\,198478), a behaviour that may extend to the F0\,Ia HD\,7927 at lower levels. By analogy with historical studies, clumpy winds are the most likely polarigenic mechanism, but pulsation might also play a role and a dedicated observational program and investigation is recommended.

Later type supergiants and other ordinary stars that happen to be extremely reddened make more reliable standards. The best standards are HD\,84810, HD\,147084, HD\,154445, HD\,161056 and HD\,161471, for which we attribute variability $\leq 0.123^\circ$. However, polarization variability may also be present, at lower levels, where we make 2-$\sigma$ detections. There is also evidence for variability in the literature or extended data for other stars in our sample (i.e. HD\,43384 and HD\,149757). The $g^\prime$ data analysed might be too thin to capture episodic behaviour in these cases; \citet{Hayes75} came to a similar conclusion after finding no variability for HD\,149757.

The companion to HD\,23512 may be a large amplitude polarimetric variable. HD\,187929 displays position angle variability not correlated with the Cepheid phase, an observation that favours a companion in a face-on orbit. 

Our results largely vindicate the findings of \citet{Bastien88} who were criticised for a lack of statistical rigor by \citet{Clarke94} in combining data from multiple sources. The methodology we apply alleviates the problems associated with combining data from disparate observing runs. Key to this is the application of a Co-ordinate Difference Matrix which works by amalgamating difference measurements of pairs of points.

By combining a decade's worth of data we have been able to make more accurate and precise determinations of $\theta$, as well as estimates of its variability, $e_{\star\theta}$ which are tabulated in Table \ref{tab:rec}. The other standard properties, as given in Tables~\ref{tab:rec} and \ref{tab:basic}, have been derived from literature we have curated to minimise the impacts of disagreement between different researchers; these should also be adopted. Along with these improvements, which are quantitatively assessed in Sec.~\ref{sec:impact}, we provide specific recommendations for observers in Sec.~\ref{sec:recommendations} that will allow for better than 0.1$^\circ$ calibration in telescope position angle from a handful of observations in most circumstances. 

%\vspace{-0.25cm}
\section*{Acknowledgements}

This research has made use of the SIMBAD database, operated at CDS, Strasbourg, France; NASA's Astrophysics Data System; and the VizieR catalogue access tool, CDS, Strasbourg, France \mbox{(\doi{10.26093/cds/vizier})}. Based on spectral data retrieved from the ELODIE archive at Observatoire de Haute-Provence (OHP). We acknowledge with thanks the variable star observations from the AAVSO International Database contributed by observers worldwide and used in this research.

JPM acknowledges support by the National Science and Technology Council of Taiwan under grant NSTC 112-2112-M-001-032-MY3. Funding for the construction of HIPPI-2 was provided by UNSW through the Science Faculty Research Grants Program. We thank the Friends of MIRA for their support.

We thank the former Director of the Australian Astronomical Observatory, Prof. Warrick Couch, the current Director of Siding Spring Observatory, Prof. Chris Lidman, and all of the staff at the AAT for their support of the HIPPI and HIPPI-2 projects. We thank Prof. Miroslav Filipovic for providing access to the Penrith Observatory. We thank all of the additional student volunteers at UNSW, MIRA, WSU and elsewhere, who assisted with observations at the various observatories. We thank Dr. Wm. Bruce Weaver for useful comments on the manuscript. We thank Dr. Sarbani Basu for help acquiring a reference. 

Based in part on data obtained at Siding Spring Observatory. We acknowledge the traditional owners of the land on which the AAT stands, the Gamilaraay people, and pay our respects to elders past and present. 

Based in part on data obtained at Western Sydney University, Penrith Observatory. We acknowledge the traditional owners of the land on which the WSU Penrith Observatory stands, the Dharug people, and pay our respects to elders past and present. 

\section*{Data Availability}

All of the data for this work is available in a catalogue described in Section \ref{sec:obs}.

%\vspace{-0.25cm}
\bibliographystyle{mnras}
\bibliography{refs}
%%%%%%%%%%%%%%%%%%%%%%%%%%%%%%%%%%%%%%%%%%%%%%%%%%
\clearpage
\pagebreak
\appendix

\section{Literature Position Angle Data}
\label{apx:lit_PA}

In this work a comparison to literature measurements is used to determine the position angle zero point. We predominantly work in the SDSS $g^\prime$ band, which for a typically reddened standard observation has $\lambda_{\rm eff}\approx470$\,nm. Most literature determinations of $\theta$ have been made in Johnson filters. The closest is $B$ band, which has $\lambda_{\rm eff}\approx440$\,nm. Here we predominantly rely on $\theta$ determinations made in/for that band by three sources \citep{Serkowski75, Hsu82, Bagnulo17}. Combining the work from these sources gives the values in Table \ref{tab:lit_pa}. HD\,161471 is not listed in these works, instead we use \citet{Serkowski69}. 

The change in position angle with time due to precession, $\Delta\theta/\Delta t$, is described by \citet{vdKamp67} as, \begin{equation}\Delta\theta/\Delta t = 0.0056\times\sin{\alpha_0}\sec{\delta_0},\label{eq:prec}\end{equation} where $\Delta t$ is in years and $(\alpha_0, \delta_0)$ are the initial co-ordinates in degrees. In Table \ref{tab:lit_pa} we adjust the value to a 2020 equinox. 

\citet{Hsu82} and \citet{Bagnulo17} determine the change in position angle with wavelength, $\Delta\theta/\Delta\lambda$; where this is available, or estimates are possible from other sources, we have made an adjustment corresponding to 30\,nm. More rigour than this is not justified by the precision of the data, which is nominally 0.2$^\circ$ for \citet{Hsu82}'s position angle determinations, and more than this for $\Delta\theta/\Delta\lambda$ (in $^{\circ}/\mu \rm m$). We round to 0.1$^\circ$ in Table \ref{tab:lit_pa}.

\begin{table}
\caption{Standards: literature derived position angles}
\tabcolsep 7 pt
%\centering
\begin{tabular}{rrccl}
\toprule
Standard&$\theta_{g^\prime}^\dagger$&   $\Delta\theta/\Delta\lambda$&   $\Delta\theta/\Delta t$ &  References\\
(HD)            &   ($^\circ$)      &   ($^{\circ}/\mu \rm m$)      &   ($^{\circ}/100\,\rm yr$)    &                   \\
\midrule
7927            &  93.0             &   $-$5.7      &   $+$0.4  &   HB82. \\
23512           &  30.4             &   $-$3.6      &   $+$0.5  &   HB82. \\
43384           & 170.0             &   $+$2.6      &   $+$0.6  &   B17, HB82, S75.\\
80558           & 163.3             &   $+$1.4      &   $+$0.6  &   B17, S75.\\
84810           & 100.0             &\phantom{+}0.0 &   $+$0.7  &   HB82, S75.\\
111613          &  80.8             &\phantom{+}0.0 &   $-$0.2  &   HB82, S75.\\
147084          &  31.8             &\phantom{+}0.0 &   $-$0.6  &   HB82, S75.\\
149757          & 127.2             &\phantom{*}$-$5.0* &   $-$0.5  &   S75.\\
154445          &  90.0             &\phantom{+}0.0 &   $-$0.5  &   HB82, S75.\\
160529          &  20.0             &   $+$3.5      &   $-$0.7  &   HB82.\\
161056          &  67.3             &   $-$1.5      &   $-$0.6  &   B17.\\
161471          &   2.4             &\phantom{**}$-$1.1**&   $-$0.7  &   SR69.\\
183143          & 179.2             &\phantom{+}0.0 &   $-$0.5  &   HB82, S75.\\
187929          &  93.7             &   $-$7.3      &   $-$0.5  &   HB82, S75.\\
198478          &   3.0             &\phantom{+}0.0 &   $-$0.6  &   HB82\\
203532          & 126.9             &   $+$2.4      &   $-$2.7  &   B17, S75.\\
210121          & 155.1             &   $+$8.6      &   $-$0.3  &   B17.\\
\bottomrule
\end{tabular}
\begin{flushleft}
\underline{References} -- B17: \citet{Bagnulo17}, HB82: \citet{Hsu82}, S75: \citet{Serkowski75}, SR69: \citet{Serkowski69}.\\
\underline{Notes} -- $\dagger$ Based on adjusted $B$ band observations from the given sources. Precessed to a 2020 equinox using Eqn. \ref{eq:prec} to generate column 4; \mbox{* Estimated}\,from \citet{Wolff96} over the range 0.4--0.6\,$\mu\rm m$; \mbox{** Estimated}\,from $UBV$ measurements in \citet{Serkowski69}. \\
\end{flushleft}
\label{tab:lit_pa}
\vspace{-12pt}
\end{table}

\vspace{-12pt}
%\pagebreak
\section{Literature Polarization Data} 
\label{apx:lit_homog}

FLC modulator calibration in HIPPI-class polarimeters requires data be checked against the polarization of standard stars. For this we use a bandpass model that requires polarization and extinction parameters for each star. So far this data has been assembled in an ad-hoc way \citep{Bailey20}. Here we develop a more systematic approach that seeks to favour the most reliable polarization (Serkowski Law) data and homogenises the available extinction and reddening data on each standard star.

%\vspace{-24pt}

\subsection{Serkowski Law parameters}
\label{sec:serk}

The intrinsic polarization of high polarization standard stars is \textit{assumed} to be negligible, and their polarization accomplished entirely by the interstellar medium\footnote{This assumption is almost certainly never valid, as these stars are all very far away and thus extreme in some way. However, their distance ensures a large interstellar polarization and this dwarfing other polarigenics is what is relied upon.}. Interstellar polarization is described by the empirically determined Serkowski Law \citep{Serkowski68}, \begin{equation} \frac{p(\lambda)}{p_{\rm max}} =
\exp\left({-K \ln^2 \frac{\lambda_{\rm max}}{\lambda}}\right), \label{eq:serk} \end{equation} where $\lambda$ is the wavelength in $\mu$m, $p$ the polarization, $p_{\rm max}$ the maximum polarization, $\lambda_{\rm max}$ the wavelength corresponding to $p_{\rm max}$, and the dimensionless constant $K$ is the inverse half-width of the curve.

We can determine the most reliable values of $p_{\rm max}$ and $\lambda_{\rm max}$ independently of $K$, since according to \citet{Hsu82} and \citet{Bagnulo17} that parameter does not significantly impact the determination of the other two. If properly calibrated, a spectropolarimeter offers the best precision for this, so long as $\lambda_{\rm max}$ is within its operating range. For this reason we prefer determinations made by \citet{Martin99, Bagnulo17} and \citet{Wolff96} (in that order) where available. \citet{Bagnulo17} use the FORS2 instrument to take data over the range 0.375 -- 0.940 $\mu$m, whereas \citet{Wolff96} obtain data from 0.400 -- 0.700 $\mu$m and then supplement this with broadand infrared data from \citet{Bailey82} and \citet{Nagata90}. Similarly, \citet{Martin99} made fits to data from the FOS instrument on the \textit{Hubble Space Telescope} supplemented with infrared ground based data (of which only \citealp{Wilking80} is relevant here).

Where a Serkowski Law fit is instead made to broadband measurements the accuracy of the determinations depends to a greater extent on the quality of the bandpass model used, the reddening law adopted and the bands in which data is obtained. Because $p \propto 1/\lambda$ in the Serkowski Law, infrared data is particularly important to getting a good fit. In acknowledgement of these factors we next prefer the work of \citet{Wilking80, Wilking82, Hsu82} and \citet{Serkowski75} in that order.

The work of \citet{Serkowski75} lists the most extensive set of observations. They apply a universal reddening law and are careful to report the details of their bandpass model. However, the longest passband used is $R$, and their final reported values are actually a straight average of their own work and the earlier determinations of others, who were not always as thorough. \citet{Hsu82} obtained multiple observations of each star in $UBVR$ and a longer 0.75 $\mu$m band. They conducted a very careful study -- their bandpass model took account of airmass and they also applied a more modern universal reddening law. \citet{Wilking80, Wilking82} obtained $JHK$ band data and added it to optical and UV data of a variety of earlier workers to achieve the widest wavelength range of the broadband works considered here. 

The third Serkowski parameter, $K$, has in some reported cases been fit, and in others assumed. \citet{Serkowski75} assumed 1.15 -- which was the best mean fit to their data and that of \citet{Coyne74} -- this value was also assumed by \citet{Hsu82}. \citet{Wolff96} and \citet{Martin99} applied the relation found by \citet{Whittet92},\begin{equation} K = 0.01 + 1.66\times\lambda_{\rm max}.\label{eqn:whitt}\end{equation} This was the mean determination from fitting $UBVRIJHK$ data obtained for 105 stars. Earlier \citet{Wilking80}, in fitting $K$ for each individual object, identified a very similar mean relationship, \begin{equation} K = -0.002 + 1.68\times\lambda_{\rm max}, \label{eqn:wilk80}\end{equation} which they revised in later work \citep{Wilking82} to, \begin{equation} K = -0.10 + 1.86\times\lambda_{\rm max}. \label{eqn:wilk82}\end{equation} Given the $\lambda_{\rm max}$ values for our standards, and the typical errors, there is no practical difference between equations \ref{eqn:whitt}, \ref{eqn:wilk80} and \ref{eqn:wilk82}.

\citet{Bagnulo17} also fit $K$ individually for each object; in doing so they obtained quite different values to earlier workers. This could be due to calibration discrepancies or differences in the regions of the ISM probed. However, their favoured explanation is a lack of infrared data, in line with the findings of \citet{Clarke83}, who showed that the relationship recovered by \citet{Whittet92} depended on the wavelength range probed. This may point to a real, if subtle, feature of interstellar polarization that is masked by fitting a Serkowski curve to data that includes infrared bands.

This presents a difficulty for us in selecting appropriate values of $K$ to use in our calibration. Not all the standards have individually fit values available. Of those that do, the wavelength range of the fit data varies. Re-fitting all of the data in a consistent way is desirable but beyond the scope of this work. Another complicating factor is that different reddening laws, and reddening parameters have been adopted by different workers, which will also impact $K$. 

Our goal is consistency. For this reason we have decided to adopt the relationship of \citet{Whittet92} as given in Eqn. \ref{eqn:whitt}, except where $K$ has been fit using a similar wavelength range -- in practice, for the standards we have, this means only the \citet{Wilking80} determinations are applicable. The difference between these values is negligible in half the cases, and very little data has been taken at infrared wavelengths with HIPPI-class instruments anyway, so the difference between the possible approaches here may be academic.

The final Serkowski parameters we adopt for the standards used in this paper, and more broadly are given in table \ref{tab:std_serk}.

\begin{table}
\caption{Standards: adopted Serkowski Law parameters}
\tabcolsep 5.65 pt
\centering
\begin{tabular}{rrrlrr}
\toprule
Standard        &   $p_{\rm max}$   &   $\lambda_{\rm max}$     &   Reference       &   \multicolumn{2}{c}{$K$}     \\
(HD)            &   (\%)            &   ($\mu$m)                &                   &   Calc.$^a$   &  Fit$^b$ \\
\midrule
7927            & 3.31              &   0.507                   & \citet{Wolff96}   &   0.85    \\
23512           & 2.29              &   0.600                   & \citet{Hsu82}     &   1.01    \\
43384           & 3.06              &   0.566                   & \citet{Bagnulo17} &       &   0.97 \\
80558           & 3.34              &   0.597                   & \citet{Bagnulo17} &   1.00    \\
84810           & 1.62              &   0.570                   & \citet{Hsu82}     &   0.96    \\
111613          & 3.14              &   0.560                   & \citet{Hsu82}     &   0.94    \\
147084          & 4.41              &   0.684                   & \citet{Martin99}  &       &   1.15 \\
149757          & 1.45              &   0.602                   & \citet{Wolff96}   &       &   1.17 \\
154445          & 3.66              &   0.569                   & \citet{Wolff96}   &       &   0.95 \\
160529          & 7.31              &   0.543                   & \citet{Hsu82}     &   0.91    \\
161056          & 4.01              &   0.584                   & \citet{Martin99}  &       &   0.96 \\
161471          & 2.28              &   0.560                   & \citet{Serkowski75} & 0.94    \\
183143          & 6.16              &   0.550                   & \citet{Wilking80} &       &   1.15 \\
187929          & 1.73              &   0.552                   & \citet{Wolff96}   &   0.93    \\
198478          & 2.75              &   0.515                   & \citet{Wolff96}   &       &   0.88 \\
203532          & 1.39              &   0.574                   & \citet{Bagnulo17} &   0.86    \\
210121          & 1.38              &   0.434                   & \citet{Bagnulo17} &   0.73    \\
\bottomrule
\end{tabular}
\begin{flushleft}
$^a$ According to \citet{Whittet92} (Eqn. \ref{eqn:whitt}).\\
$^b$ From \citet{Wilking80}.
\end{flushleft}
\label{tab:std_serk}
\end{table}

\begin{table*}
\caption{Standards: adopted spectral types and extinction parameters}
\tabcolsep 4.5 pt
\centering
\begin{tabular}{rllclll}
\toprule
Standard        &   SpT     &   Plx &   $E_{\rm (B-V)}$\,$\pm$\,$\sigma$ &  References                             &    \multicolumn{1}{c}{$R_{\rm V}$\,$\pm$\,$\sigma$}  & References    \\
(HD)            &           &(mas)&   (mag)               &                                                  &   \multicolumn{1}{c}{(mag)}   &               \\
\midrule
7927            & F0        &   0.2142  &   0.51 $\pm$ 0.02   & EH03, B96, W96, S75.                                                                 & 3.11              & H61.\\
23512           & A0        &   7.3345  &   0.37 $\pm$ 0.01   & FM07, S75.                                                                           & 3.27 $\pm$ 0.13   & FM07, G87, W81$^\dagger$.\\
43384           & B3        &   0.5467  &   0.57 $\pm$ 0.02   & VH10, R09, S75, L68.                                                                      & 3.06 $\pm$ 0.16   & R09, W03. \\
80558           & B6        &   0.5375  &   0.59 $\pm$ 0.03   & W03, WB80, S75.                                                                      & 3.25 $\pm$ 0.15   & W03, WB80.\\
84810           & G5        &   1.9842  &   0.18 $\pm$ 0.01   & B07, B96, HB89, ($\mathsection{}$).                                                    & 3.06              & B07. \\
111613          & A1        &   0.4534  &   0.40 $\pm$ 0.01   & E22, S75.                                                                            & 3.72 $\pm$ 0.32   & E22, dG89*.\\
147084          & A4        &   3.71    &   0.75 $\pm$ 0.03   & G01, dG89, WB80, S75.                                                                & 3.67 $\pm$ 0.26   & V13*, dG89*, RL85*, WB80.\\
149757          & O9.5      &   8.91    &   0.32 $\pm$ 0.01   & PK22, FM07, V04, W03, W96, OD94,                                                     & 2.93 $\pm$ 0.20   & PK22, FM07, V04, W03, OD94, C89, dG89*, WB80.\\ 
                &           &           &        & C89, dG89 WB80, S75. \\
154445          & B1        &   4.0229  &   0.40 $\pm$ 0.03   & FM07, V04, W03, W96, OD94, C89,                                                      & 3.03 $\pm$ 0.18   & FM07, V04, W03, OD94, C89, A88, WB80.\\ 
                &           &           &        & A88, WB80, S75. \\
160529          & A2        &   0.5366  &   1.29 $\pm$ 0.01   & WB80, S75.                                                                           & 2.94              & WB80. \\
161056          & B1.5      &   2.4404  &   0.60 $\pm$ 0.05   & W03, OD94, WB80, S75.                                                                & 3.11 $\pm$ 0.02   & W03, OD94, WB80. \\
161471          & F2        &   1.69    &   0.26 $\pm$ 0.07   & B96, C86, H79, S75.                                                                  & 2.42              & L14**. \\
183143          & B7        &   0.4296  &   1.24 $\pm$ 0.03   & E22, W03, A88, S75.                                                                  & 3.16 $\pm$ 0.22   & E22, W03, A88. \\
187929  & F6$^\mathparagraph$&  3.6715  &   0.16 $\pm$ 0.02   & B96, W96, E91, HB89, T87, S75.                                                       & 3.10              & C89$^\ddagger$.\\
198478          & B3        &   0.5435  &   0.54 $\pm$ 0.02   & V04, W03, W96, A88, S75.                                                             & 2.89 $\pm$ 0.27   & V04, W03, A88.\\
203532          & B3        &   3.4402  &   0.32 $\pm$ 0.03   & FM07, V04, W03, WB80, S75.                                                           & 3.05 $\pm$ 0.25   & FM07, V04, W03, WB80. \\
210121          & B7        &   2.9971  &   0.35 $\pm$ 0.05   & FM07, V04.                                                                           & 2.22 $\pm$ 0.29   & FM07, V04. \\
\bottomrule
\end{tabular}
\begin{flushleft}
\underline{References} -- A88: \citet{Aiello88}, B07: \citet{Benedict07}, B96: \citet{Bersier96}, C86: \citet{Cholakyan86}, C89: \citet{Cardelli89}, dG89: \citet{deGeus89}, E22: \citet{Ebenbichler22}, E91: \citet{Evans91}, EH03: \citet{Evans03}, FM07: \citet{Fitzpatrick07}, G01: \citet{Gray01} via \citet{Cox17}, G87: \citet{Guthrie87}, H61: \citet{Hoag61}, H79: \citet{Hobbs79}, HB89: \citet{Hindsley89}, L14: \citet{Luck14}, L68: \citet{Lesh68} via \citet{Cox17}, OD94: \citet{ODonnell94}, PK22: \citet{Piccone22}, R09: \citet{Rachford09}, RL85: \citet{Rieke85}, S75: \citet{Serkowski75}, \mbox{T87: \citet{Turner87},} V04: \citet{Valencic04}, V13: \citet{Voshchinnikov13}, VH10: \citet{Voshchinnikov10}, W03: \citet{Wegner03}, W81: \citet{Witt81}, W96: \citet{Wolff96}, WB80: \citet{Whittet80}. \\
\underline{Notes} -- $\dagger$ We have averaged all the values in \citet{Witt81}'s Table 2 except Serkowski's based on polarization, which are outliers; $\ddagger$ No specific value for this star could be found, we have adopted \citet{Cardelli89}'s  Galactic average (also found by e.g. \citealp{Wegner03}); * These references give only $A_{\rm V}$, the value for $E_{\rm (B-V)}$ in column 4 has been used to calculate $R_{\rm V}$; ** \citet{Luck14} calculate $E_{\rm (B-V)}$ for HD\,161471, but it is a significant outlier compared to the other measurements, so we neglect it and use column 4 and their $A_{\rm V}$ calculation for determining $R_{\rm V}$; $\mathsection{}$ The value given in \citet{Serkowski75} is an outlier and is neglected; $\mathparagraph$ HD\,187929 is an F6\,Ia\,$+$\,B9.8\,V binary. 
\end{flushleft}
\label{tab:redden}
\end{table*}

\vspace{48pt}

\subsection{Extinction parameters}

As described in \citet{Bailey20}, our bandpass model applies a \citet{Castelli03} atmosphere model based on spectral type\footnote{The model grid is coarse, and figures for an intermediate spectral type are linearly interpolated after complete calculation of the bracketing types.}. For distant stars the spectral energy distribution can be reddened using the relationship described by \citet{Cardelli89}. For stars as distant as our standards, this is necessary for a thorough treatment.

Extinction and reddening are related, \begin{equation}R_V \equiv A_V/E_{\rm (B-V)},\end{equation} where $R_{\rm V}$ is the normalised extinction, $A_{\rm V}$ is the extinction in the $V$ band, and $E_{\rm (B-V)}$ the selective extinction. Our algorithm requires $R_{\rm V}$ and $E_{\rm (B-V)}$.

Sometimes the extinction parameters have been reported with Serkowski parameter determinations. However, we have found these to be inconsistent, sometimes superseded by better measurements, and in others simply assumed. We therefore conducted a thorough, though non-exhaustive, search of the literature with the aim of determining appropriate mean values from a reliable sample. We give this data in Table \ref{tab:redden}. In some cases only $A_{\rm V}$ has been reported, in such cases we have calculated $R_{\rm V}$ using the mean $E_{\rm (B-V)}$ we determine; these are marked with an asterisk is the table. The spectral types given in Table \ref{tab:redden} are the most common indicated in the cited extinction literature.

Table \ref{tab:redden} gives the standard deviation of $E_{\rm (B-V)}$ and $R_{\rm V}$ as the 1$\sigma$ value. With few exceptions, $E_{\rm (B-V)}$ is well defined. The range of $R_{\rm V}$ values reported in the literature is much wider; there are also fewer determinations and we have been forced to use whatever we can find. In particular there are few determinations of $R_{\rm V}$ for the Cepheid variables (HD\,84810 and HD\,187929). Nevertheless, many of the adopted means match expectations well. The galactic mean is known to be 3.10 \citep{Cardelli89} and 12 of 17 stars fall in the range 2.89--3.29. We adopt 3.10 for HD\,187929 for which we found no $R_{\rm V}$ value in the literature. 

Of the $R_{\rm V}$ outliers, HD\,147084 and HD\,111613 have high values. \citet{Whittet80} report a mean $R_{\rm V} =$ 3.9 for the Sco-Cen association, of which both these stars are members (Upper Sco and Lower Cen-Cru respectively; \citeauthor{Whittet80}'s measurement for HD\,147084 is 3.82). So, the adopted values are reasonable. The other two outliers, HD\,210121 and HD\,161471, have lower values. Both are amongst the closer high polarization standards at about 450 and 600 pc respectively (parallax values, as currently listed in SIMBAD, are given in Table \ref{tab:redden} for interest's sake; these values are not used in our calculations). We might expect nearer stars to typically have less extinction since the diffuse Local Hot Bubble will make up more of the sight line to them, but many of the other standards are just as close. The determination for HD\,210121 is based on two recent measurements from reliable sources (that are both low), so is likely to be robust. 

On the other hand, the only $A_{\rm V}$ value for HD\,161471 comes from \citet{Luck14}, whose $E_{\rm (B-V)}$ value of 0.47 we had to exclude as an outlier. We therefore examined the extinction maps of \citet{Lallement19}, which indicate lower than typical extinction on HD\,161471's sight line, graphically integrating the extinction density along that sight line, while imprecise, gives answers consistent with \citet{Luck14}'s $A_{\rm V}$ determination. 

\vspace{36pt}

%\pagebreak
\section{Modulator Re-calibration}
\label{apx:mod_recal}

\subsection{Original calibration efficiency check}

\begin{table}
\caption{Initial Efficiency Check}
\tabcolsep 4.5 pt
\centering
\begin{tabular}{lllcr}
\toprule
Modulator Manufacturer      &   Desig.  &  &$p_{\rm obs}/p_{\rm pred}$  &   $N_o$   \\
\midrule
Meadowlark                  &  ML$^\dagger$ &           &   0.945       &   198 \\
Boulder Nonlinear Systems   &  BNS*     &               &   1.049       &   60  \\
Micron Technologies         &  MT       & HIPPI         &   1.026       &   6   \\
                            &           & Mini-HIPPI    &   0.955       &   140 \\
                            &  \textit{MTE3$^\ddagger$}     & \textit{HIPPI-2}       &   \textit{1.002}       &   \textit{13}   \\
\bottomrule
\end{tabular}
\begin{flushleft}
\underline{Notes} -- * All BNS performance Eras combined. $\dagger$ Calculated prior to 2023APR run. $\ddagger$ Calculated prior to N2023FEB run.
\end{flushleft}
\label{tab:eff_chk}
\end{table}

In this work we aim to assess the long term variability of established polarization standard stars as observed by HIPPI-class instruments. A number of different modulators, with different characteristics, have been used with these instrument. Consequently, we carried out a preliminary analysis (Table \ref{tab:eff_chk}) to check the predicted polarization against our observations. The predictions depend on the literature adopted parameters of the stars. The determination of the observed polarization depends on our bandpass model, particularly the modulator efficiency. 

The data presented in Table \ref{tab:eff_chk} is derived from modulator calibrations presented in \citet{Bailey15, Bailey20} and \citet{Cotton22a} and from standard star data indicated in our past works \citep{Bailey15, Bailey17, Bailey20, Cotton22b, Bailey23}\footnote{Data for HD\,43384 and HD\,198478 has been taken with the ML and MT modulators (during era 3), but as we have not previously indicated favoured parameters nor used them for calibration purposes, those observations are not included in Table \ref{tab:eff_chk}.}. For this purpose only data taken with the `B' modulator in either the SDSS $g^\prime$ band or with no filter (Clear), which has a similar effective wavelength, has been used -- this is the data relevant to this study. Additionally, because these bands are used for $\theta$ calibration, there are many more such observations, and limiting this analysis to them is more robust for our purpose here.

In the Table \ref{tab:eff_chk} we have grouped together the different performance eras of the Boulder Nonlinear Systems (BNS) modulator, which each have a separate calibration but in some cases represent only a handful of $g^\prime$ or Clear observations. Runs with the Micron Technologies (MT) modulator have been divided by instrument, since these show a clear difference. Table \ref{tab:eff_chk} reveals a mismatch between the different modulators. The final line in the table represents data taken with the MT modulator with HIPPI-2 at MIRA, as descibed in \citet{Cotton22b} and subsequently, where our calibration method varied, as described in the next section (\ref{apx:recal_method}).

The efficiency differences represent minor sources of error in high precision observations involving small polarizations. Yet, if combining data on high polarization targets as in this work, the impact is significant. It motivated us to homogenise the standard data (Appendix \ref{apx:lit_homog}), and perform new modulator calibrations. 

\subsection{Recalibration methodology}
\label{apx:recal_method}

Ferro-electric Liquid Crystal (FLC) modulators, as used in HIPPI-class polarimeters, have an efficiency as a function of wavelength, $e(\lambda)$, that can be described by the terms: $\lambda_{\rm 0}$, the wavelength of peak efficiency; $e_{\rm max}$, the maximum efficiency; and $Cd$, the product of terms describing the crystal birefringence ($C$), and the plate thickness ($d$). These terms are related by one of two expressions\footnote{See the Appendix of \citet{Bailey20} for a derivation.}; near total polarization by, \begin{equation} e(\lambda) \approxeq \frac{e_{\rm max}}{2} \left ( 1 + \frac{1-\cos{2\pi\Delta/\lambda}}{3+\cos{2\pi\Delta/\lambda}} \right ), \label{eq:e_pHigh}\end{equation} and at smaller polarizations ($\lessapprox$10\%) by, \begin{equation} e(\lambda) \approxeq e_{\rm max} \left ( \frac{1-\cos{2\pi\Delta/\lambda}}{2} \right ), \label{eq:e_pLow}\end{equation} where in both equations, \begin{equation} \Delta = \frac{\lambda_{\rm 0}}{2} + Cd \left ( \frac{1}{\lambda^2}-\frac{1}{\lambda_{\rm 0}^2}\right ). \label{eq:Del}\end{equation}

In past work \citep{Bailey15, Bailey20} we have used lab based data derived from injecting a polarized source beam through broad and narrow-band filters and applying Eqn.\ \ref{eq:e_pHigh} to derive $e_{\rm max}$; subsequently Eqn.\ \ref{eq:e_pLow} was used to fit the other parameters, based on multi-band observations of high polarization standard stars. 

In \citet{Cotton22b} we switched to fitting $e_{\rm max}$ along with the other parameters to the observed data. This was possible because more multi-band standard data was taken specifically for this purpose. Whereas shorter runs prior to calibration did not allow such complete data sets in the other instances. We now have a greater amount of standard data and are in a position to re-evaluate the earlier calibrations based on a similar approach.

The filters used are fully described in \citep{Bailey15}, \citet{Bailey20} and \citet{Cotton22a}; they include SDSS $g^\prime$ and $r^\prime$, Johnson-Cousins $U$, $V$ (or a similar substitute), along with a 425~nm and 500~nm short pass filters (425SP and 500SP, respectively), and a 650~nm long pass filter (650LP). Each was paired with either the B or R PMTs described in Section \ref{sec:obs}. Note also that HIPPI used SDSS $g^\prime$ and $r^\prime$ filters manufactured by Omega Optics, whereas the other instruments used Astrodon Generation 2 equivalents. The two \mbox{HIPPI-2} instruments used different $V$ band filters, with the southern instrument using a filter with a typical Johnson-like profile, and the northern instrument using a square profiled filter approximating the band, labelled $V_p$ \mbox{(see \citealp{Cotton22a})}.

We use the Python \textsc{SciPy} routine curve\_fit to match the observed polarizations to predictions using our bandpass model, with $\lambda_{\rm 0}$, $Cd$ and $e_{\rm max}$ as the fit parameters. Previously we have error weighted the data, but here equal weighting (nominal 200 ppm error) is given to each observation in recognition of stellar variability likely being a greater source of error than shot noise.

\subsection{Analysis and adopted parameters}

\subsubsection{Meadowlark}

The original Meadowlark calibration (EA\footnote{Referred to as E1 in \citet{Bailey20}.}) was carried out with data from the Gemini run (N2018JUN) and two runs from early 2019 (2019FEB, 2019MAR). To check for any evolution of the modulator we first fit both the original data set (EA) and that collected since, broken down into two periods (EB and EC) for which we had sufficient multi-band data, with the results presented in Table \ref{tab:ML}. The EA determination of $Cd$ is different to that previously found from the lab based method \citep{Bailey20}, but fitting $e_{\rm max}$ makes the most difference; our lab-based method previously gave us a value close to 1, now we find only 0.936.

% E1 references in this paragraph were E1+ originally.
Table \ref{tab:ML} appears to show evolution of parameters from era EA to EB to EC. However, when we removed two $U$ band observations from the 2019FEB run and refit N2018JUN as E0 and everything else as (E1), a satisfying fit was obtained for E1. The $U$ band data has a significant position angle rotation compared to all the other bands, which almost certainly indicates rotation within the band, suppressing the observed polarization. With those two observations removed, that data is no longer such a good match for the N2018JUN run, which has a redder and quite different fit. The telescope polarization at Gemini North was very large at blue wavelengths and not well fit in position angle, which probably accounts for this. Alternatively, the reflectance of the mirrors with wavelength might not be so well determined, and could be wrapped into the modulator parameters.

\begin{table}
\caption{ML modulator parameters}
\tabcolsep 7 pt
\begin{tabular}{lrccc}
\toprule
Desig.    &  $N_o$  & $\lambda_{\rm 0}$&   $Cd$              &   $e_{\rm max}$  \\
          &       &   (nm)    &   ($\times10^7$ nm$^3$)    &                    \\
\midrule
\multicolumn{3}{l}{\textit{From \citet{Bailey20}.}} \\
EA       &       &   455.2 $\pm$ 1.9 &	2.677 $\pm$	0.103  & 1.000 \phantom{$\pm$ 0.000} \\
\midrule
\multicolumn{3}{l}{\textit{Initial fits.}} \\
EA       &   31  &   455.1 $\pm$ 1.4   & 2.035 $\pm$ 0.112 & 0.936 $\pm$ 0.010 \\
EB       &  180  &   449.8 $\pm$ 1.8   & 1.813 $\pm$ 0.089 & 0.937 $\pm$ 0.003 \\
EC       &   28  &   447.2 $\pm$ 1.9   & 1.506 $\pm$ 0.076 & 0.925 $\pm$ 0.005 \\
\midrule
\multicolumn{3}{l}{\textit{Adopted values.}} \\
E0       &   17  &   454.6 $\pm$ 1.9   & 2.121 $\pm$ 0.161 & 0.930 $\pm$ 0.014 \\
E1       &  220  &   447.2 $\pm$ 1.1   & 1.651 $\pm$ 0.053 & 0.935 $\pm$ 0.002 \\
\bottomrule
\end{tabular}
\begin{flushleft}
\underline{Notes} --  Eras E0 and E1 are split up differently to EA, EB and EC.\\
EA: N2018JUN, 2019FEB (incl. $U$ band), 2019MAR. \\ 
EB: 2019APR, 2019JUN, 2019JUL, 2019AUG, 2019OCT, 2019DEC, 2020FEB, 2020JUN, 2020DEC, 2021JAN, 2021FEB, 2021APR, 2021DEC, 2022MAR, 2022APR, 2022JUN, 2022JUN2, 2023APR, 2023MAY.\\
EC: 2023APR, 2023MAY. \\
E0: N2018JUN \\
\end{flushleft}
\label{tab:ML}
\end{table}

\subsubsection{Boulder Nonlinear Systems}

\begin{table}
\caption{BNS modulator parameters}
\tabcolsep 7.5 pt
\begin{tabular}{lrccc}
\toprule
Desig.    &  $N_o$  & $\lambda_{\rm 0}$&   $Cd$            &   $e_{\rm max}$  \\
        &       &   (nm)    &   ($\times10^7$ nm$^3$)   &                    \\
\midrule
\multicolumn{3}{l}{\textit{From \citet{Bailey20}.}} \\
E1  &   &   494.8 $\pm$ 1.6 &   1.738 $\pm$	0.060   &	0.977  \phantom{$\pm$ 0.000} \\
E2  &   &   506.3 $\pm$ 2.9	&   1.758 $\pm$ 0.116	&   0.977  \phantom{$\pm$ 0.000} \\
E3  &   &   512.9 $\pm$ 3.9	&   2.367 $\pm$ 0.177   &	0.977  \phantom{$\pm$ 0.000} \\
E4  &   &   517.5 $\pm$16.1 &	2.297 $\pm$ 0.924   &	0.977  \phantom{$\pm$ 0.000} \\
E5  &   &   546.8 $\pm$ 6.0	&   2.213 $\pm$ 0.261   &	0.977  \phantom{$\pm$ 0.000} \\
E6  &   &   562.7 $\pm$ 4.7	&   2.319 $\pm$ 0.193   &	0.977  \phantom{$\pm$ 0.000} \\
E7  &   &   595.4 $\pm$ 4.8	&   1.615 $\pm$ 0.145   &	0.977  \phantom{$\pm$ 0.000} \\
\midrule
\multicolumn{3}{l}{\textit{Initial fits.}} \\
E1  & 35    &   492.8 $\pm$ 1.5 &	1.956 $\pm$ 0.086   &	0.985 $\pm$ 0.006 \\
E2  & 21    &   501.0 $\pm$ 3.6 &	1.715 $\pm$ 0.229   &	0.976 $\pm$ 0.013 \\
E3  & 22    &   515.9 $\pm$ 2.1 &	2.187 $\pm$ 0.133   &	1.000 $\pm$ 0.010 \\
E4  & 9     &   525.0 $\pm$11.7 &   1.985 $\pm$ 0.503   &	1.000 $\pm$ 0.020 \\
E5  & 12    &   547.4 $\pm$ 5.9 &	1.954 $\pm$ 0.277	&   0.956 $\pm$ 0.018 \\
E6  & 11    &   564.0 $\pm$ 6.7 &   2.309 $\pm$ 0.262   &	0.986 $\pm$ 0.016 \\
E7  & 20    &   583.6 $\pm$ 3.8 &   1.920 $\pm$ 0.131   &	0.981 $\pm$ 0.008 \\
\midrule
\multicolumn{3}{l}{\textit{Adopted values*.}} \\
E1  & 35    &   492.3 $\pm$ 1.0 &	1.995 \phantom{$\pm$ 0.000} &   0.985 \phantom{$\pm$ 0.000}\\
E2  & 21    &   501.9 $\pm$ 3.2 &	1.995 \phantom{$\pm$ 0.000} &   0.985 \phantom{$\pm$ 0.000}\\
E3  & 22    &   516.6 $\pm$ 1.9 &	1.995 \phantom{$\pm$ 0.000} &   0.985 \phantom{$\pm$ 0.000}\\
E4  & 9     &   520.4 $\pm$ 3.6 &	1.995 \phantom{$\pm$ 0.000} &   0.985 \phantom{$\pm$ 0.000}\\
E5  & 12    &   551.3 $\pm$ 5.9 &	1.995 \phantom{$\pm$ 0.000} &   0.985 \phantom{$\pm$ 0.000}\\
E6  & 11    &   571.3 $\pm$ 4.0 &	1.995 \phantom{$\pm$ 0.000} &   0.985 \phantom{$\pm$ 0.000}\\
E7  & 20    &   582.3 $\pm$ 1.8 &	1.995 \phantom{$\pm$ 0.000} &   0.985 \phantom{$\pm$ 0.000}\\
\bottomrule
\end{tabular}
\begin{flushleft}
\underline{Notes} -- * $Cd$ and $e_{\rm max}$ based on error weighted average of initial fits. \\
E1: 2014AUG, 2015MAY, 2015JUN, 2015OCT, 2015NOV. \\
E2: 2016FEB, 2016JUN, 2016DEC, 2017JUN, 2017AUG. \\
E3: 2018JAN, 2018FEB, 2018MAR, 2018MAY. \\
E4: 2018JUL. \\
E5: 2018AUG 2018-08-16 -- 2018-08-23. \\
E6: 2018AUG 2018-08-24 -- 2018-08-27. \\
E7: 2018AUG 2018-08-29 -- 2018-09-02. \\
\end{flushleft}
\label{tab:BNS}
\end{table}

As previously reported \citep{Bailey20}, the performance of this modulator evolved over time, particularly rapidly during the 2018AUG run, after which it was removed from service. 

We initially fit the data as per the previously delineated performance eras, as shown in Table \ref{tab:BNS}. From this it was noted that only the $\lambda_{\rm 0}$ values were more than 1.5$\sigma$ from the error-weighted mean. Only a small number of stars were available for some of the Eras, reducing the robustness of the determinations, so it was decided to refit $\lambda_{\rm 0}$ for each, but keeping $Cd$ and $e_{\rm max}$ fixed to the error-weighted means. In general the reduced $\chi^2$ statistic was improved by this, which validates the approach. 

The adopted values result in an $e_{\rm max}$ that is greater than previously established for the BNS modulator. The opposite is true for the ML unit, as expected from Table \ref{tab:eff_chk}.

% To be updated.
\subsubsection{Micron Technologies}

\begin{table}
\caption{MT modulator parameters}
\tabcolsep 6.5 pt
\begin{tabular}{lrccc}
\toprule
Desig.    &  $N_o$  & $\lambda_{\rm 0}$&   $Cd$            &   $e_{\rm max}$  \\
        &       &   (nm)    &   ($\times10^7$ nm$^3$)   &                    \\
\midrule
\multicolumn{5}{l}{\textit{From \citet{Bailey15}* and \citet{Cotton22b}.}} \\
E0      &   &   505\phantom{.0} $\pm$ 5\phantom{.0} &   1.75\phantom{0} $\pm$ 0.05\phantom{0}   &   0.98\phantom{0 $\pm$ 0.000}\\
E3      &   &   502.6 $\pm$ 2.0 &   1.920 $\pm$ 0.101   &   0.977 $\pm$ 0.005   \\
\midrule
\multicolumn{5}{l}{\textit{Initial fits.}} \\
E0      & 7     &   502.9 \phantom{$\pm 0.0$} & 1.913 \phantom{$\pm$ 0.000} & 1.000 $\pm$ 0.009 \\
E1      & 36    &   494.9 $\pm$ 3.7 &   2.005 $\pm$ 0.235   &   0.928 $\pm$ 0.012   \\
E2      & 109   &   494.9 \phantom{$\pm 0.0$} & 2.005 \phantom{$\pm$ 0.000} & 0.908 $\pm$ 0.002 \\
E3      & 59    &   502.9 $\pm$ 2.1 &   1.913 $\pm$ 0.147   &   0.979 $\pm$ 0.005   \\
\midrule
\multicolumn{5}{l}{\textit{Secondary fits.}} \\
E1 \& E2 & 145  &   501.7 $\pm$ 2.1 & 1.730 $\pm$ 0.119 & 0.912 $\pm$ 0.005 \\
E0 \& E3 & 66   &   505.4 $\pm$ 2.1 & 1.815 $\pm$ 0.149 & 0.983 $\pm$ 0.006 \\
\midrule
\multicolumn{5}{l}{\textit{Adopted values$^\dagger$.}} \\
E1 \& E2 & 141  &   503.6 \phantom{$\pm 0.0$} & 1.747 \phantom{$\pm$ 0.000} & 0.917 $\pm$ 0.002 \\
E0 \& E3 & 39   &   503.6 \phantom{$\pm 0.0$} & 1.747 \phantom{$\pm$ 0.000} & 0.977 $\pm$ 0.003 \\
\bottomrule
\end{tabular}
\begin{flushleft}
\underline{Notes} -- * MT parameters based on observation were derived for \citet{Bailey20} but the lab based data from \citet{Bailey15} has been preferred until now. $\dagger$ $\lambda_{\rm 0}$ and $Cd$ based on error weighted average of secondary fits.\\
E0: 2014MAY.\\
E1: All UNSW runs. \\
E2: All Pindari runs. \\
E3: All MIRA OOS runs (N2022AUG, N2023FEB, N2023MAY to 2023-09-01). \\
\end{flushleft}
\label{tab:MT}
\end{table}

The Micron Technologies modulator is the oldest unit and has seen service on many hundreds of nights. It has operated reliably throughout that time. Yet Table \ref{tab:eff_chk} foreshadows the issue we investigate here -- divergent performance between the earliest (2014MAY) run and later Mini-HIPPI runs (both at UNSW and Pindari Observatory).

Our analysis is presented in Table \ref{tab:MT}. The difference between the prior fit to run N2022AUG (E3) and the initial fit results from a combination of the newly adopted standard star parameters and the addition of data from the N2023FEB and N2023MAY runs -- which are wholly in the $g^\prime$ band. We also refitted, but did not tabulate, just the N2022AUG run ($N_o=15$) to gauge the effect of the updated stellar parameters in isolation: $\lambda_{\rm 0} = 503.0 \pm 1.7$, $Cd = 1.949 \pm 0.113 \times10^{7} nm^{3}$, $e_{max} = 0.981 \pm 0.005$ -- within error of those reported in \citet{Cotton22b}.

Our analysis of other eras is complicated by the data available. Only $g^\prime$ standard observations were made during the Pindari observatory runs (E2). Only a single $r^\prime$ standard observation was made during the 2014MAY HIPPI run (E0), the other half dozen observations were either in $g^\prime$ or Clear. In general there is a dearth of multi-band data amongst these runs, and the situation is further complicated by the restricted number of standards available to the two small telescopes -- the smallest used at Pindari Observatory in particular. This means we cannot reliably fit $Cd$ nor $\lambda_{\rm 0}$ for era E0 nor E2 in isolation. Consequently, we initially fixed those values at those derived from the E1 (UNSW Mini-HIPPI runs) and E3 (MIRA HIPPI-2 runs) to gauge $e_{\rm max}$, as shown in Table \ref{tab:MT}. We fit E0 with both E3- (shown) and E1-seeds (not shown); $e_{\rm max}$ was similar but the E3 seed resulted in a $\chi^2$ half the value.

Mini-HIPPI used a Glan-Taylor prism (Thorlabs GT5-A; \citealp{Bailey17}), rather than the Wollaston prism employed by HIPPI/-2, the second beam of which is much less efficient\footnote{According to Thorlabs, some of the ordinary component escapes through the side-port along with all of the extraordinary component of the input beam, and for this reason they do not recommend utilising the secondary beam of their Glan-Taylor prism.}; this is not otherwise accounted for and is likely the reason for lower $e_{\rm max}$ for E1 and E2. The similarly high values of $e_{\rm max}$ for E0 and E3 indicates the modulator itself is unchanged.

Subsequently, we fit eras E1 and E2 together, and eras E0 and E3 together. These secondary fits with larger data sets showed no significant difference between them in terms of $\lambda_{\rm 0}$ and $Cd$. Therefore we determined their error-weighted averages and fixed those parameters to determine two values to adopt for $e_{\rm max}$ for the MT modulator, when used with either Mini-HIPPI or HIPPI/-2.

Assuming the primary Glan-Taylor beam to have the same polarization efficiency as each beam in the Wollaston prism therefore implies $e_{\rm max}$ for the secondary beam is 0.857. Or, put another way, the polarizing efficiency of the secondary beam is 87.7\% that of the primary beam -- in line with expectations.

\subsection{Accuracy Assessment}
\label{apx:accuracy}

\begin{table}
\caption{Deviation from literature by instrument and band}
\tabcolsep 2.5 pt
\begin{tabular}{lcrrccrrr}
\toprule
Filter  &   PMT &   $\lambda_{\rm eff}$ & $N_o$   &   \multicolumn{2}{c}{$p_{\rm obs}/p_{\rm pred}$}    &   \multicolumn{2}{c}{$|p_{\rm obs} - p_{\rm pred}|$} &    \multicolumn{1}{c}{$e_p$}\\
        &       &       &       &   $\bar{x}$   &   $\sigma$ &   \multicolumn{1}{c}{$\bar{x}$}   & \multicolumn{1}{c}{$\eta$}   & \multicolumn{1}{c}{$\bar{x}$}\\
        &       & (nm)  &       &      &    &   \multicolumn{1}{c}{(ppm)}   & \multicolumn{1}{c}{(ppm)}   & \multicolumn{1}{c}{(ppm)}\\
\midrule
\textit{HIPPI/-2} \\
$U$*    &   B                   & 382.1   &    2    & 0.833     & 0.037     &   4767    &   4767    &   95 \\   
425SP   &   B                   & 406.1   &   29    & 0.973     & 0.077     &   1661    &   1020    &   95 \\
500SP   &   B                   & 445.7   &   26    & 0.995     & 0.030     &    657    &    641    &   37 \\
$g^{\prime}$ &$\nicefrac{B}{R}$ & 471.5   &  294    & 0.995     & 0.027     &    576    &    472    &   27 \\
Clear   &   B                   & 484.6   &   15    & 1.011     & 0.041     &    604    &    566    &   28 \\
$V^\dagger/V_p$    &$\nicefrac{B}{R}$ & 540.2   &   16    & 0.977     & 0.045     &    855    &    475    &   45 \\
500SP$^\ddagger$   &   R        & 549.1   &    2    & 0.965     & 0.021     &    750    &   750     &   18 \\
$r^{\prime}$ & B                & 604.6   &   26    & 1.016     & 0.033     &    968    &    677    &   45 \\
$r^{\prime}$  & R               & 625.6   &   13    & 0.978     & 0.024     &    784    &    513    &   33 \\
425SP$^\ddagger$  &   R         & 712.2   &    3    & 0.894     & 0.040     &   3691    &   3274    &   45 \\
650LP   &   R                   & 727.4   &   11    & 1.017     & 0.026     &    984    &    697    &   30 \\
\midrule
\textit{Mini-HIPPI} \\
425SP   &   B                   & 399.6   &    2    &   1.013   &   0.045   &    868    &    868    &  196 \\
$g^\prime$ & B                  & 475.0   &  112    &   0.994   &   0.030   &    458    &    431    &   73 \\
Clear & B                       & 488.3   &   30    &   1.026   &   0.032   &    758    &    754    &   70 \\
$r^{\prime}$ & B                & 603.7   &    2    &   0.947   &   0.024   &   1195    &   1195    &  154 \\
\bottomrule
\end{tabular}
\begin{flushleft}
\underline{Notes} -- * Observations excluded from fits; large $\theta$ rotation likely indicates suppressed polarization. $\dagger$ Some observations made with a marked $V$ band filter. $\ddagger$ Filters have red leaks, see \citet{Bailey20}.
\end{flushleft}
\label{tab:acc}
\end{table}

Some measures of the correspondence between the standard observations and bandpass predictions after the modulator recalibration are given in Table \ref{tab:acc}. The agreement is best for $g^\prime$ observations and the bands nearest to it, owing to the much greater number of observations made in $g^\prime$. In aggregate the deviation from the predictions is no more than a few per cent\footnote{Here we mean the fractional difference given as a percentage.} in any of the regularly used bands, and typically less than one percent for HIPPI and HIPPI-2. These figures compare favourably to those presented for HIPPI-2 in \citet{Bailey20}. 

The greatest deviation occurs for $U$ band, which we excluded from our fitting routines because a position angle rotation indicated the polarization was likely suppressed -- the tabulated data supports this. The three observations made with the 425SP and 500SP filters (which both have a red leak we have characterised) with the R PMT are also in poor agreement. These were all made when the BNS modulator was evolving quickly, which probably explains this. Deficiencies in the components of the bandpass model will also show up here. For instance, the efficiency curves of the PMTs are based on manufacturer data and not well defined at the extremes of the wavelength range. Aside from the obvious ($U$, 425SP, 650LP) we also expect the $r^\prime$ band paired with the B PMT to be slightly less accurate, since it includes a large contribution from the edge of the B PMT's sensitivity range. 

The standard deviation in $p_{\rm obs}/p_{\rm pred}$ is typically a few percent; this is better than our previous calibration \citep{Bailey20} by a factor of two. Contributors to the scatter include inaccuracy and imprecision in the literature determinations -- which have only been reported to the nearest 100 ppm (0.01\%); see Table \ref{tab:std_serk} -- as well as instrumental effects and any intrinsic variability of the stars.

Some of our $V$ band observations are clearly discrepant, and this manifests in a larger standard deviation. Inspection of the filter revealed a small mark near its centre which probably explains this.

The standard deviation is noticeably larger for the 425SP band. For hot and/or luminous stars with significant electron scattering in their atmospheres a greater magnitude of intrinsic polarization is to be expected from some mechanisms. However, bluer wavelengths are also most impacted by seeing noise and airmass changes during an observation. In the case of the instruments used here, the comparative steepness of the modulator efficiency curves in the blue will also be a factor. For most standards $\lambda_{\rm max}$ is around 0.5 to 0.6 $\mu$m, which means inaccuracy in its determination will be magnified in the shortest wavebands. 

Mean and median values for the difference between $p_{\rm obs}$ and $p_{\rm pred}$ are also given for comparison to the nominal measurement errors in Table \ref{tab:acc}. These are only intended as representative, since this metric will be very sensitive to stars with larger polarizations, and the mix of stars is heterogeneous. However, it does show the scale of unaccounted for scatter is similar for HIPPI/-2 and Mini-HIPPI, despite utilising completely different telescopes and sites and, gives a rough indication of what level of improvement in standard determinations is desirable and/or what intrinsic variability might be present. In particular, it is noteworthy that the median disagreement in the most reliable central bands is about 460 ppm (the number weighted average of the two medians).

\section{Selected Symbols}
\label{apx:symbols}
\begin{tabular}{ll}
Symbol & Description \\
\toprule
\multicolumn{2}{l}{\textit{Stokes parameters and related quantities}} \\
$I$ & Intensity \\
$Q$, $U$ & Linear Stokes parameters, Equatorial frame \\
$q$, $u$ & Normalized linear Stokes parameters (NLSP) \\
$q_i$, $u_i$ & NLSP in instrument reference frame \\
$q^\prime$, $u^\prime$ & NLSP rotated such that $q^\prime$ is in the direction of $p$ \\
$p$ & Polarization magnitude \\
$p_{\rm lit}$ & Literature (nominal) value for $p$ \\
$p_{\rm obs}$ & Observed $p$ value \\
$p_{\rm pred}$ & Predicted $p$ value \\
$\Delta p$ & $p_{\rm pred} - p_{\rm obs}$ \textit{or} change from earlier/literature value\\
$p_{\rm ISM}$ & Interstellar polarization \\
$\theta$ & Polarization position angle measured N over E \\
$\theta_{\rm lit}$ & Literature (nominal) value for $\theta$ \\
$\theta_{\rm obs}$ & Observed $\theta$ value \\
$\theta_{\rm pred}$ & Predicted $\theta$ value \\
$\theta_{\rm diff}$ & Difference to arbitrary zero \\
$\theta_t$ & Telescope position angle \\
$\theta_0$ & Reference axis position angle (N for Equ.) \\ 
$\Delta\theta$ & Change in $\theta$ relative to literature/earlier value \\
$\Delta\theta_t$ & Change in $\theta_t$ relative to past calibration \\
$\zeta$ & Offset between arbitrary reference frame and $\theta_0$ \\
\midrule
\multicolumn{2}{l}{\textit{Errors, averages and uncertainties, etc.}} \\
$\bar{x}$ & Average of x, where x is $q$, $u$, $\theta$, $p$, etc. \\
$e_{x}$ & Error in x \\
$e_{\rm m}$ & Measured internal statistical error \\
$e_{\rm i}$ & Error in instrumental/telescope set-up \\
$e_{\rm t}$ & Error in telescope zero point \\
$e_\star$ & Error associated with stellar variability \\
$\sigma_x$ & [Weighted] standard deviation of x \\
$\eta(x)$ & Median of x \\
$N_o$ & Number of observations \\
$N_r$ & Number of subruns \\
$N_S$ & Number of sets (runs w/ multiple repeat obs.) \\
%\midrule
%\multicolumn{2}{l}{\textit{Serkowski Law and reddening}} \\
%$p_{\rm max}$ & Maximum polarization \\
%$\lambda_{\rm max}$ & Wavelength of maximum polarization \\
%$K$ & Serkowski constant \\
%$R_V$ & normalized extinction \\
%$A_V$ & 
\bottomrule
\end{tabular}

% Don't change these lines
\bsp	% typesetting comment
\label{lastpage}
\end{document}